\documentclass[floatfix,twocolumn,showpacs,preprintnumbers,amsmath,amssymb,superscriptaddress,longbibliography,aps,pre]{revtex4-1}
\usepackage{color}
\usepackage[usenames,dvipsnames,svgnames,table]{xcolor}
\usepackage[colorlinks=true,linkcolor=blue,urlcolor=blue,citecolor=blue]{hyperref}

\usepackage{amssymb}

\usepackage{amsmath}

\usepackage{mathtools}
\usepackage{graphicx}
\usepackage{dcolumn}
\usepackage{array}
\usepackage{lipsum}
\usepackage{bm}
\usepackage{subfigure}
\usepackage{multirow}
\usepackage{tabularx}
\usepackage{braket}
\graphicspath{{plots/}}

\begin{document}
%\underline{Draft version:}
\title{
Attenuating the fermion sign problem in path integral Monte Carlo simulations using the Bogoliubov inequality and thermodynamic integration
}

\author{Tobias Dornheim}
\email{t.dornheim@hzdr.de}

\affiliation{Center for Advanced Systems Understanding (CASUS), D-02826 G\"orlitz, Germany}

\author{Michele Invernizzi}
\email{michele.invernizzi@phys.chem.ethz.ch}
\affiliation{Institute of Computational Sciences, Universit\`a della Svizzera italiana, 6900 Lugano, Switzerland}

\affiliation{National Centre for Computational Design and Discovery of Novel Materials MARVEL, Universit\`a della Svizzera italiana, 6900 Lugano, Switzerland}

\affiliation{Department of Physics, ETH Zurich, 8092 Zurich, Switzerland}

\author{Jan Vorberger}

\email{j.vorberger@hzdr.de}

\affiliation{Helmholtz-Zentrum Dresden-Rossendorf (HZDR), D-01328 Dresden, Germany}

\author{Barak Hirshberg}
\email{barakh@ethz.ch}

\affiliation{Institute of Computational Sciences, Universit\`a della Svizzera italiana, 6900 Lugano, Switzerland}

\affiliation{Department of Chemistry and Applied Biosciences, ETH Zurich, 8092 Zurich, Switzerland}

\begin{abstract}
Accurate thermodynamic simulations of correlated fermions using path integral Monte Carlo (PIMC) methods are of paramount importance for many applications such as the description of ultracold atoms, electrons in quantum dots, and  warm-dense matter. The main obstacle is the fermion sign problem (FSP), which leads to an exponential increase in computation time both with increasing the system-size and with decreasing temperature. Very recently,
Hirshberg et al.~[\textit{J.~Chem.~Phys.}~\textbf{152}, 171102 (2020)] have proposed to alleviate the FSP based on the Bogoliubov inequality.
In the present work, we extend this approach by adding a parameter that controls the perturbation, allowing for an extrapolation to the exact result. In this way, we can also use thermodynamic integration to obtain an improved estimate of the fermionic energy.
As a test system, we choose electrons in $2D$ and $3D$ quantum dots and find in some cases a speed-up exceeding $10^6$, as compared to standard PIMC, while retaining a relative accuracy of $\sim0.1\%$.
Our approach is quite general and can readily be adapted to other simulation methods.

\end{abstract}

\maketitle

\section{Introduction\label{sec:introduction}}

The accurate estimation of electronic properties is of paramount importance for many fields such as quantum chemistry, physics, and material science~\cite{quantum_theory}. The most accurate results can be obtained using quantum Monte Carlo (QMC) methods which, in principle, allow for a \emph{quasi-exact} description.
Unfortunately, QMC simulations of fermions are severely hampered by the notorious fermion sign problem (FSP)~\cite{Loh_sign_problem_PRB_1990,troyer,dornheim_sign_problem}, which leads to an exponential increase in the computation time with increasing the system-size or decreasing temperature, and has been shown to be $NP$-hard for a specific class of Hamiltonians~\cite{troyer}.

In the ground-state, the seminal QMC study of the uniform electron gas by Ceperley and Alder~\cite{ceperley_alder_UEG} has facilitated the  success of density functional theory (DFT) regarding the description of real materials~\cite{perdew_zunger_PRB,perdew1996generalized,Burke_perspective_JCP_2012}. These results were obtained on the basis of the fixed-node approximation~\cite{Ceperley1991,Anderson_fixed_node}, where the sign problem is avoided by an \emph{a priori} decomposition of the wave-function into a positive and a negative region. Although formally exact, the true nodal structure of the wave function is not known, and one has to rely on approximations. This limitation, however, can be alleviated as the ground-state energy is variational with respect to the nodes, which can be exploited for optimization~\cite{Foulkes_QMC_RMP,backflow,Needs_2009}.
At the same time, there is a broad consensus among the QMC community that the fixed-node approximation has severe limitations in many cases, and alternative methods~\cite{Booth_FCIQMC_2009,Auxiliary,Auxiliary2,PhysRevX.5.041041} are desirable~\cite{Booth2013}.

In addition, a surge of activity has recently emerged in the field of fermionic QMC simulations at finite temperature~\cite{Brown_ethan,Schoof_CPP_2011,Dornheim_NJP_2015,dornheim_POP,Blunt_PRB_2014,Rubenstein_auxiliary_finite_T,Malone_JCP_2015,Malone_PRL_2016,Clark_PRB_2017,Dornheim_CPP_2019,yilmaz2020restricted,2020arXiv200403229D,Driver_PRE_2018,dornheim_dynamic}. This has been motivated mainly by interest in warm dense matter (WDM)---an exotic state at the interface of plasma and solid state physics~\cite{wdm_book,new_POP,review,fortov_review}. For example, thermal DFT simulations~\cite{Pribram-Jones2014, Smith2018, Cytter2018} of WDM require the construction of exchange--correlation functionals that explicitly take into account the temperature~\cite{karasiev_importance,kushal}, which can be realized on the basis of QMC data for electrons at these conditions~\cite{dornheim_prl,groth_prl}. See Ref.~\cite{review} for a review  on recent developments.

Other fields for the application of fermionic QMC methods include dipolar systems such as ultracold atoms or Rydberg dressed states~\cite{Dornheim_PRA_2020,dornheim_sign_problem}, bilayer-systems~\cite{dynamic_Alex_2,Schleede_bilayer_2012}, electrons in quantum dots~\cite{Dornheim_CPP_2016,Ilkka_PRB_2017,Egger_PRL_1999,Egger_2001,Filinov_PRL_2001}, and even semi-relativistic quark-gluon plasmas~\cite{Filinov_PRC_2013,Filinov_quark_2015}. These systems offer a plethora of interesting effects such as an abnormal superfluid fraction~\cite{Blume_PRL_2014,Dornheim_PRA_2020}, Wigner crystallization~\cite{Filinov_PRL_2001,Filinov_crystal2}, the BCS-BEC transition~\cite{Fehske_BCS_2012,BCS_2002}, and collective excitations~\cite{dornheim_dynamic,dynamic_folgepaper,dornheim_HEDP,hamann2020dynamic}.

Despite this progress, there are still many thermodynamic conditions that are not accessible to QMC methods~\cite{yilmaz2020restricted,status} and their development remains an active topic of research. 
In this work, we present an extension of the standard path-integral Monte Carlo (PIMC) method~\cite{cep} which is motivated by the behavior of the sign for different interaction potentials and is justified by the well-known Bogoliubov inequality~\cite{verbeure2010many}. More specifically, Hirshberg et al.~\cite{Hirshberg_JCP_2020} have recently proposed to carry out a path-integral Molecular Dynamics (PIMD) simulation~\cite{Hirshberg_PIMD} of an auxiliary system where the FSP is less severe and obtain an accurate estimate for the energy of a computationally more challenging system using the Bogoliubov inequality. Here, we adapt this idea to the PIMC method. We also extend this approach by adding to the original system a repulsive two-body term that phenomenologically mimics the effect of the Pauli repulsion between fermions and allows for a controlled extrapolation towards the exact result.
Moreover, we show that it is possible to accurately estimate the energy difference between the original and the auxiliary system using thermodynamic integration~\cite{FrenkelBook}, which further increases the reliability of the method. Our approach results in a speed-up of up to $10^6$ as compared to standard PIMC, while retaining a relative accuracy of $\sim0.1\%$, which is fully sufficient for practical applications.

The paper is organized as follows: In Sec.~\ref{sec:theory}, we introduce the theoretical background including the PIMC method and the related FSP (Sec.~\ref{sec:PIMC}), our approach and how it can be justified from the Bogoliubov inequality (\ref{sec:bogoliubov}), and the possibility of obtaining accurate estimates from thermodynamic integration (\ref{sec:invernizzi}). In Sec.~\ref{sec:results}, we present extensive results for electrons in two-dimensional (2D) quantum dots~\cite{Egger_2001}, starting with a brief introduction of the model Hamiltonian (\ref{sec:Hamiltonian}). We investigate in detail the extrapolation of an auxiliary system, where the sign problem is substantially less severe, to the original system of interest in Sec.~\ref{sec:extrapolation}. The method is further improved by using thermodynamic integration in Sec.~\ref{sec:invernizzi_results}. Finally, we briefly extend our considerations to electrons in a $3D$ harmonic trap in Sec.~\ref{sec:3D_results}.
The paper is concluded by a concise summary and discussion (Sec.~\ref{sec:summary}), where we also indicate possible future directions.

\section{Theory\label{sec:theory}}

\subsection{Path-integral Monte Carlo\label{sec:PIMC}}

The basic idea of the PIMC method~\cite{cep} is to stochastically sample the thermal density matrix of the canonical ensemble
\begin{eqnarray}\label{eq:density_matrix}
\rho(\mathbf{R},\mathbf{R'},\beta) = \bra{\mathbf{R}} e^{-\beta\hat H} \ket{\mathbf{R'}}\ ,
\end{eqnarray}
where $\mathbf{R}=(\mathbf{r}_1,\dots,\mathbf{r}_N)^T$ contains the coordinates of all $N$ particles, $\beta=(k_\textnormal{B}T)^{-1}$ is the inverse temperature and $\hat H$ denotes the Hamiltonian.  The path-integral expression is obtained by performing a Trotter decomposition~\cite{Trotter}, leading to each particle being expressed as an entire path at $P$ discrete positions in imaginary-time $\tau\in[0,\beta]$. The collection of the paths of all $N$ particles is known as a configuration $\mathbf{X}=(\mathbf{R}_0,\dots,\mathbf{R}_{P-1})^T$. Each configuration contributes to the full partition function according to its corresponding weight $W(\mathbf{X})$, which is a function that can be readily evaluated~\cite{Hirshberg_JCP_2020},
\begin{eqnarray}\label{eq:Z}
Z = \int \textnormal{d}\mathbf{X}\ W(\mathbf{X})\ .
\end{eqnarray}
In practice, one uses the metropolis algorithm~\cite{metropolis} to generate a Markov chain of configurations $\mathbf{X}$ which are distributed as $P(\mathbf{X})=W(\mathbf{X})/Z$.

For indistinguishable particles, one has to explicitly sum over all possible permutations of particle coordinates~\cite{dornheim_permutation_cycles}. For bosons, the thermal density matrix is symmetric under the exchange of particle coordinates, and all terms remain positive. Thus, modern sampling algorithms~\cite{boninsegni1,boninsegni2} allow for quasi-exact simulations of up to $10^4$ particles, which has facilitated profound insights into phenomena such as superfluidity~\cite{Filinov_PRL_2010,Boninsegni_supersolid,dornheim_superfluid,Pollet_PRL_superfluid} and collective excitations~\cite{Boninsegni1996,Filinov_PRA_2012,dornheim_dynamic,Dornheim_Vorberger_finite_size_2020,dynamic_Alex_2,Boninsegni_PRB_2018}. Recently, it became possible to simulate large bosonic systems also using PIMD~\cite{Hirshberg_PIMD}.

For fermions, on the other hand, the density matrix is anti-symmetric under particle-exchange, which leads to sign changes in  $W(\mathbf{X})$ for each pair exchange. Therefore, $P=W/Z$ cannot be interpreted as a probability distribution.
At this point, one introduces a modified partition function
\begin{eqnarray}\label{eq:z_prime}
Z' = \int \textnormal{d}\mathbf{X}\ |W(\mathbf{X})|\ ,
\end{eqnarray}
where the configurations are generated according to the absolute value of $W(\mathbf{X})$, i.e., $P'(\mathbf{X})=|W(\mathbf{X})|/Z'$.
The exact fermionic expectation value of an observable $\hat A$ is then computed as
\begin{eqnarray}\label{eq:ratio}
\braket{\hat A} = \frac{\braket{\hat A \hat S}'}{\braket{S}'}\ ,
\end{eqnarray}
where $S(\mathbf{X})=W(\mathbf{X})/|W(\mathbf{X})|$ denotes the sign associated with a particular configuration, and the denominator of Eq.~(\ref{eq:ratio}) is the so-called
average sign $S$.

$S$ is a  measure for the amount of cancellation of positive and negative terms in $Z$, and exponentially decreases both with the system-size $N$ and the inverse temperature $\beta$,
\begin{eqnarray}\label{eq:sign}
S = \textnormal{exp}\left(
-\beta N (f-f')
\right)\ ,
\end{eqnarray}
where $f$ and $f'$ denote the free energy density of the original and modified system, respectively. Furthermore, the statistical error in estimating the ratio in Eq.~(\ref{eq:ratio}) is inversely proportional to $S$,
\begin{eqnarray}\label{eq:exponential}
\frac{\Delta A}{A}\sim \frac{1}{S\sqrt{N_\textnormal{MC}}}\sim \frac{\textnormal{exp}\left(\beta N (f-f')\right)}{\sqrt{N_\textnormal{MC}}}\ .
\end{eqnarray}
The resulting exponential increase in the Monte Carlo error bar with increasing $N$ or decreasing temperature can only be compensated for by increasing the number of  samples $N_{MC}$. This inevitably becomes unfeasible and one runs into an exponential wall, which is known as the fermion sign problem~\cite{dornheim_sign_problem,Loh_sign_problem_PRB_1990,troyer}. Methods to overcome the FSP are therefore very desirable. In the following two sections, we describe two approaches for alleviating the FSP in PIMC simulations.

\subsection{Extrapolation based on the Bogoliubov inequality\label{sec:bogoliubov}}

Let $\hat H$ denote the original Hamiltonian that we want to simulate using fermionic PIMC,
\begin{eqnarray}
\hat H = \hat K + \hat V_\textnormal{ext} + \hat W\ ,
\end{eqnarray}
with $\hat K$, $\hat V_\textnormal{ext}$ and $\hat W$ being the kinetic, external potential, and interaction contribution to the total energy.
We further assume that we are interested in the properties of this system at relatively low temperature, and that the manifestation of quantum degeneracy effects results in a low value of the average sign $S$. In a recent paper, Hirshberg et al.~\cite{Hirshberg_JCP_2020} have shown that it is possible to accurately approximate the energy $E_{\hat H}=\braket{\hat H}$, by simulating an auxiliary system where $\hat W$ is replaced by a different pair potential $\hat R$ that more effectively separates the particles,
\begin{eqnarray}\label{eq:HR}
\hat H_R &=& \hat K + \hat V_\textnormal{ext} + \hat R .
\end{eqnarray}
This resulted in a substantially less severe manifestation of the sign problem~\cite{dornheim_sign_problem}, and simulations became feasible at lower temperatures than for the original system. Then, they used the Bogoliubov inequality~\cite{verbeure2010many}
\begin{eqnarray}\label{eq:bogoliubov}
F_{\hat H} - F_{\hat H_R} \leq \braket{\hat H - \hat H_R}_{\hat H_R}\ ,
\end{eqnarray}
and assumed that the free energy can be approximated by the energy at low temperatures, to obtain an upper bound on the energy of the original system 
\begin{eqnarray}\label{eq:bound}
E_{\hat H} \lesssim \braket{\hat H}_{\hat H_R} \ .
\end{eqnarray}
The subscript indicates that the expectation value of the original Hamiltonian is evaluated in the ensemble of $\hat H_R$.
Since the sign problem is most severe at low temperature [cf.~Eq.~(\ref{eq:exponential})], this approximation is expected to hold, and the scheme is highly valuable as $R$ could in principle be optimized variationally.

In the present work, we extend this approach in terms of a coupling parameter $\eta$, by re-writing Eq.~(\ref{eq:HR}) as
\begin{eqnarray}\label{eq:Hamiltonian_eta}
\hat H_\eta = \hat H + \eta \hat\phi \ ,
\end{eqnarray}
where $\hat\phi$ is a pair potential that should mimic the effective repulsion due to the fermionic degeneracy, as discussed at the end of this section.
Clearly, the PIMC energies computed for the Hamiltonian in Eq.~(\ref{eq:Hamiltonian_eta}) are $\eta$-dependent, and for any  observable $\hat A$, it holds that
\begin{eqnarray}
\braket{\hat A}=\lim_{\eta\to0} \braket{\hat A}_\eta = \lim_{\eta\to0} \frac{\braket{\hat A\hat S}_\eta}{\braket{\hat S}_\eta}\ .
\end{eqnarray}
 This results in two key advantages: i) The difference between $E_{\hat H}$ and $\braket{\hat H}_{\eta}$ vanishes as $\eta\to0$ and the energy of the original system can be obtained by extrapolation. ii) The energy difference between the original and auxiliary systems can be readily estimated from the PIMC data using thermodynamic integration, as described in the next section.
 
 In practice, we use the modified Hamiltonian with the additional repulsive term to carry out fermionic PIMC calculations for various values of $\eta$. Due to the added repulsion, the simulations converge faster as $\eta$ is increased. Then, we evaluate $E(\eta) \equiv \braket{\hat H}_{\eta}$ for each one and extrapolate it towards $\eta \to 0$ where the simulations are not feasible. This should converge to the exact result from above, at least at low temperatures when the neglect of entropic contributions is justified.
It is important to note that extrapolation of QMC results is a notoriously difficult task, see e.g. Refs.~\cite{Groth_PRB_2016,sign_blessing,Malone_PRL_2016} for three examples from alternative QMC methods for fermions. Thus, any additional information about the functional behavior of $\braket{\hat A}_\eta$ with respect to $\eta$ is highly valuable. We show in Sec.~\ref{sec:extrapolation} that a simple empirical extrapolation scheme works well and the results are not very sensitive to the range of $\eta$ used in the fitting.

Throughout this work, we restrict ourselves to two-body correlations and write $\hat\phi$ as the sum over an effective pair potential,
\begin{eqnarray}
\hat\phi = \frac{1}{2}\sum_{k\neq l}^N \Psi(\hat{\mathbf{r}}_l,\hat{\mathbf{r}}_k)\ .
\end{eqnarray}
Following an observation from Ref.~\cite{dornheim_sign_problem} for electrons in a $2D$ harmonic confinement, we choose a dipolar short-range repulsion~\cite{Dornheim_PRA_2020},
\begin{eqnarray}\label{eq:dipole}
\Psi(\mathbf{r}_1,\mathbf{r}_2) = \frac{1}{|\mathbf{r}_2-\mathbf{r}_1|^3}\ ,
\end{eqnarray}
which was shown to have a higher average sign in comparison to Coulomb repulsion between the fermions. This choice for electrons in quantum dots guarantees that the fermions feel the Coulomb repulsion at long range, but that the repulsion due to the dipolar interaction is dominant at short range. This additional short range term is what mimics an increased Pauli repulsion and results in a larger average sign. We speculate that any short-range potential that is more repulsive than Coulombic interaction at small separations would be appropriate.

%A second choice from physical intuition is given by a simple Gaussian of the form
%\begin{eqnarray}\label{eq:gaussian}
%R(\mathbf{r}_1,\mathbf{r}_2) = \frac{1}{\sqrt{2\pi}c\lambda_\beta} \textnormal{exp}\left(
%-\frac{(\mathbf{r}_2-\mathbf{r}_1)^2}{2(c\lambda_\beta)^2}
%\right) \ ,
%\end{eqnarray}
%where the variance $\sigma$ is given by the thermal wavelength scaled by the free parameter $c$.

%The third type of artificial potential considered in this work is inspired by the exact solution of the two-body problem in the noninteracting case. More specifically,
%we introduce the ancilla function \textcolor{red}{maybe an illustrative figure?}
%\begin{eqnarray}\label{eq:ancilla}
%A(\mathbf{r}_1,\mathbf{r}_2) = \rho_0(\mathbf{r}_1,\mathbf{r}_1,\beta)^2-\rho_0(\mathbf{r}_1,\mathbf{r}_2,\beta)^2\ ,
%\end{eqnarray}
%corresponding to the antisymmetrized two-body density matrix of two noninteracting free fermions.
%Eq.~(\ref{eq:ancilla}) then directly yields the two-body potential
%\begin{eqnarray}\label{eq:DetPot}
%R(\mathbf{r}_1,\mathbf{r}_2) = - \frac{1}{\epsilon}\textnormal{log} A(\mathbf{r}_1,\mathbf{r}_2)\ ,
%\end{eqnarray}
%which we denote as \emph{determinant potential} in the following.

\subsection{Thermodynamic integration\label{sec:invernizzi}}
As an alternative to direct extrapolation, the free energy difference between the original and auxiliary system can be evaluated using thermodynamic integration, a widely used free-energy method for atomistic simulations~\cite{FrenkelBook}.
Contrary to the Bogoliubov inequality that leads to an upper bound [cf.~Eq.~(\ref{eq:bogoliubov})], using thermodynamic integration we obtain an equality.
Given a Hamiltonian of the form of Eq.~(\ref{eq:Hamiltonian_eta}), the difference in the free energy can be estimated as
\begin{eqnarray}\label{eq:delta_F}
F_{\hat H} - F_{\hat H_\eta} = - \int_0^\eta \textnormal{d}\eta'\ \braket{\hat\phi}_{\eta'}\ .
\end{eqnarray}

%Again, we note that $F=E-TS$ (with $S$ being the entropy), which  gives
%\begin{eqnarray}\label{eq:invernizzi_estimate}
%E_{\hat H} &=& E_{\hat H_\eta} - \int_0^\eta \textnormal{d}\eta'\ \braket{\hat\phi}_{\eta'}\\ \nonumber & &+ T\left(
%S_{\hat H}-S_{\hat H_\eta}
%\right)\ .
%\end{eqnarray}
Assuming again that at low temperatures the free energy can be approximated by the energy, we obtain
\begin{eqnarray}\label{eq:invernizzi_estimate}
E_{\hat H} &\approx& E_{\hat H_\eta} - \int_0^\eta \textnormal{d}\eta'\ \braket{\hat\phi}_{\eta'}\ .
\end{eqnarray}
The integral in Eq.~(\ref{eq:invernizzi_estimate}) can be estimated from our PIMC simulation data at different values of $\eta$, whereas the entropic contribution, hereafter denoted as $\Delta S(\eta)$, remains unknown. Still, it is reasonable to assume that the integral term constitutes the dominant contribution at low temperatures. This is precisely where the FSP is most severe and, consequently, our approach is needed the most.
Finally, we notice that also when using thermodynamic integration we need to perform an extrapolation for $\eta=0$, but this time only of the perturbing potential $\phi$ and not of the whole energy as when using the Bogoliubov inequality.
An extensive discussion of the practical aspects regarding the application of Eq.~(\ref{eq:invernizzi_estimate}) is given in in Sec.~\ref{sec:invernizzi_results}.

\section{Results\label{sec:results}}

\subsection{Model system and speed-up factor\label{sec:Hamiltonian}}
We consider the Hamiltonian of $N$ spin-polarized electrons in a harmonic confinement, a commonly employed model for quantum dots, 
\begin{eqnarray}\label{eq:Hamiltonian_trap}
\hat H = - \frac{1}{2} \sum_{k=1}^N \nabla_k^2 + \frac{1}{2} \sum_{k=1}^N \mathbf{\hat r}_k^2 + \sum_{k>l}^N \frac{ \lambda }{ |\mathbf{\hat r}_l - \mathbf{\hat r}_k| } \quad ,
\end{eqnarray}
where we assume oscillator units, corresponding to the characteristic length $l_0=\sqrt{\hbar/m\Omega}$ (with $\Omega$ being the trap frequency) and energy scale $E_0=\hbar\Omega$. The first term corresponds to the kinetic contribution $\hat K$ and the last two terms to the external potential and the Coulomb interaction, $\hat V_\textnormal{ext}$ and $\hat W$, respectively. The constant $\lambda$ is the ratio between the screened Coulomb repulsion in the quantum dot and $E_0$~\cite{Ellenberger96}. All simulation results in this work have been obtained for strictly two- and three-dimensional systems.

All the PIMC results in this work have been obtained using a canonical adaption~\cite{mezza} of the worm algorithm presented in Refs.~\cite{boninsegni1,boninsegni2}. We use a primitive factorization of the density matrix (see Refs.~\cite{Sakkos_JCP_2009,Brualla_JCP_2004} for a detailed discussion of different factorization schemes) with $P\in[200,500]$ imaginary-time propagators. This is sufficient for convergence within the respective error bars, see the appendix of Ref.~\cite{dornheim_sign_problem} for a practical demonstration for a similar system. 

 In the following, we will refer to the results for $\eta = 0$, obtained using Eq.~\ref{eq:Hamiltonian_trap}, as standard PIMC. They will be compared with results obtained with an added dipolar repulsion, given by Eq.~\ref{eq:dipole}, for different values of the coupling parameter $\eta$.
For each of the studied systems, we report the speed-up factor $T(\eta)$ obtained with respect to a standard PIMC calculation.
This speed-up factor follows directly from Eq.~(\ref{eq:exponential}) and is defined as
\begin{eqnarray}\label{eq:speed-up}
T(\eta) = \left(
\frac{S(\eta)}{S(\eta=0)}
\right)^2 \ ,
\end{eqnarray}
where $S(\eta)$ is the average sign obtained with the additional repulsive term in Eq.~(\ref{eq:Hamiltonian_eta}).
For example, if $S(\eta)$ is ten times larger than $S(\eta=0)$, then we need two orders of magnitude less Monte-Carlo samples $N_\textnormal{MC}$ to achieve the same level of statistical uncertainty, i.e., $T(\eta)=100$. 

\subsection{Direct extrapolation scheme\label{sec:extrapolation}}

\begin{figure}\centering
\includegraphics[width=0.415\textwidth]{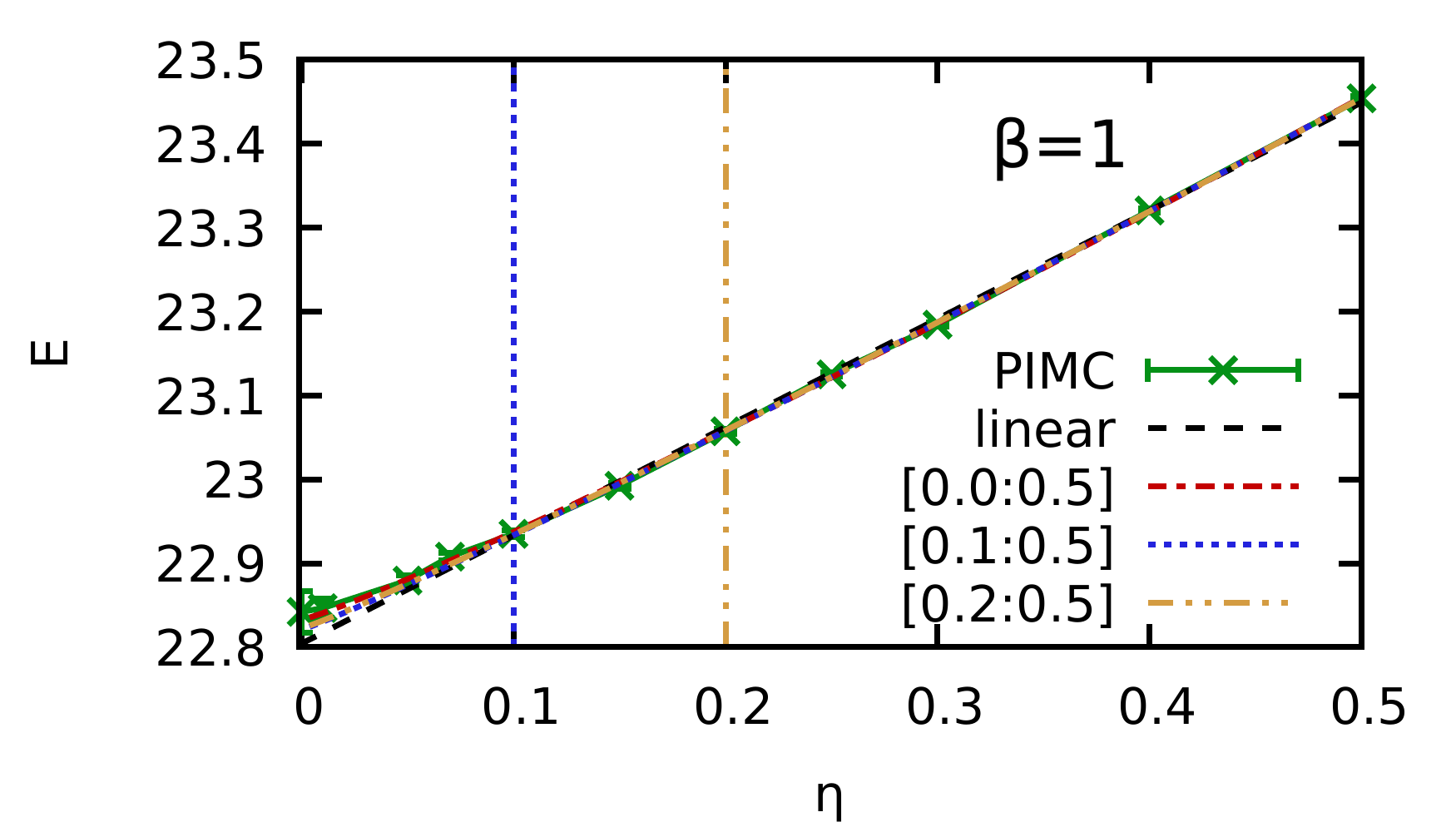}\\\vspace*{-0.8cm}\includegraphics[width=0.415\textwidth]{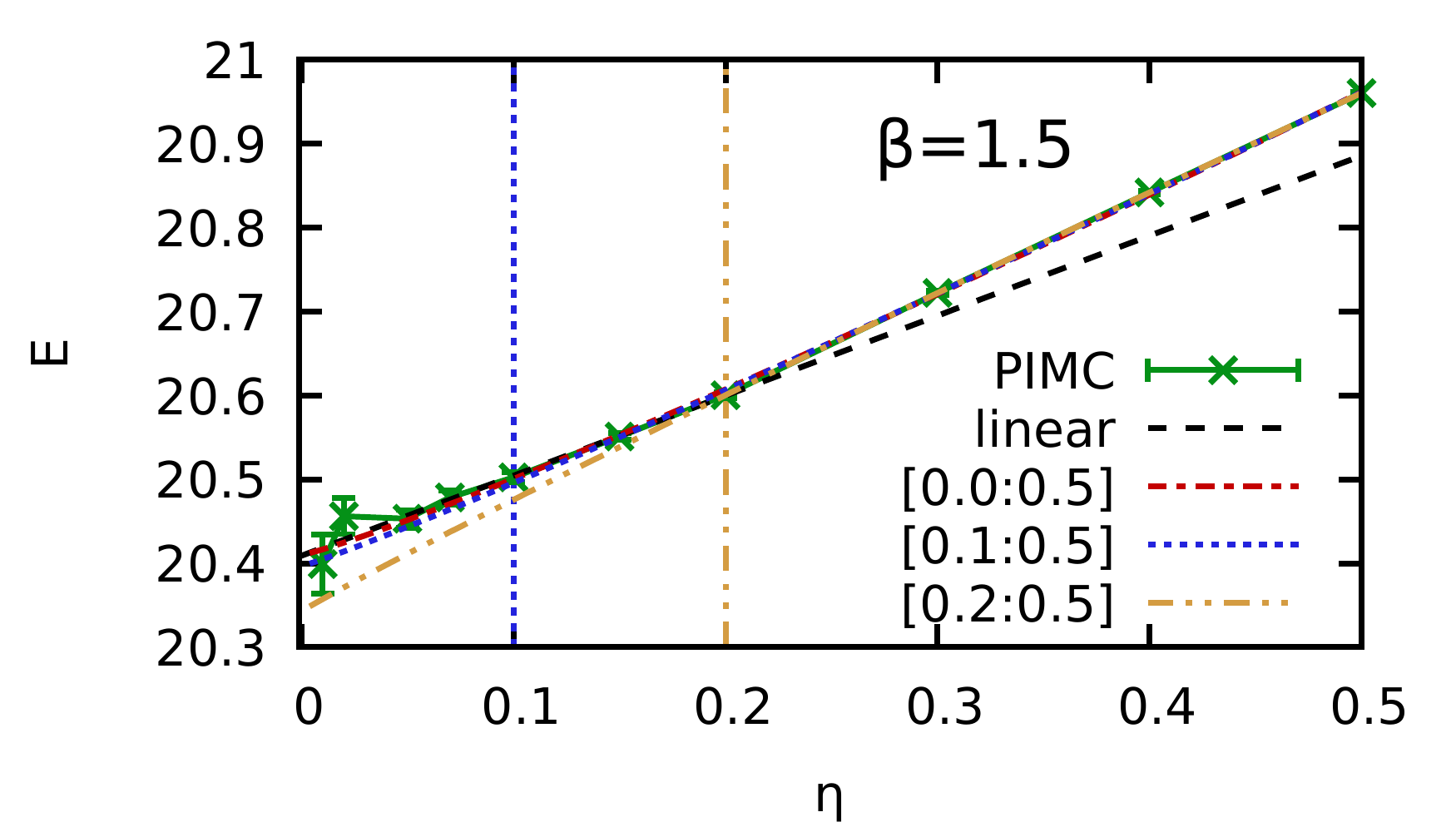}\\\vspace*{-0.8cm}\includegraphics[width=0.415\textwidth]{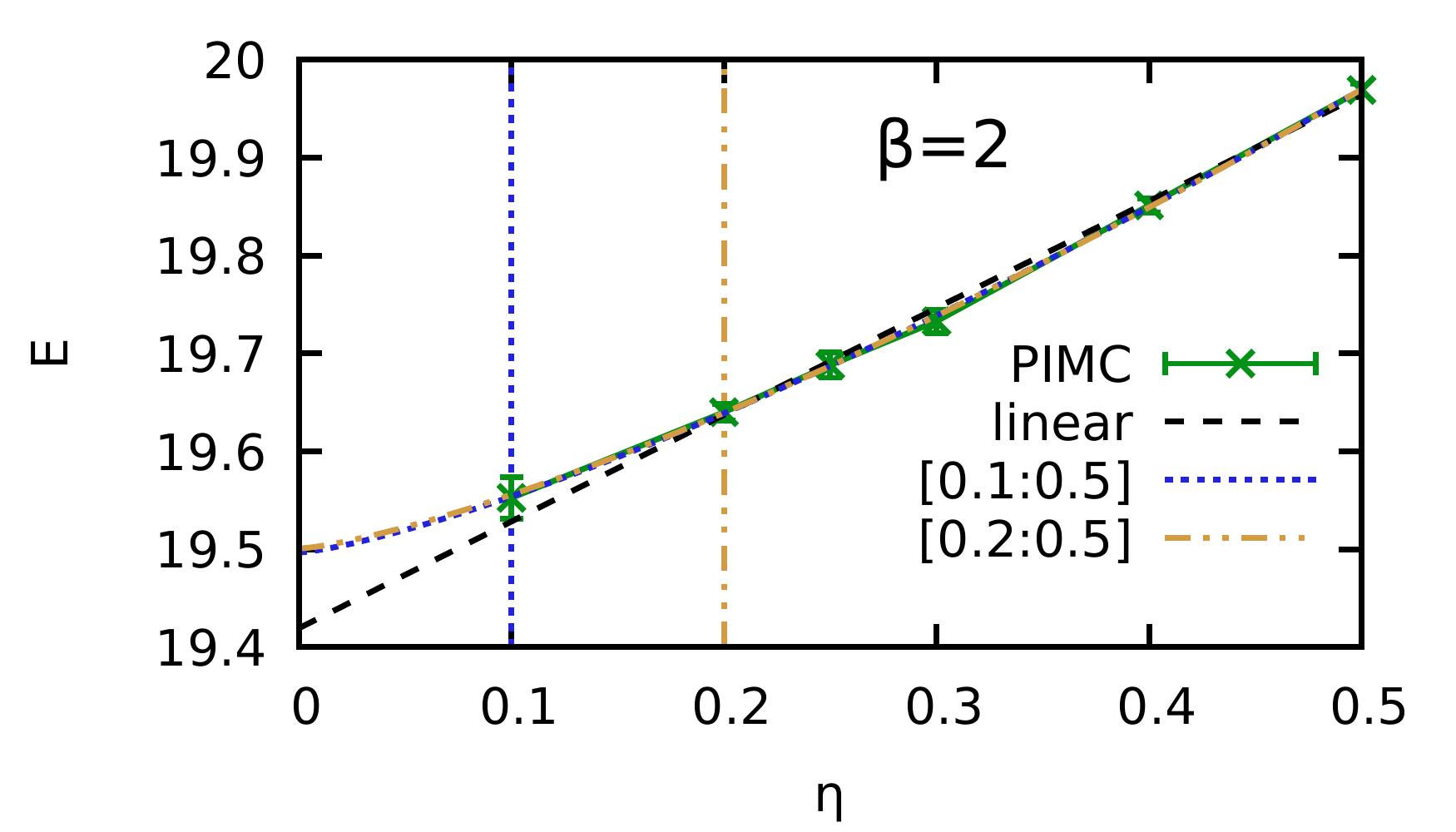}
\caption{\label{fig:trap_results}
Convergence ($\eta$-dependence) of PIMC results for $N=6$ spin-polarized electrons in a $2D$ harmonic trap for $\lambda=0.5$ using the dipole repulsion from Eq.~(\ref{eq:dipole}). The top, center and bottom panels show the expectation value of the original Hamiltonian $E(\eta) \equiv \braket{\hat H}_\eta$ for $\beta=1$, $\beta=1.5$ and $\beta=2$, respectively. The green crosses are PIMC expectation values, the dotted blue and dashed-double-dotted yellow curves fits according to Eq.~(\ref{eq:c-fit}) in different $\eta$-ranges, and the dashed black lines linear fits serving as a guide to the eye.
}
\end{figure}

\begin{figure}\centering
\includegraphics[width=0.45\textwidth]{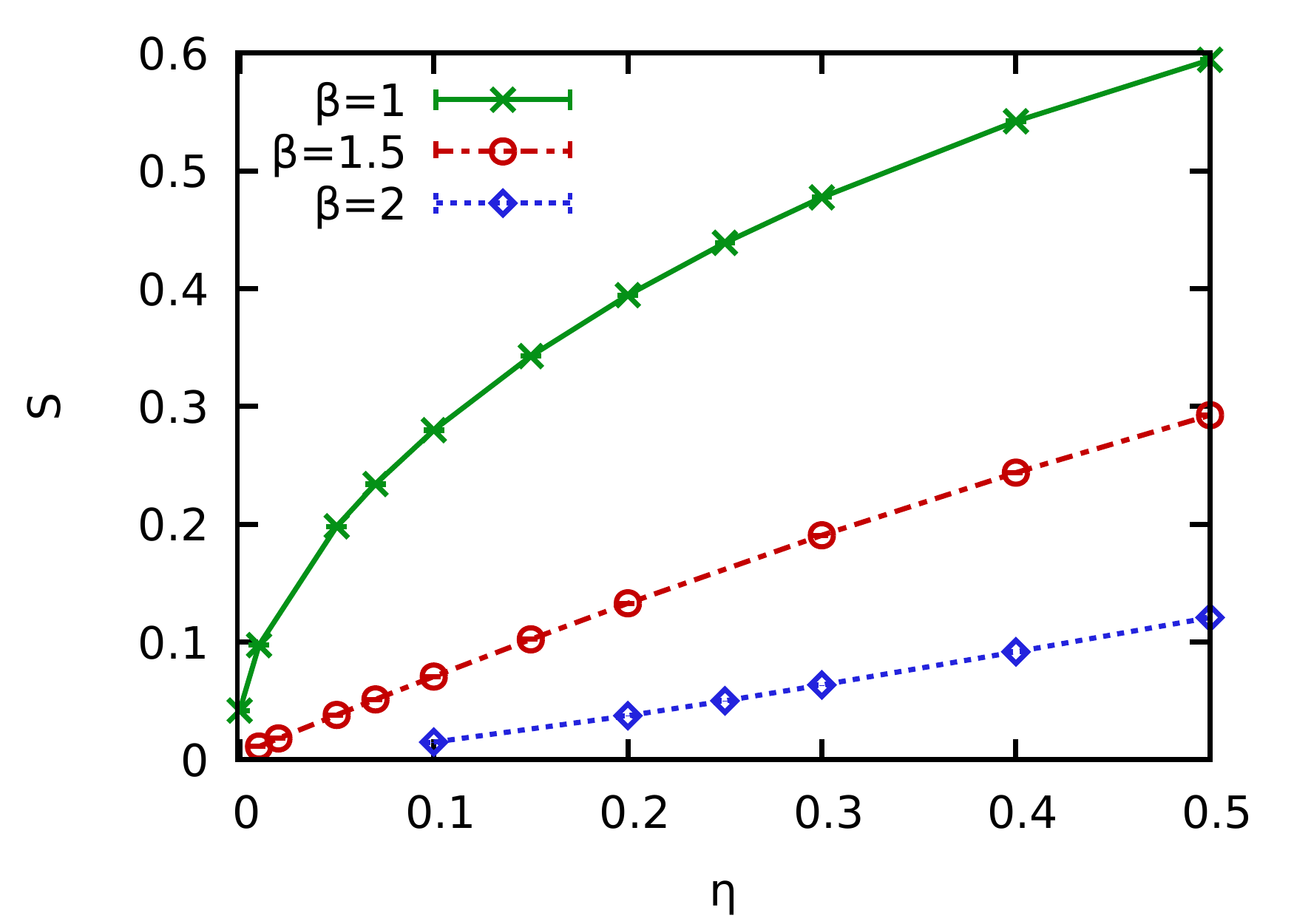}\\ \vspace*{-0.92cm} \hspace*{0.3cm}
\includegraphics[width=0.42\textwidth]{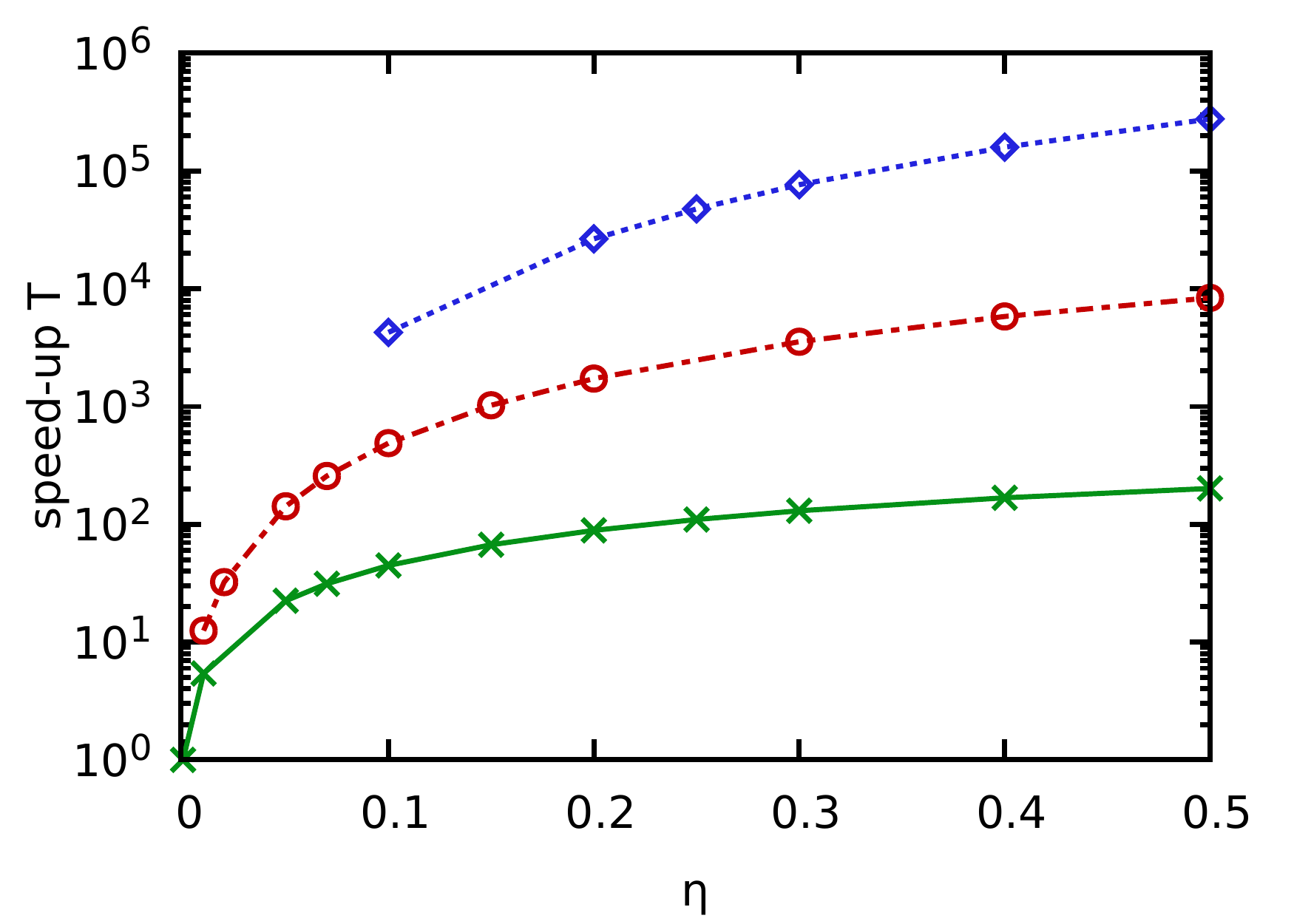}
\caption{\label{fig:speed_results}
Dependence of the average sign $S$ (top panel) and the speed-up $T$ (bottom panel, cf.~Eq.~(\ref{eq:speed-up})) on the repulsion strength $\eta$ for $N=6$ spin-polarized electrons in a $2D$ harmonic trap for $\lambda=0.5$ using the dipole repulsion from Eq.~(\ref{eq:dipole}). The solid green, dash-dotted red, and dotted blue curves show PIMC results for $\beta=1$, $\beta=1.5$, and $\beta=2$, respectively. Note the logarithmic scale of the $y$-axis in the bottom panel.
}
\end{figure}

In Fig.~\ref{fig:trap_results}, we show PIMC results for $N=6$ spin-polarized electrons in a $2D$ harmonic trap for an intermediate value of  $\lambda=0.5$, and three different temperatures, $\beta=1$ (top panel), $\beta=1.5$ (center panel), and $\beta=2$ (bottom panel). The green crosses depict the PIMC results for $\braket{\hat H}_\eta$ for different values of $\eta$. First, we note that the PIMC data do monotonically converge towards the exact $\eta=0$ limit from above, as predicted by Eq.~(\ref{eq:bound}).
For $\beta=1$, the system is substantially out of the ground-state, and we find an average sign of $S\sim10^{-2}$ for standard PIMC (i.e., $\eta=0$), cf.~Fig.~\ref{fig:speed_results}. Hence, the  PIMC simulations can be converged over the entire $\eta$-range. The dashed black line corresponds to a linear fit within the interval $\eta\in[0,0.5]$ and has been included as a guide to the eye, although we find it to be surprisingly accurate in this case. The dash-dotted red line was obtained from a fit within the same interval, but using the modified functional form
\begin{eqnarray}\label{eq:c-fit}
E(\eta) = a + b \eta^c\ ,
\end{eqnarray}
which has been found empirically, and with $a$, $b$, and $c$ being free parameters. Furthermore, the dotted blue and dash-double-dotted yellow lines have also been obtained from fits via Eq.~(\ref{eq:c-fit}), but within the intervals $\eta\in[0.1,0.5]$ and $\eta\in[0.2,0.5]$, respectively. Evidently, all three curves are in excellent agreement with the PIMC data over the full depicted $\eta$-range. This is an important intermediate result, as it indicates that Eq.~(\ref{eq:c-fit}) constitutes a suitable form to extrapolate results to $\eta=0$ when PIMC simulations are limited to some finite value of $\eta$ due to the sign problem.

We next examine the center panel, which corresponds to $\beta=1.5$, an intermediate temperature. In this case, we find $S\sim10^{-3}$ for standard PIMC, which means that simulations are computationally demanding, but still feasible over the full $\eta$-range. Firstly, we note that the linear fit is not as good as for $\beta=1$ above, so we have only used data points for $\eta\in[0,0.2]$ in the linear fitting. Secondly, we find that the blue and red curves are in excellent agreement, whereas the yellow curve somewhat deviates in the limit $\eta\to0$. Most probably, this is a  consequence of the reduced fit interval of $\eta\in[0.2,0.5]$. Still, the extrapolation to $\eta=0$ agrees with the two other curves (with the red curve basically being exact) to a relative accuracy of $0.3\%$, which is sufficient for most applications. At the same time, the yellow curve requires simulations with an average sign of $S\sim0.1$ at the lowest ($\eta=0.2$), resulting in a speed-up by a factor $T\sim10^3$, see Fig.~\ref{fig:speed_results}.

Finally, the bottom panel corresponds to $\beta=2$, where  $S\sim10^{-4}$ for standard PIMC. Thus, PIMC simulations are severely hampered by the sign problem and simulations are not feasible for $\eta\lesssim0.1$. Still, the dotted blue ($\eta\in[0.1,0.5]$) and dash-double-dotted yellow ($\eta\in[0.2,0.5]$) curves are in very good agreement with each other, which strongly suggests that the extrapolation is reliable. Evidently, the extrapolation procedure can provide accurate results where standard PIMC simulations are prohibitive because of the sign problem.

A quantitative analysis of the corresponding speed-up of our simulations is presented in Fig.~\ref{fig:speed_results}. The top panel depicts results for the average sign $S(\eta)$ while the bottom panel shows the corresponding speed-up factor $T(\eta)$, as defined in Eq.~(\ref{eq:speed-up}).

The green crosses, red circles, and blue diamonds in the top panel of Fig.~\ref{fig:speed_results} show the $\eta$-dependence of $S$ for $\beta=1$, $\beta=1.5$, and $\beta=2$, respectively, for the same simulations reported in Fig.~\ref{fig:trap_results}. All three data sets exhibit a qualitatively similar progression and monotonically increase with $\eta$. This growth is more pronounced for the lower temperature, $\beta=2$, where the sign increases by more than two orders of magnitude at $\eta=0.5$ as compared to $\eta=0$.
Consequently, the corresponding speed-up (Fig.~\ref{fig:speed_results}, bottom panel) is the highest for $\beta=2$ (dotted blue line), exceeding $10^5$ for $\eta=0.5$. From the bottom panel of Fig.~\ref{fig:trap_results}, it is evident that simulations for $\eta\geq0.2$, which were sufficient to obtain the energy of the original system with an accuracy of  $\sim0.1\%$ at this temperature, also provide a speed-up exceeding $10^4$.

For $\beta=1.5$ (dash-dotted red), the sign problem is less severe, and we find a speed-up of $T\sim10^3$ for $\eta=0.1$, where no bias in the extrapolation to $\eta=0$ was resolved. Finally, the solid green line corresponds to $\beta=1$, where we find a speed-up of $T\sim10^2$ for  $\eta \gtrsim 0.2$.

\begin{figure}\centering
\includegraphics[width=0.415\textwidth]{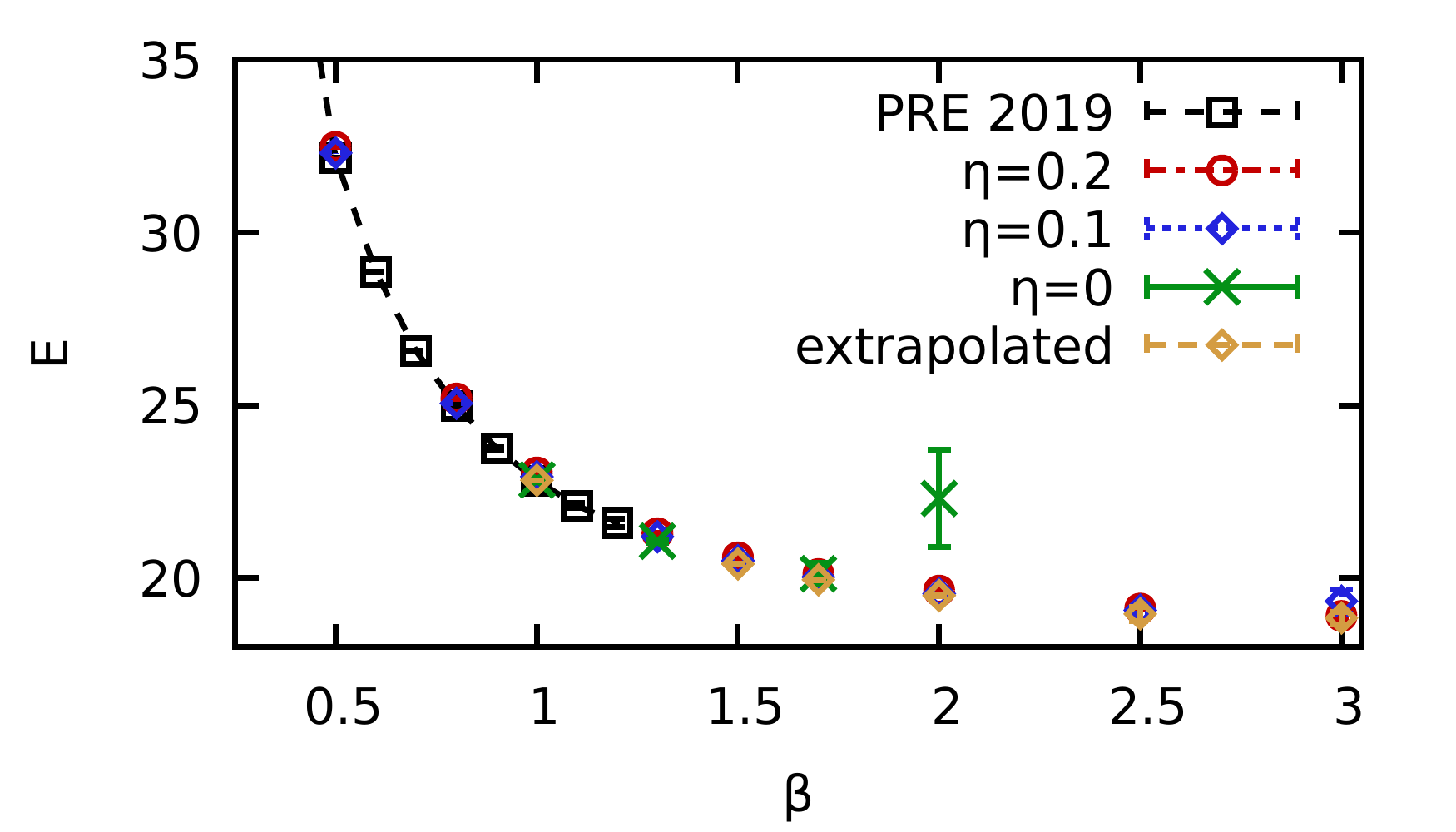}\\\vspace*{-0.8cm}\includegraphics[width=0.415\textwidth]{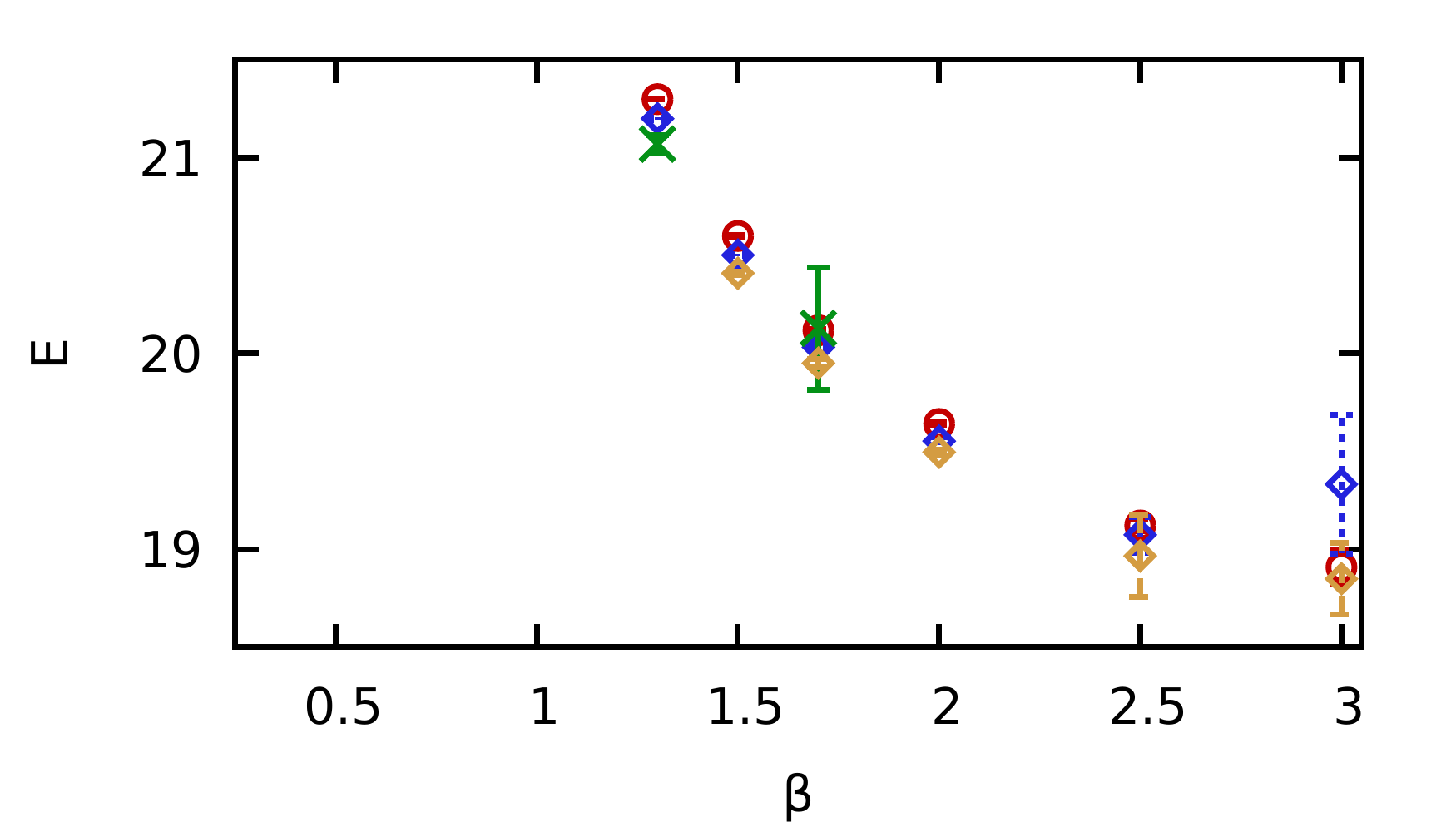}\\\vspace*{-0.8cm}\includegraphics[width=0.415\textwidth]{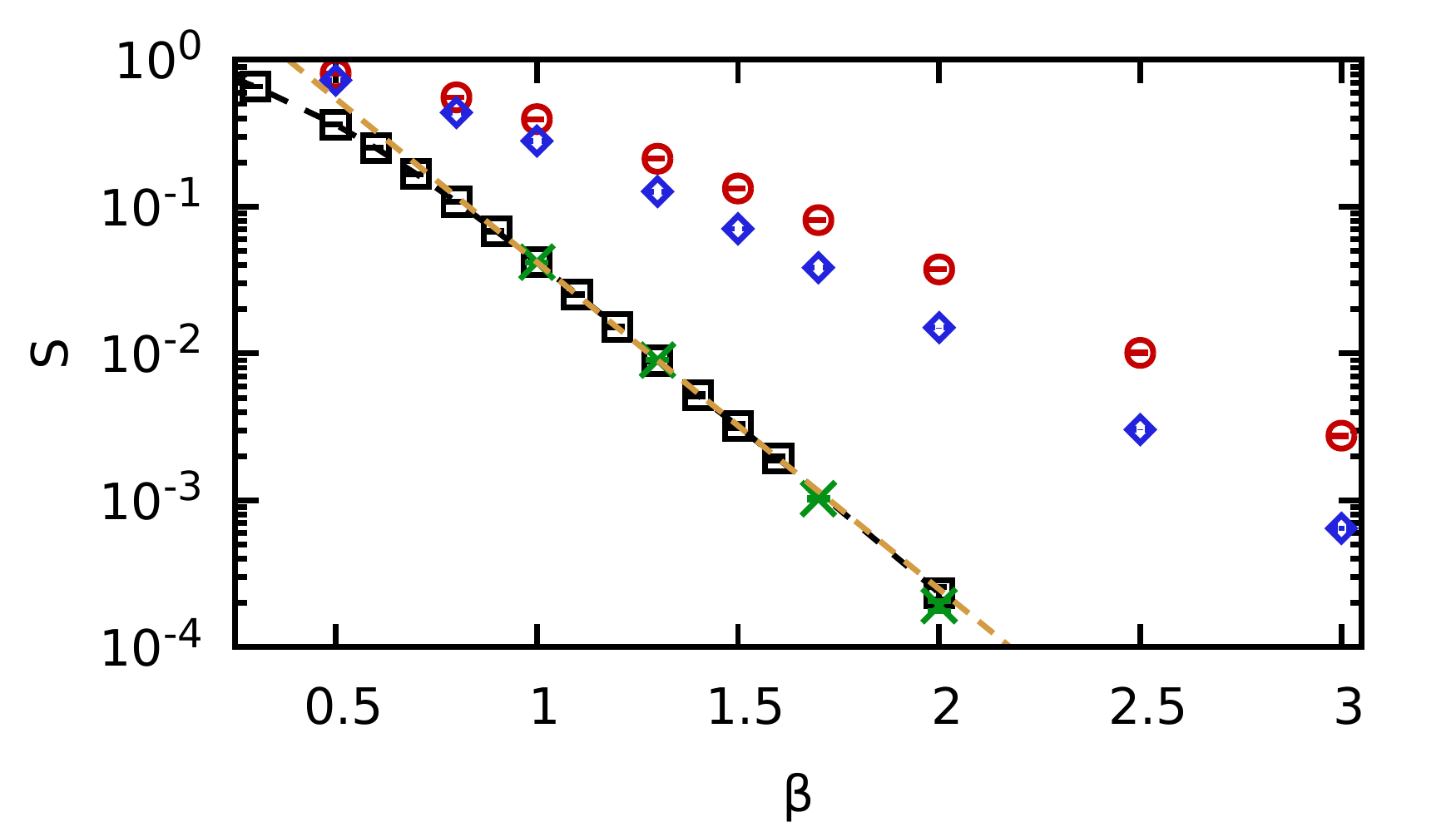}
\caption{\label{fig:trap_results_beta}
Temperature dependence of the energy for $N=6$ spin-polarized electrons in a $2D$ harmonic trap. The top panel shows the total energy $E$ versus $\beta$, with the following key: black squares are standard PIMC data from Ref.~\cite{dornheim_sign_problem}; green crosses are standard PIMC from this work; red circles and blue diamonds have been obtained from Eq.~(\ref{eq:bound}) for $\eta=0.2$ and $\eta=0.1$, respectively. The center panel shows a magnified zoom of the top panel around the lowest temperature points. The bottom panel shows the corresponding data for the average sign, with the yellow curve depicting an exponential fit according to Eq.~(\ref{eq:Sign_fit}). Note that the red, blue, and green data points have been obtained for the same amount of Monte Carlo samples and, thus, can be directly compared regarding efficiency.
}
\end{figure}

We conclude the examination of this system by investigating the behaviour of our new approach upon decreasing the temperature. The results of this analysis are shown in Fig.~\ref{fig:trap_results_beta}, where the top and bottom panel show the $\beta$-dependence of the total energy $E$ and the average sign $S$. The black squares correspond to the standard PIMC data from Ref.~\cite{dornheim_sign_problem}, and accurate results for $E$ are available for $\beta\lesssim1.3$. Data points at lower temperatures present very large error bars, and have been omitted for better visibility. Looking at $S$ itself, we find a steep decay which is of an exponential form for large $\beta$, see Eq.~(\ref{eq:exponential}) and the corresponding analysis in Ref.~\cite{dornheim_sign_problem}.
The yellow curve depicts a fit of the form
\begin{eqnarray}\label{eq:Sign_fit}
S(\beta) = a_S e^{-\beta b_S}\ ,
\end{eqnarray}
obtained for $\beta\in[1,3]$ and fully confirms this trend.

The green crosses in Fig.~\ref{fig:trap_results_beta} represent new standard PIMC results for $\eta=0$, but obtained at a substantially increased computational cost. Therefore, results are available for $\beta\lesssim1.5$, but at $\beta=2$ the statistical uncertainty substantially increases due to the sign problem.

We next examine the performance of our new approach based on the Bogoliubov inequality and the modified Hamiltonian from Eq.~(\ref{eq:Hamiltonian_eta}). The red circles and blue diamonds represent $\braket{\hat H}_\eta$ for $\eta=0.2$ and $\eta=0.1$, respectively, and the simulations can be converged down to $\beta=3$. This can be seen particularly well in the center panel of Fig.~\ref{fig:trap_results_beta}, showing a magnified segment around the low-temperature points. The yellow diamonds show the results which have been extrapolated to $\eta=0$ as described in the discussion of Fig.~\ref{fig:trap_results}. These results show that our scheme allows to double the feasible $\beta$-range despite the exponential wall in compute time given by the FSP.

\begin{figure}\centering
\includegraphics[width=0.415\textwidth]{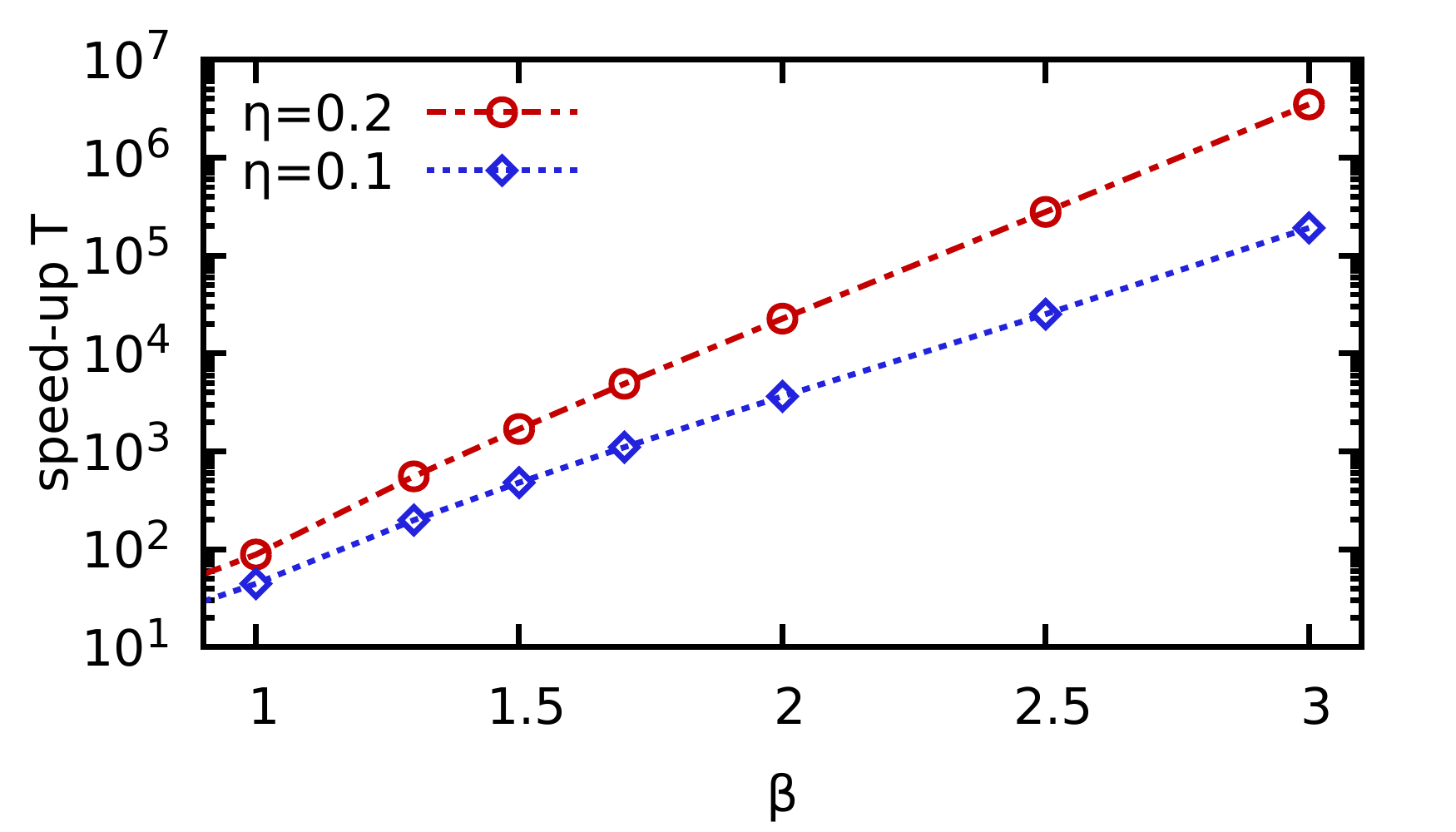}
\caption{\label{fig:trap_speed_beta}
Relative speed-up $T$ [cf.~Eq.~(\ref{eq:speed-up})] of our simulations at $\eta=0.2$ (red circles) and $\eta=0.1$ (blue diamonds) compared to standard PIMC (i.e., $\eta=0$) for the temperature scan from Fig.~\ref{fig:trap_results_beta}. Note the logarithmic scale of the $y$-axis.
}
\end{figure}  

  The speed-up $T(\eta)$ [cf.~Eq.~(\ref{eq:speed-up})] is shown in Fig.~\ref{fig:trap_speed_beta}, with the red circles and blue diamonds depicting the $\beta$-dependence for $T(0.2)$ and $T(0.1)$, respectively. Firstly, we observe that the speed-up monotonically increases with decreasing temperature for both values of $\eta$. Moreover, this increase appears to be of an exponential form for large $\beta$, which helps to explain the remarkable extension of the parameter space that can be covered with this method. In particular, we find $T\sim10^6$ ($T\sim10^5$) for $\eta=0.2$ ($\eta=0.1$) for the lowest depicted temperature, $\beta=3$.

At the same time, it is important to note that this exponentially growing speed-up is still not sufficient to fully counter the sign problem, since $S$ does still monotonically (and, indeed, exponentially) decrease with $\beta$ for every fixed value of $\eta$, cf.~the bottom panel of Fig.~\ref{fig:trap_results_beta}. Therefore, a full solution of the fermion sign problem (which we would define as a simulation scheme without an exponential increase in compute time with decreasing temperature) would require that the minimum value of $\eta$ that is needed for the extrapolation to $\eta\to0$ increases towards low temperature. This, however, is not supported by the results of our current study.

\begin{figure*}\centering
\hspace*{-0.295cm}\includegraphics[width=0.44\textwidth]{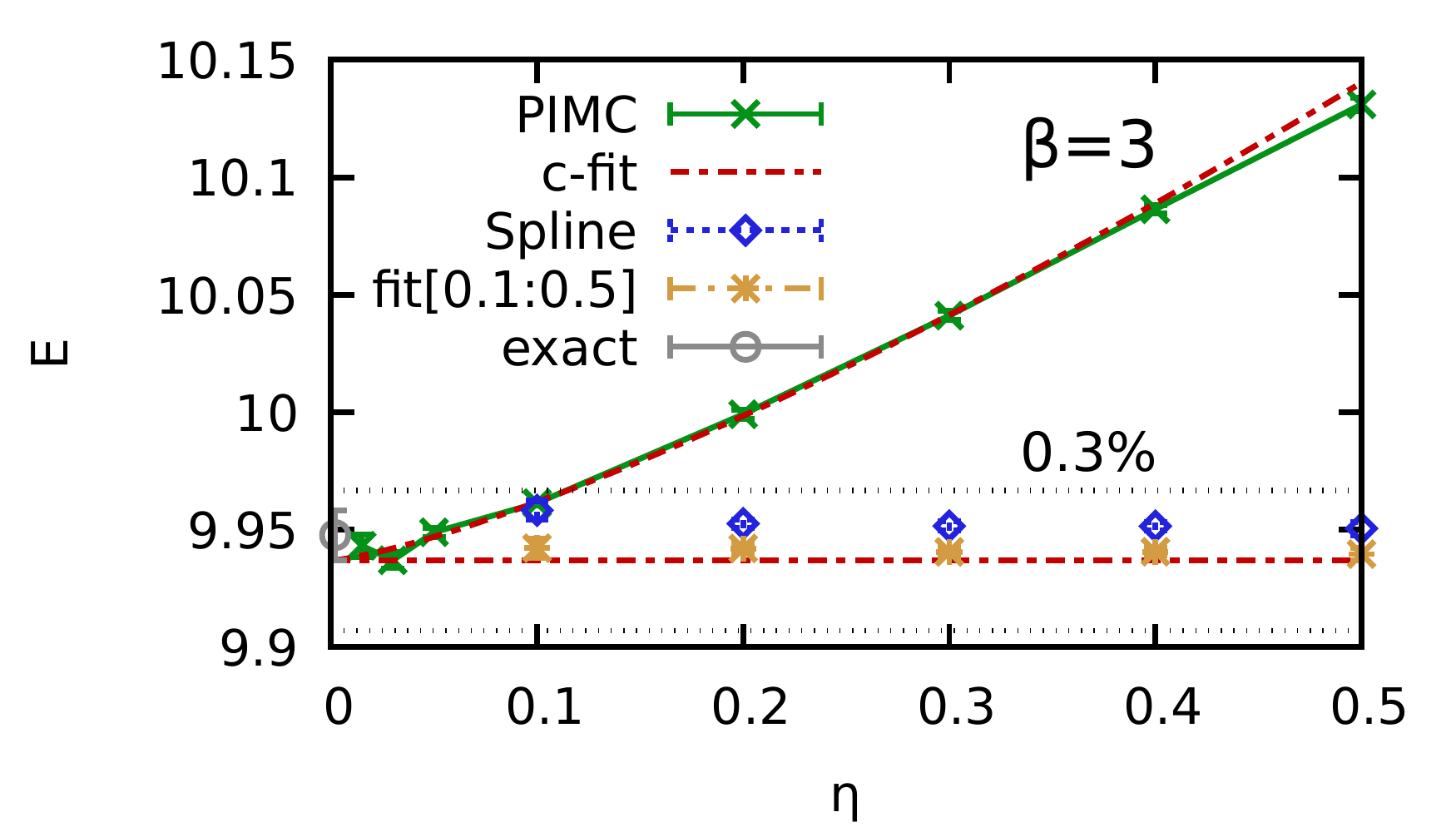}\includegraphics[width=0.44\textwidth]{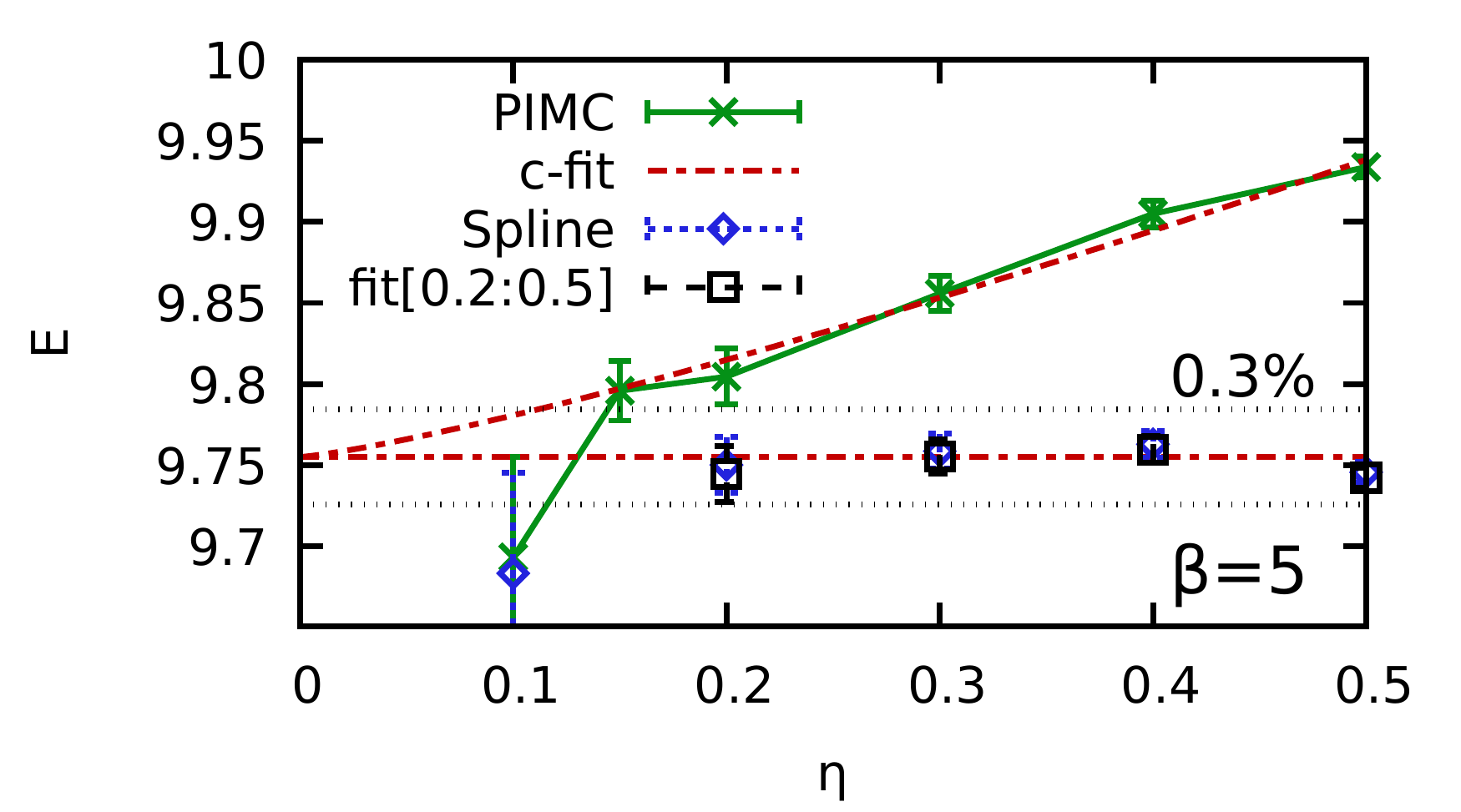}\\ \vspace*{-0.99cm}
\hspace*{0.12cm}\includegraphics[width=0.412\textwidth]{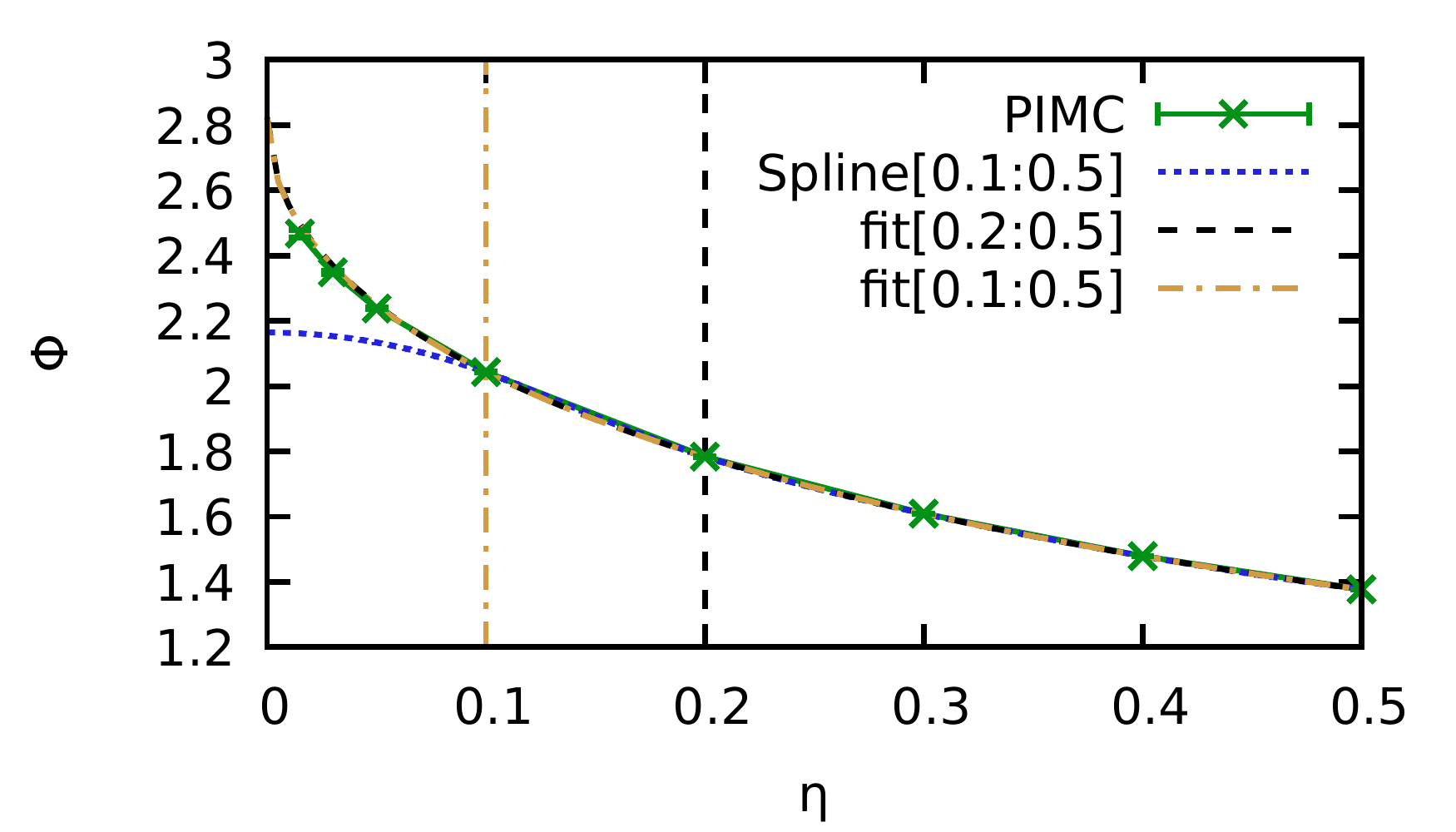}\hspace*{0.3cm}\includegraphics[width=0.4265\textwidth]{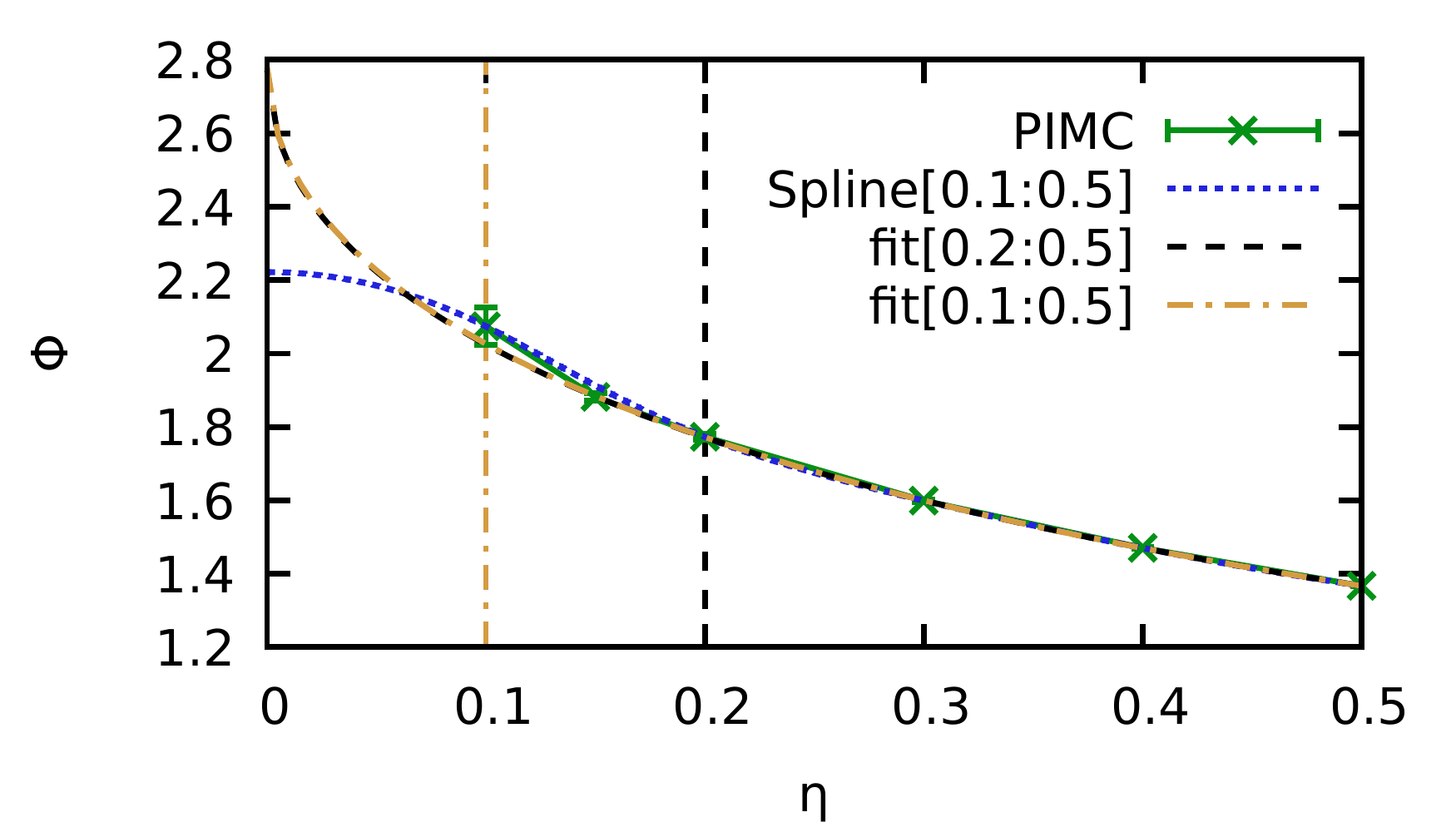}
\caption{\label{fig:Invernizzi_analysis_N4_low}
Results for the thermodynamic integration correction from Sec.~\ref{sec:invernizzi} for $N=4$ spin-polarized electrons in a $2D$ harmonic trap with $\lambda=0.5$ and $\beta=3$ (left), $\beta=5$ (right). Top row: energy estimates $E$. Bottom row: fit of $\braket{\hat\phi}_\eta$ to evaluate the integral in Eq.~(\ref{eq:invernizzi_estimate}). The dashed black and dash-dotted yellow lines have been fitted according to Eq.~(\ref{eq:eta_fit}), and the dash-dotted red line corresponds to a spline-fit to $\eta\braket{\hat\phi}_\eta$ for $\eta\in[0:0.5]$ (without the data point at $\eta=0.05$).
}
\end{figure*}

\subsection{Thermodynamic integration scheme\label{sec:invernizzi_results}}

We estimate the energy of the original system by thermodynamic integration using Eq.~(\ref{eq:invernizzi_estimate}), as an alternative to extrapolation. The results are shown in the left column of Fig.~\ref{fig:Invernizzi_analysis_N4_low} for $N=4$ spin-polarized electrons in a  $2D$ quantum dot with $\lambda=0.5$ and $\beta=3$. The top panel shows data for the total energy $E$, and the bottom panel the expectation value $\phi(\eta) \equiv \braket{\phi}_\eta$, the argument of the thermodynamic integration. 
%In other words, the $\eta\to0$ limit of $\phi(\eta)$ corresponds to the estimation of $\braket{\hat\phi}$ with respect to the unmodified Hamiltonian without the additional $\eta$-term.
The green crosses depict the PIMC data for $E_\eta=\braket{\hat H}_\eta$ (top) and $\phi(\eta)$ (bottom), that are available at discrete $\eta$ values. Since the evaluation of Eq.~(\ref{eq:invernizzi_estimate}) requires the computation of the area under $\phi(\eta')$ for $\eta'\in[0,\eta]$, two practical obstacles have to be overcome:
i) In order to avoid performing many simulations, the computation of the integral would benefit from a continuous representation of $\phi(\eta)$ by fitting a modest number of data points and ii) we need to know $\phi(\eta)$ in the limit of small $\eta$, where PIMC simulations might no longer be feasible due to the sign problem.

Overcoming the first problem by itself is relatively easy, and the dotted blue curve in the bottom panel of Fig.~\ref{fig:Invernizzi_analysis_N4_low} corresponds to a cubic spline fit to the PIMC data in the interval $\eta\in[0.1,0.5]$. The spline fit is capable to smoothly interpolate the PIMC data, but clearly fails outside the interval where input data are provided. Thus, a more controlled extrapolation to the $\eta\to0$ limit of $\phi$ is needed for the solution to ii).
We find empirically that a suitable choice is 
\begin{eqnarray}\label{eq:eta_fit}
\phi(\eta) = a_\phi + b_\phi \eta + c_\phi \eta^{1/2} \ ,
\end{eqnarray}
where $a_\phi$, $b_\phi$, and $c_\phi$ are the free parameters. The resulting fits are shown in Fig.~\ref{fig:Invernizzi_analysis_N4_low} as the dash-dotted yellow and dashed black curves, which have been obtained taking into account PIMC data for $\eta\in[0.1,0.5]$ and $\eta\in[0.2,0.5]$, respectively.

We can see that both curves a) nicely reproduce the PIMC input data, b) are in excellent agreement with each other, and c) almost perfectly match the PIMC data that are outside of the fit interval. Therefore, the fit function Eq.~(\ref{eq:eta_fit}) is capable to provide an accurate representation of $\phi(\eta)$ over the entire $\eta$-range given as input only four data points at $\eta\in[0.2,0.5]$.

Next, we use these findings to  estimate the energy-correction according to Eq.~(\ref{eq:invernizzi_estimate}). The results are shown in the top panel of Fig.~\ref{fig:Invernizzi_analysis_N4_low}, where the green crosses correspond to the uncorrected PIMC simulation data for $E_\eta=\braket{\hat H}_\eta$. For completeness, we also include an extrapolation of these data to $\eta=0$ according to Eq.~(\ref{eq:c-fit}, as described in Sec.~\ref{sec:extrapolation} (dash-dotted red line). The grey circle corresponds to a standard PIMC simulation ($\eta=0$) that is exact within the given error bars.

Using the representation of $\phi(\eta)$ according to Eq.~(\ref{eq:eta_fit}) to estimate the correction leads to the yellow stars. As expected, the entire $\eta$-dependence has been removed by the correction, and the data are in perfect agreement with the exact standard PIMC results for all depicted values of $\eta$. In addition, we also show the corrected PIMC data that have been obtained by using the spline as a representation of $\phi(\eta)$ instead, see the blue diamonds in the top panel. 
%First and foremost, we note that there is a small but significant difference between the blue and yellow points, which is of the order of $0.1\%$ (see also the light dotted grey line around the horizontal red line, corresponding to a deviation of $0.3\%$ around the extrapolation). This is certainly remarkable given the poor quality of the extrapolation of the spline for small $\eta$, and the substantial difference to both the fit and the PIMC data in this regime, cf.~the bottom panel. Yet, this pronounced difference is only of minor consequence for the estimation of Eq.~(\ref{eq:invernizzi_estimate}), which can be understood as follows: for the direct extrapolation of $E(\eta)$ to $\eta\to0$ as discussed in Sec.~\ref{sec:extrapolation}, any inaccuracy in the functional form directly impacts the quality of the final result for the energy of the original system, $E=E(\eta=0)$. For the estimation of the correction given by Eq.~(\ref{eq:invernizzi_estimate}), on the other hand, we only have to extrapolate $\phi(\eta)$, and, thus, to estimate a comparably small correction to the total value. Further, the main contribution to $E(\eta)-E(\eta=0)$ is accurately estimated from the PIMC data for $\phi$, and we only extrapolate a small fraction of the actual correction term that is due to the area under $\phi(\eta)$ for very small values of $\eta$.
It is important to notice that even when using such a poor extrapolation scheme, the systematic error introduced in the energy estimate is only around $0.1\%$.
This shows that any error in the extrapolation of $\phi(\eta)$ will only account for a small contribution to the overall correction obtained via Eq.~(\ref{eq:invernizzi_estimate}), while most of it comes from interpolation of PIMC data.
This is not the case when using instead the direct extrapolation method, where any inaccuracy in the chosen functional form might more strongly impact the quality of the final result.

Finally, we mention that at the conditions considered in the left column of Fig.~\ref{fig:Invernizzi_analysis_N4_low}, the entropic contribution to Eq.~(\ref{eq:invernizzi_estimate}) does indeed vanish within the given level of accuracy, as expected. This changes only for higher temperatures, see the discussion of Fig.~\ref{fig:Invernizzi_analysis_N4_high} below.

We have also used this approach to tackle a harder example, shown in the right column of Fig.~\ref{fig:Invernizzi_analysis_N4_low}, where we have investigated a substantially lower temperature, $\beta=5$, for which $S\lesssim10^{-5}$. Therefore, standard PIMC is not available in this case, and PIMC simulations are only feasible for $\eta\gtrsim0.1$.

The bottom panel shows the estimation of $\phi(\eta)$ and, also in this case, the fits from Eq.~(\ref{eq:eta_fit}) are indistinguishable for the two different intervals of input data, which substantiates the high quality of this representation. The spline, on the contrary, significantly deviates at low $\eta$ values.

The corresponding energies are shown in the top right panel of Fig.~\ref{fig:Invernizzi_analysis_N4_low}, and the dash-dotted red line depicts the direct extrapolation of the PIMC data to $\eta\to0$ according to Eq.~(\ref{eq:c-fit}). In addition, the black squares and blue diamonds have been obtained from the estimation of the thermodynamic integration correction using the fit from Eq.~(\ref{eq:eta_fit}) and the spline, respectively. Firstly, we find that both data sets can hardly be distinguished at these conditions, so that the extrapolation of $\phi(\eta)$ only plays a minor role for the overall level of accuracy of the energy. Secondly, the corrected energies do not exhibit any residual dependence on $\eta$ and fluctuate around the horizontal red line, with an uncertainty level of 0.1\%.

\begin{figure}\centering
\includegraphics[width=0.429\textwidth]{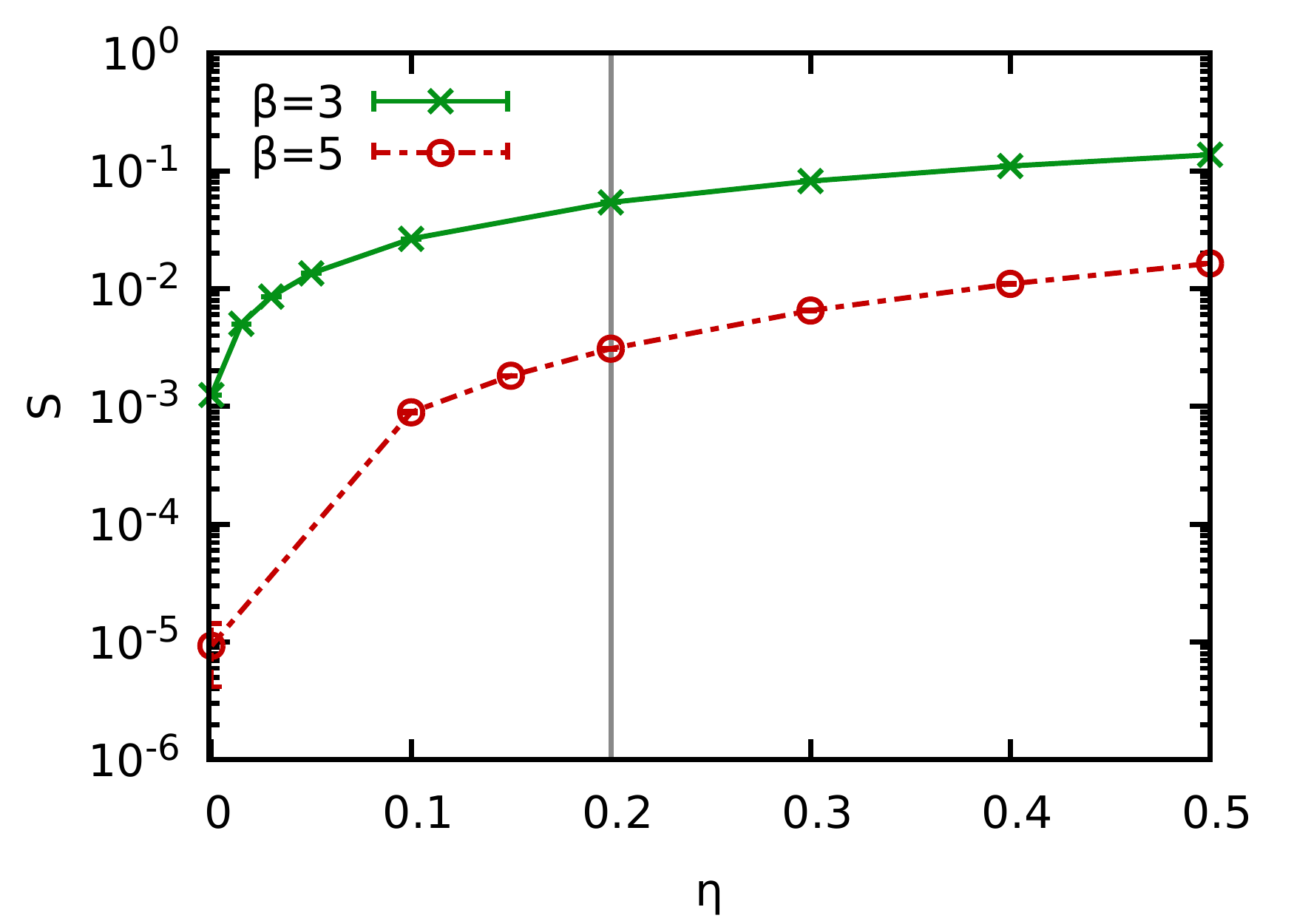}\\ \vspace*{-0.92cm} \hspace*{0.1cm}
\includegraphics[width=0.419\textwidth]{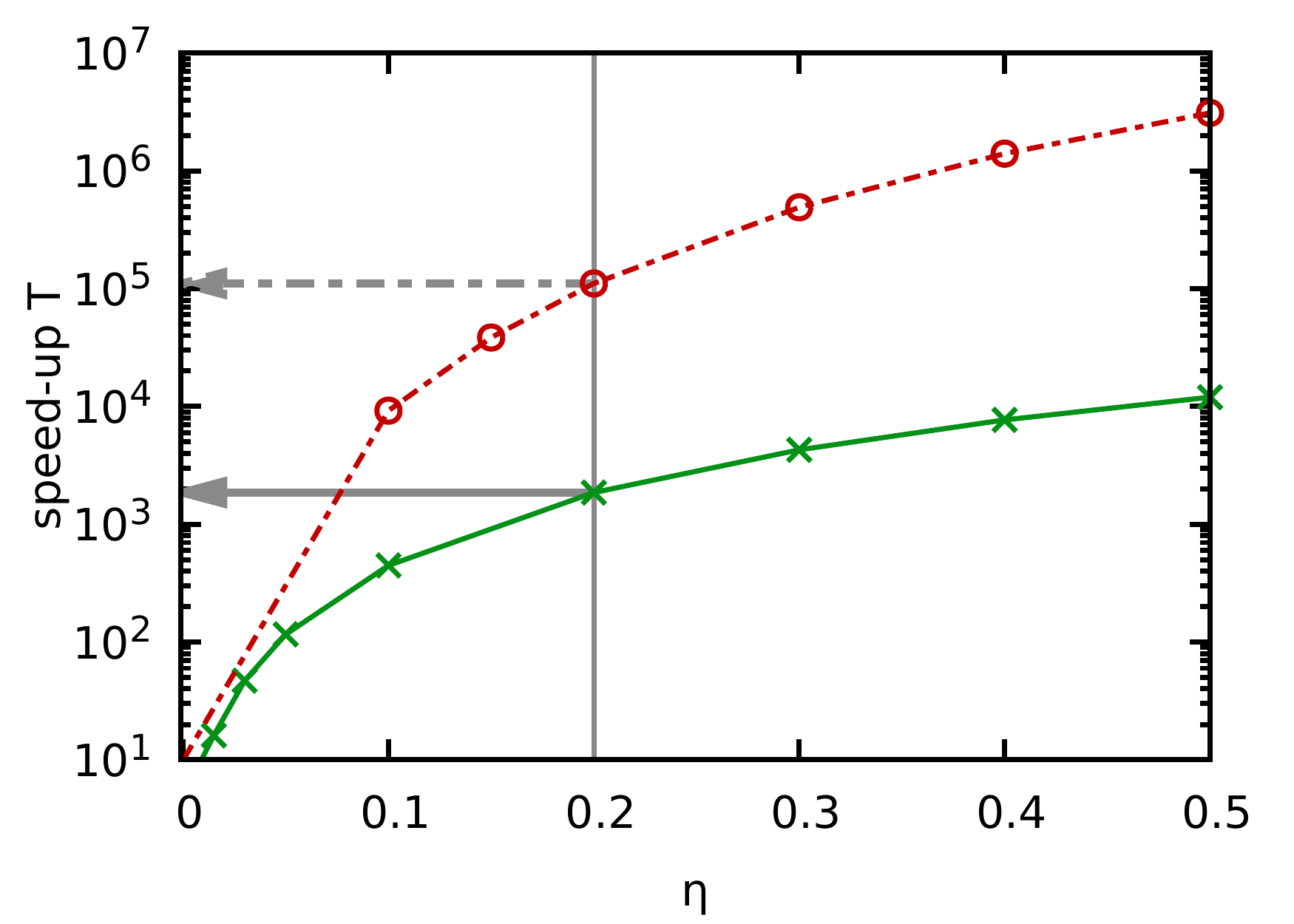}
\caption{\label{fig:Invernizzi_speedup_N4}
PIMC results for the average sign $S$ (top panel) and the respective speed-up $T$ (bottom panel) for $N=4$ spin-polarized electrons in a $2D$ harmonic trap with $\lambda=0.5$. The solid green and dash-dotted red curves correspond to $\beta=3$ and $\beta=5$, respectively. The arrows point to the speed-up for $\eta=0.2$.
}
\end{figure}

\begin{figure*}\centering
\hspace*{-0.295cm}\includegraphics[width=0.44\textwidth]{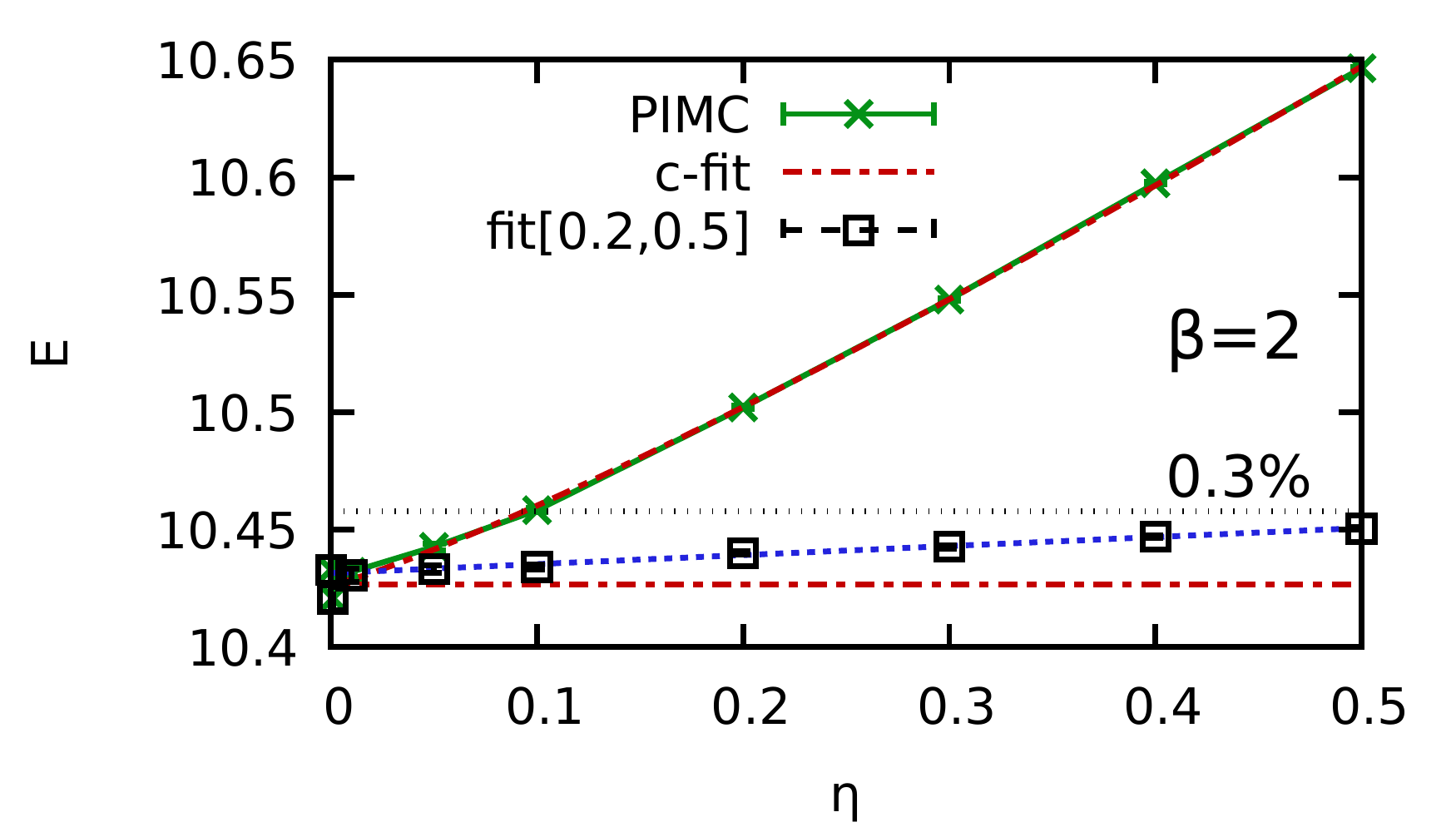}\includegraphics[width=0.44\textwidth]{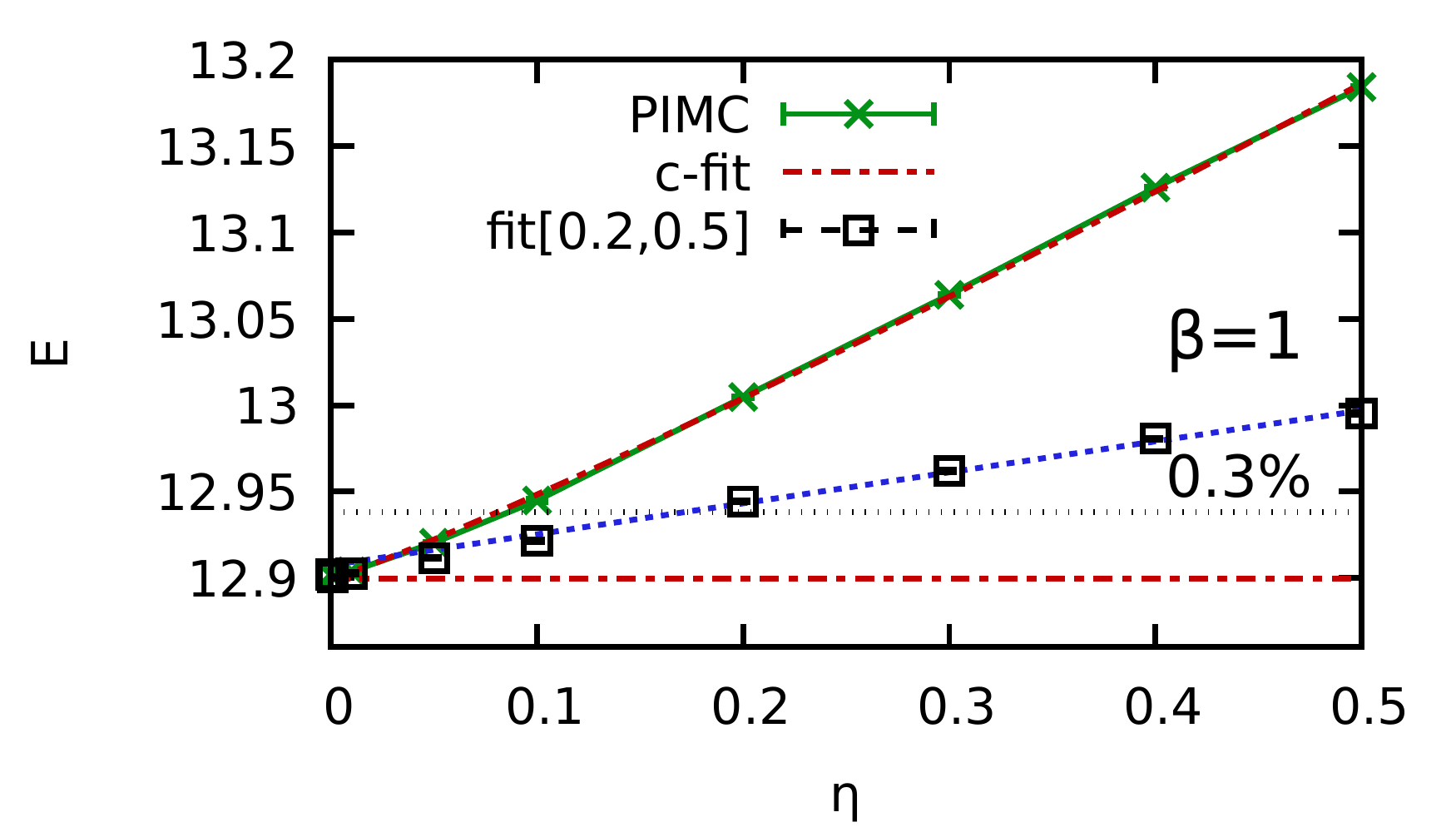}\\ \vspace*{-0.94cm}
\hspace*{0.02cm}\includegraphics[width=0.415\textwidth]{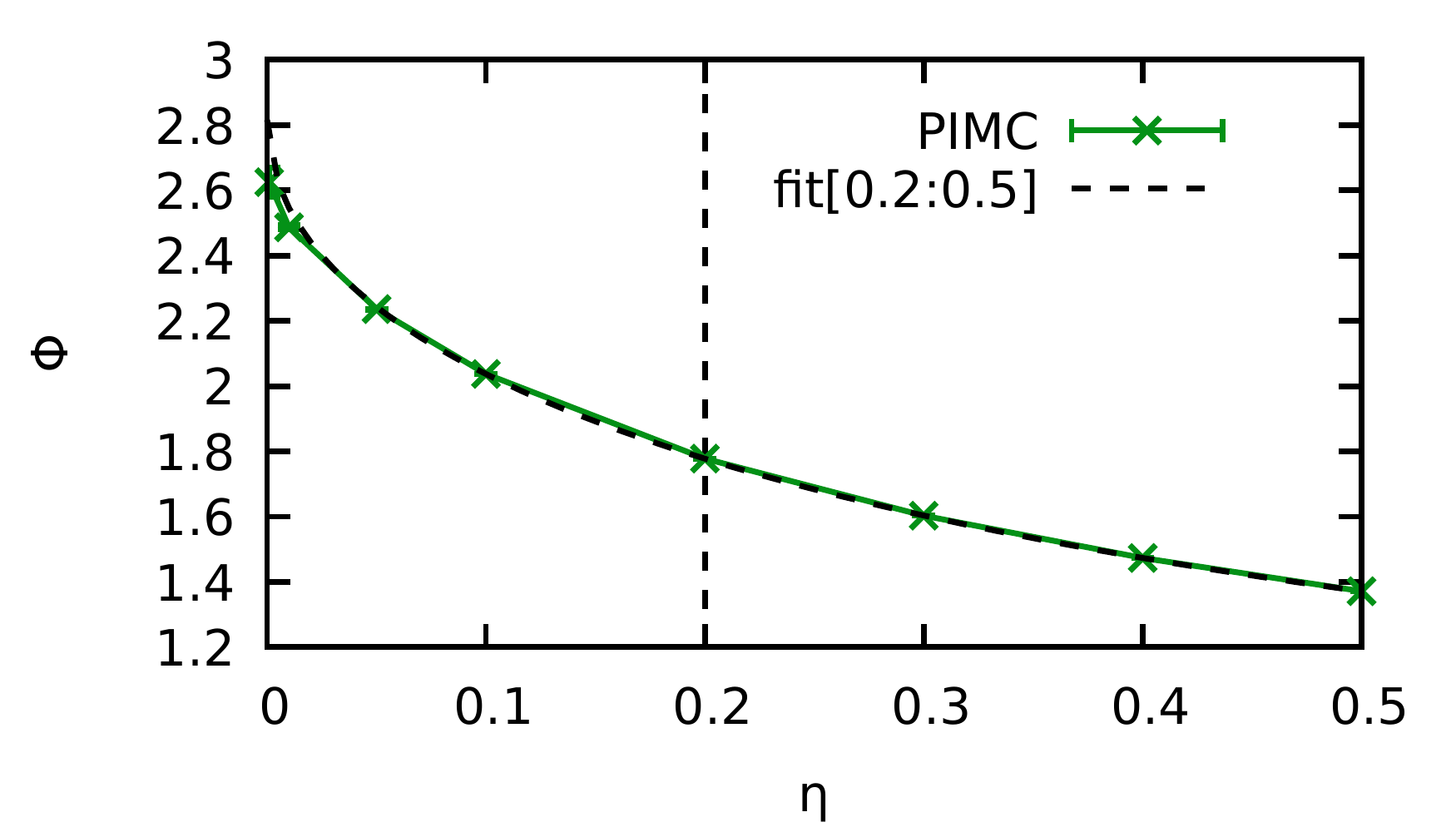}\hspace*{0.52cm}\includegraphics[width=0.412\textwidth]{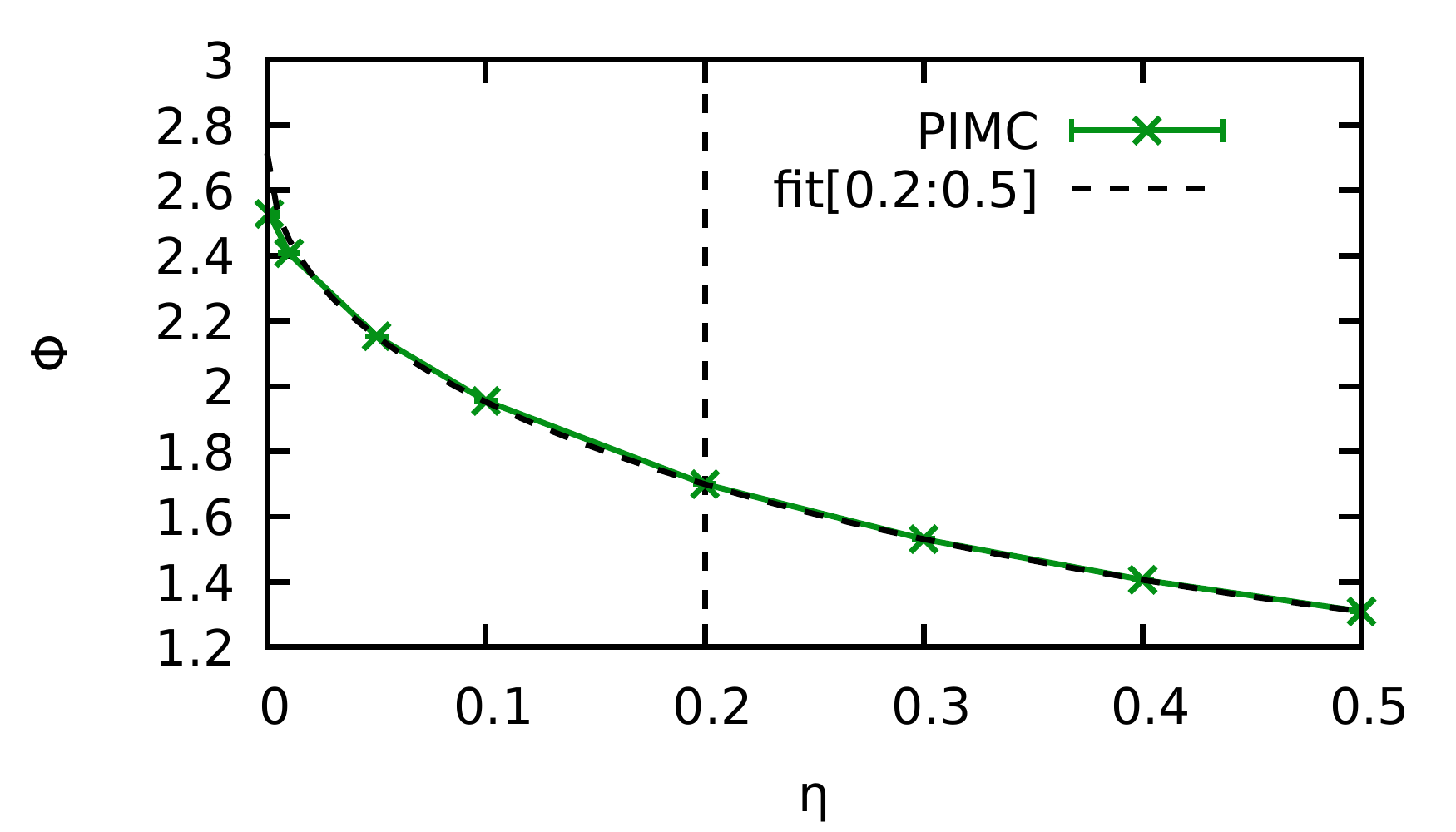}
\caption{\label{fig:Invernizzi_analysis_N4_high}
Results for the thermodynamic integration correction from Sec.~\ref{sec:invernizzi} for $N=4$ spin-polarized electrons in a $2D$ harmonic trap with $\lambda=0.5$ and $\beta=2$ (left), $\beta=1$ (right). Top row: energy estimates $E$. Bottom row: fit of $\braket{\hat\phi}_\eta$ to evaluate the integral in Eq.~(\ref{eq:invernizzi_estimate}). The dashed black and dash-dotted yellow lines have been fitted according to Eq.~(\ref{eq:eta_fit}), and the dash-dotted red line corresponds to a spline-fit to $\eta\braket{\hat\phi}_\eta$ for $\eta\in[0:0.5]$ (without the data point at $\eta=0.05$).
}
\end{figure*}

The increase of the average sign and the corresponding speed-up for both $\beta=5$ and $\beta=3$ is shown in Fig.~\ref{fig:Invernizzi_speedup_N4}.
For the higher temperature (green crosses), we observe an increase in $S$ (top panel) by two order of magnitude between $\eta=0$ and $\eta=0.5$, which results in a speed-up of up to $T\sim10^4$ (bottom panel). For $\beta=5$ (red circles), the relative gain in the sign is even larger, leading to a speed-up (dash-dotted red) exceeding $T\sim10^6$ at the largest value of $\eta$.

A realistic application for our method performs simulations down to $\eta\gtrsim0.2$, as this does suffice for a quasi-exact extrapolation to $\eta\to0$. This boundary is marked as the vertical grey line in Fig.~\ref{fig:Invernizzi_speedup_N4}, and the two horizontal arrows point to the respective speed-up on the $y$-axis. For $\beta=3$, standard PIMC simulations are, in principle, possible, but our scheme results in a speed-up by a factor of $T\sim10^3$. For $\beta=5$, standard PIMC is not feasible, and it is only our speed-up by a factor of $T\sim10^5$ that makes it possible to obtain accurate data.

\begin{figure}\centering
\hspace*{-0.21cm}\includegraphics[width=0.429\textwidth]{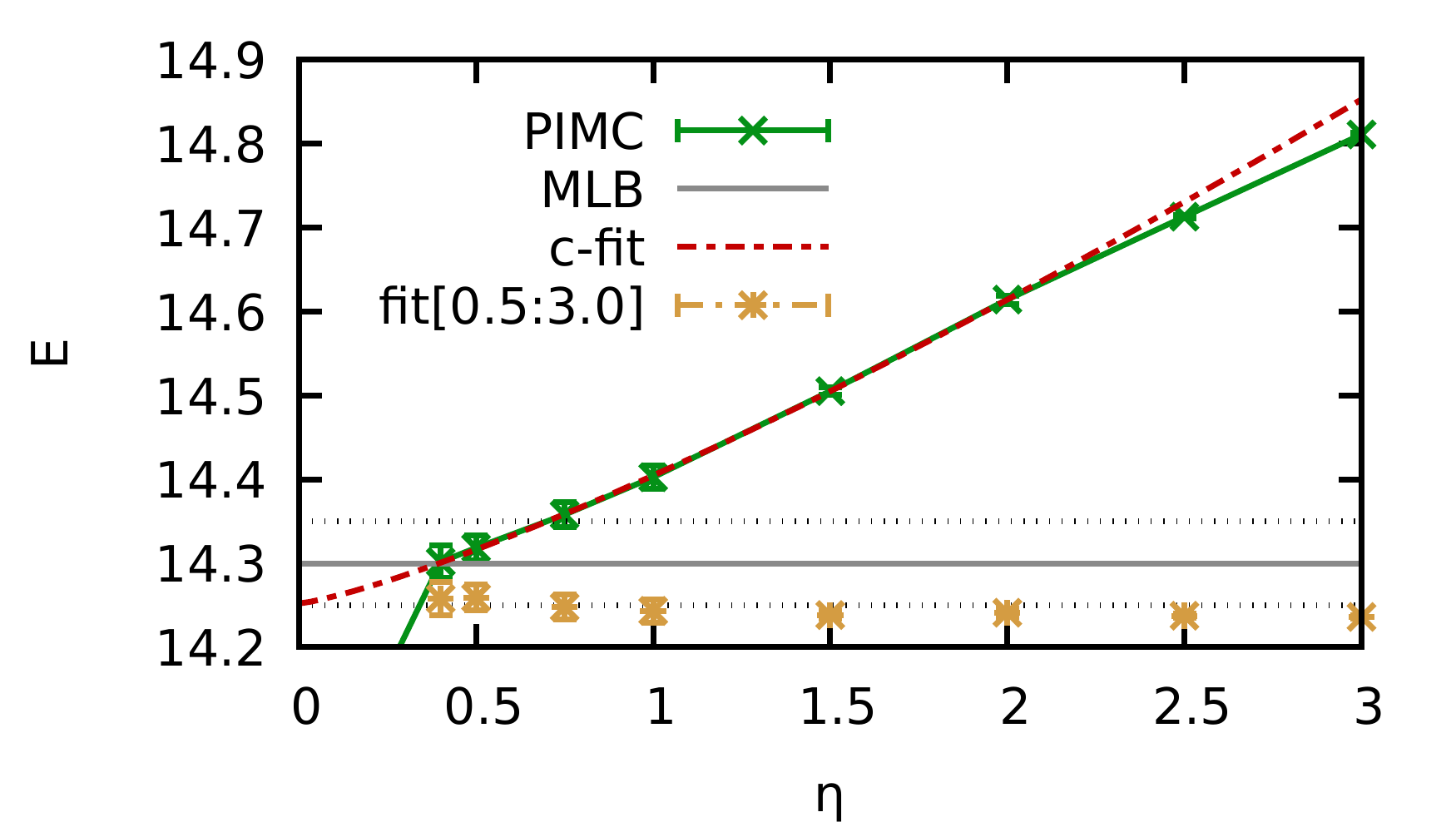}\\ \vspace*{-0.9cm}
\includegraphics[width=0.415\textwidth]{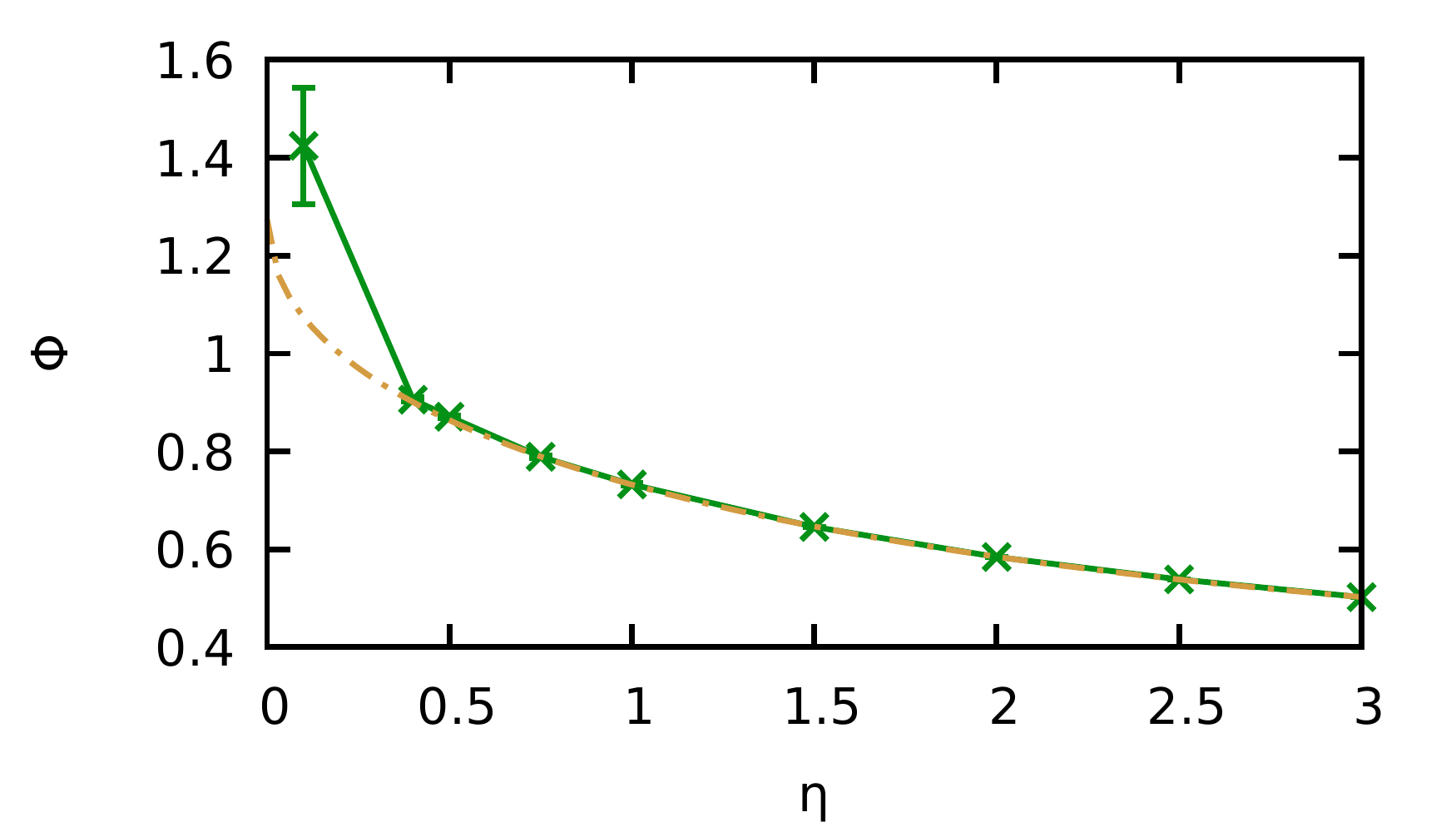}
\caption{\label{fig:Egger}
Results for the thermodynamic integration correction from Sec.~\ref{sec:invernizzi} for $N=4$ spin-polarized electrons in a $2D$ harmonic trap with $\lambda=2$ and $\beta=10$. Top panel: energy estimates $E$. The solid grey line (dotted grey line) corresponds to the MLB result (error bar) from Ref.~\cite{Egger_PRL_1999}. Bottom panel: fit of $\braket{\hat\phi}_\eta$ to evaluate the integral in Eq.~(\ref{eq:invernizzi_estimate}). The dash-double-dotted yellow line has been fitted according to Eq.~(\ref{eq:eta_fit}).
%, 
%and the dotted blue line corresponds to a spline-fit to $\eta\braket{\hat\phi}_\eta$ 
%for $\eta\in[0.5:3]$.
}
\end{figure}

For completeness, we also examine the application of the correction from Eq.~(\ref{eq:invernizzi_estimate}) for higher temperatures, which is shown in Fig.~\ref{fig:Invernizzi_analysis_N4_high}
for $\beta=2$ (left column) and $\beta=1$ (right column). The sign problem is not severe at these parameters and we find $S\approx0.02$ ($S\approx0.25$) for $\beta=2$ ($\beta=1$). Thus, PIMC simulations are computationally feasible over the entire $\eta$-range for both cases. Still, we find that the functional form from Eq.~(\ref{eq:eta_fit}) allows for an accurate representation of $\phi(\eta)$ (see the dashed black curve in the bottom row) taking only into account the four data points at $\eta=0.2,0.3,0.4,0.5$.

The corresponding energies are shown in the top row of Fig.~\ref{fig:Invernizzi_analysis_N4_high}, with the green crosses depicting the PIMC estimates for $\braket{\hat H}_\eta$, and the red curves have been obtained from a direct fit to these data according to Eq.~(\ref{eq:c-fit}). Finally, the black squares have been obtained from the thermodynamic integration correction Eq.~(\ref{eq:invernizzi_estimate}). 
In contrast to the previous results in Fig.~\ref{fig:Invernizzi_analysis_N4_high}, we find a significant dependence of the corrected data points on $\eta$, which is due to the entropic contribution $\Delta S(\eta)$ to Eq.~(\ref{eq:invernizzi_estimate}).

Interestingly, this function can be perfectly reproduced by a linear fit,
\begin{eqnarray}\label{eq:entropy_fit}
\Delta S(\eta) = a_S + b_S\eta \ ,
\end{eqnarray}
and the results are depicted by the dotted blue lines in Fig.~\ref{fig:Invernizzi_analysis_N4_high}.

For $\beta=2$, the entropic contribution is quite small and does not exceed $0.3\%$ even for $\eta=0.5$. In contrast, $\Delta S$ attains a maximum value of $\sim0.7\%$ for $\beta=1$. To put these findings into the proper context, we find it useful to briefly recall the following points: i) while we cannot directly estimate $\Delta S(\eta)$ from our PIMC results, the influence on the correction from Eq.~(\ref{eq:invernizzi_estimate}) decreases for low temperature, when the sign problem is most severe; ii) even when $\Delta S(\eta)$ does have an  impact on the corrected energies, the residual dependence on the parameter $\eta$ is much smaller than the direct dependence of $\braket{\hat H}_\eta$, which makes a potential extrapolation to $\eta\to0$ much less uncontrolled; iii) empirically, we find a simple linear dependence of $\Delta S$ on $\eta$, which is an additional advantage over the direct extrapolation of the uncorrected energy, where the functional dependence is more complicated, cf.~Sec.~\ref{sec:extrapolation}.

We thus conclude that the correction introduced in Sec.~\ref{sec:invernizzi} using thermodynamic integration constitutes a distinct improvement over the direct extrapolation explored in Sec.~\ref{sec:extrapolation}.

We conclude this section with an application of the correction approach at higher $\lambda$  and lower temperature. In Fig.~\ref{fig:Egger}, we show PIMC results for the energy (top panel) and $\phi$ (bottom panel) for $N=4$ spin-polarized electrons at $\lambda=2$ and $\beta=10$. First, we mention that standard PIMC is not available at these conditions, and the sign vanishes within a statistical uncertainty of $10^{-5}$. We have chosen this particular set of parameters because it was previously studied by Egger \textit{et al.}~\cite{Egger_PRL_1999} using the approximate multi-level blocking (MLB) method~\cite{Egger_2001,MLB_1,MLB_2}. While being potentially biased~\cite{Schoof_PHD}, such a data point still constitutes a valuable reference for the development of a new method.

As usual, the PIMC results for $\braket{\hat H}_\eta$ are depicted as the green crosses, and the dash-dotted red curve corresponds to a direct extrapolation of these data according to Eq.~(\ref{eq:c-fit}). The solid grey horizontal line depicts the MLB value from Ref.~\cite{Egger_PRL_1999}, and the two light dotted grey lines depict the corresponding statistical uncertainty, which is given by $\Delta E/E\approx0.35\%$.
Evidently, the direct extrapolation of our data falls into the bottom of the uncertainty interval from the MLB method.

We next consider the thermodynamic integration correction, for which we need a representation of $\phi(\eta)$, shown in the bottom panel of Fig.~\ref{fig:Egger}. 
%The blue dotted curve corresponds to a spline fit, which smoothly interpolates the PIMC data, where they are available, but underestimates $\phi$ for small values of $\eta$. 
The dash-double-dotted yellow curve has been obtain using Eq.~(\ref{eq:eta_fit}) as a fit function for the PIMC data, which was empirically shown to be accurate over the entire relevant $\eta$-range, see above. Using this representation to evaluate Eq.~(\ref{eq:invernizzi_estimate}) results in the yellow stars in the top panel.

Firstly, we note that these data do not exhibit any residual dependence on $\eta$, as the entropic contribution is negligible at such a low temperature.
%Secondly, we find that they exhibit systematic deviations of the order of $\Delta E/E\sim0.1\%$, which is a direct result of the different extrapolation of $\phi(\eta)$ to $\eta\to0$. 
%Still, this difference is substantially smaller than the MLB error bar.
Moreover, we find that the yellow stars are in excellent agreement to the result from the direct extrapolation of the PIMC data, and, thus, also in agreement within the given uncertainty interval of the MLB results.

Therefore, both the MLB method and our new approach have been successfully validated against each other for this system.

\subsection{Electrons in a 3D quantum dot\label{sec:3D_results}}

\begin{figure}\centering
\hspace*{-0.58cm}\includegraphics[width=0.44\textwidth]{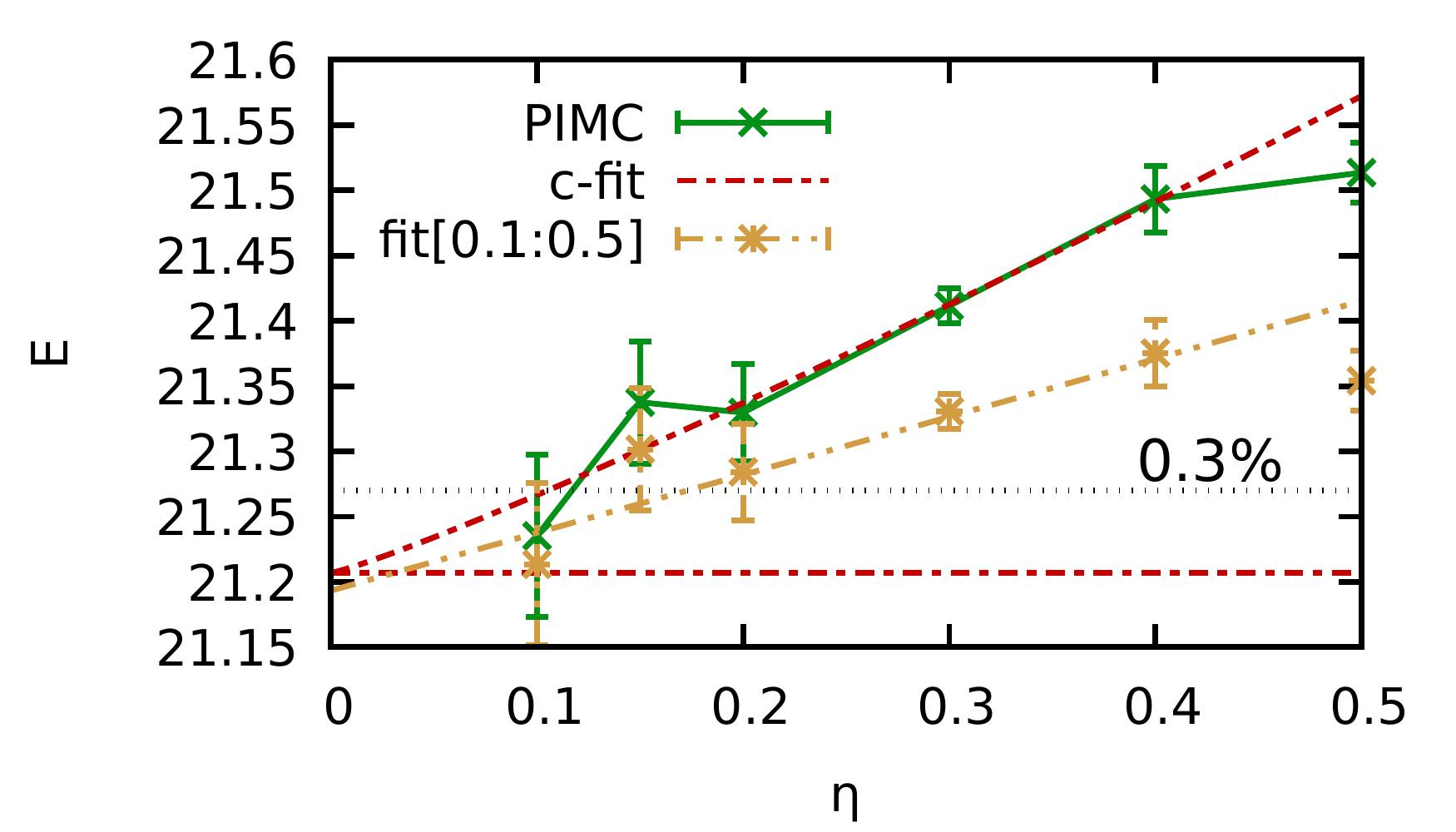}\\ \vspace*{-0.925cm}
\hspace*{0.1cm}\includegraphics[width=0.415\textwidth]{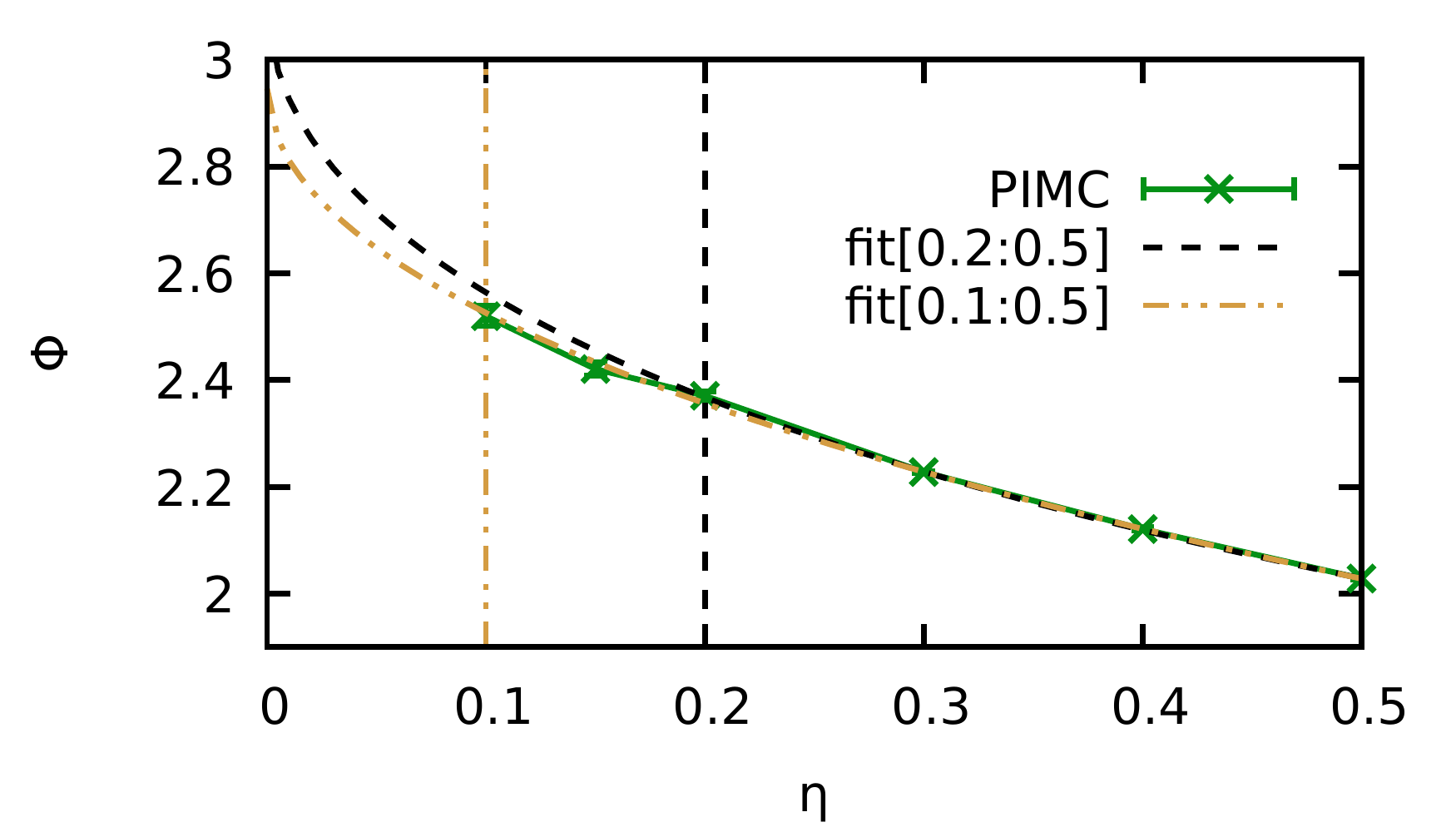}\hspace*{0.3cm}
\caption{\label{fig:trap_results_3D}
Top: Convergence ($\eta$-dependence) of PIMC results for $N=6$ spin-polarized electrons in a $3D$ harmonic confinement with $\lambda=0.5$ and $\beta=2$. The green crosses depict PIMC data for $E_\eta=\braket{\hat H}_\eta$ and the dash-dotted red line a fit according to Eq.~(\ref{eq:c-fit}). The yellow stars have been obtained using the thermodynamic integration correction from Eq.~(\ref{eq:invernizzi_estimate}), and the dashed-double-dotted yellow line an extrapolation of the entropic contribution, cf.~Eq.~(\ref{eq:entropy_fit}). Bottom: PIMC data for $\phi$ (green crosses) and fits via Eq.~(\ref{eq:eta_fit}) (dashed black and dashed-double-dotted yellow) for two different $\eta$-ranges.
}
\end{figure}

\begin{figure}\centering
\includegraphics[width=0.44\textwidth]{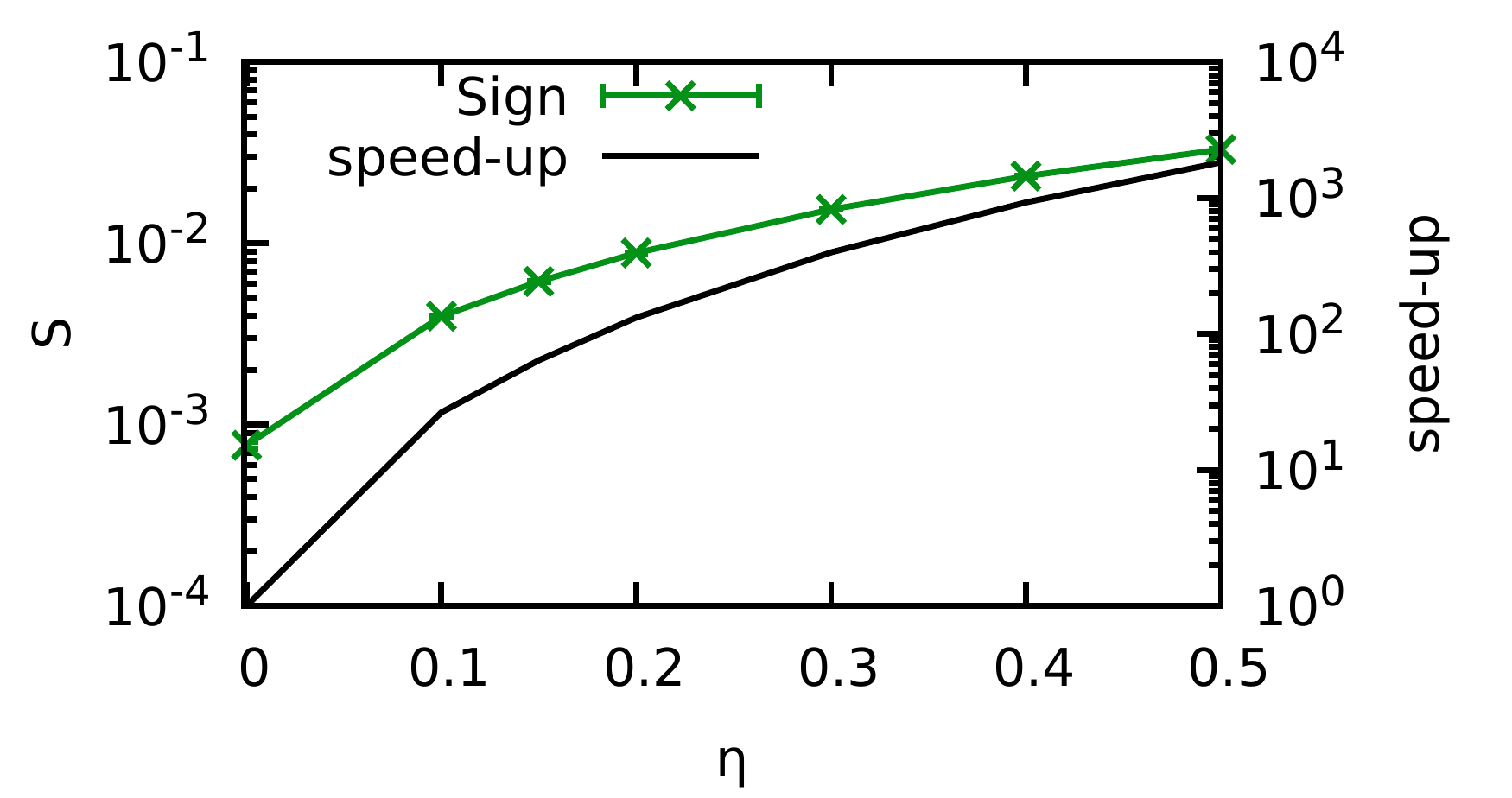}
\caption{\label{fig:speed_results_3D}
Dependence of the average sign $S$ (green crosses, left $y$-axis) and the speed-up T (cf.~Eq.~(\ref{eq:speed-up}), solid black line, right $y$-axis) on the repulsion strength $\eta$ for $N=6$ spin-polarized electrons in a $3D$ harmonic trap for $\lambda=0.5$ and $\beta=2$.
}
\end{figure}

The final example to be investigated in this work is the application of our method to a $3D$ system. This is shown in Fig.~\ref{fig:trap_results_3D} for $N=6$ electrons in a $3D$ harmonic confinement for $\lambda=0.5$ and $\beta=2$. As before, the green crosses depict the raw PIMC data for $\braket{\hat H}_\eta$ and the dash-dotted red-line a direct extrapolation thereof according to Eq.~(\ref{eq:c-fit}). At these conditions, we find an average sign of $S\sim10^{-4}$ [cf.~Fig.~\ref{fig:speed_results_3D}], which means that standard PIMC simulations are only feasible for $\eta\gtrsim0.1$. Yet, the fit function from Eq.~(\ref{eq:c-fit}) evidently allows for a controlled extrapolation to $\eta\to0$, which is further highlighted by the dotted light grey horizontal line, depicting an uncertainty interval of $0.3\%$ around the extrapolated value.

We next explore the estimation of the thermodynamic integration correction introduced in Sec.~\ref{sec:invernizzi}. To this end, we show $\phi(\eta)$ in the bottom panel of Fig.~\ref{fig:trap_results_3D} and the green crosses again show the PIMC data. The dashed black (dashed-double-dotted yellow) line corresponds to a fit using the functional form from Eq.~(\ref{eq:eta_fit}) for $\eta\in[0.2,0.5]$ ($\eta\in[0.1,0.5]$). Naturally, these representations are in excellent agreement for large $\eta$, whereas some deviations appear for the extrapolation of $\eta\to0$.

Using one of these representations as input to evaluate the thermodynamic integration correction from Eq.~(\ref{eq:invernizzi_estimate}) gives the yellow stars (Eq.~(\ref{eq:eta_fit}) for $\eta\in[0.1,0.5]$. For completeness, we mention that even using a spline-representation of $\phi(\eta)$ would result in almost indistinguishable energy values, which further validates the observation from the previous Sec.~\ref{sec:invernizzi_results} that the particular extrapolation of $\phi(\eta)$ to $\eta\to0$  hardly influence the quality of the  corrected energies. At the same time, we observe a distinct entropic contribution to Eq.~(\ref{eq:invernizzi_estimate}) at these conditions, and the dashed-double-dotted yellow line corresponds to a linear fit to the corrected data points, cf.~Eq.~(\ref{eq:entropy_fit}).
Evidently, the linear representation is in perfect agreement to the yellow stars over the entire depicted $\eta$-range, and the $\eta\to0$ limit nicely agrees with the direct extrapolation of the uncorrected PIMC data points.

We conclude this section with an analysis of the average sign $S$ and the corresponding speed-up $T$ [cf.~Eq.~(\ref{eq:speed-up})], shown in Fig.~\ref{fig:speed_results_3D}. The green crosses show the PIMC data for $S$ (left $y$-axis) and exhibit a monotonic increase similar to the $2D$ case. Further, the respective speed-up is depicted as the solid black curve (right $y$-axis) and attains values exceeding $T=10^3$ for $\eta=0.5$.

\section{Summary and Discussion\label{sec:summary}}

% We have presented a new approach for \textit{ab initio} path-integral Monte-Carlo simulation of fermions at finite temperature based on the well-known Bogoliubov inequality. More specifically, we have extended the idea by Hirshberg \textit{et al.}~\cite{Hirshberg_JCP_2020} by incorporating an additional repulsive term to the Hamiltonian phenomenologically modelling the Pauli repulsion, which gives us two distinct advantages: i) it allows for a controlled extrapolation to the exact energy value of the original system of interest and ii) one can compute a correction term using thermodynamic integration, which makes the method even more reliable.

We present a new approach for PIMC simulations of fermions at finite temperature, that extends and improves the idea by Hirshberg et al.~\cite{Hirshberg_JCP_2020}, that was based on the Bogoliubov inequality. We add to the Hamiltonian a repulsive term that mimics Pauli repulsion at short range and is proportional to a coupling parameter $\eta$.
This significantly improves the efficiency of the simulations, by increasing the average sign $S$.
We then propose two simple post-processing schemes to recover the energy of the original system.
The first one is based on the Bogoliubov inequality and consists of extrapolating $\braket{\hat H}_{\eta}$ obtained for various values of $\eta$ to the limit $\eta\to0$.
The second one instead is based on thermodynamic integration and also relies on an extrapolation, but only of the perturbation term $\phi$ and not of the whole energy. Combined with the fact that it is based on an exact relation and not an inequality, this makes the second scheme generally more reliable.
We believe that having two distinct schemes for evaluating the energy, starting from the same simulations, is a great advantage, and makes the method more robust against possible poor extrapolation choices. For all the systems considered here, the two schemes provided compatible energy estimates. Most importantly, they allowed evaluating accurate estimates of energies for conditions in which standard PIMC simulations were not feasible.

As a practical application, we have investigated electrons both in $2D$ and $3D$ quantum dots, and our method works very well in both of these cases. In particular, we have found that a direct extrapolation  by itself allows for a speed-up of up to $10^6$ (cf.~Fig.~\ref{fig:trap_speed_beta}), while retaining a relative accuracy of $0.1\%$, which is fully sufficient for practical purposes. This, in turn, allows to double the feasible $\beta$-range as compared to standard PIMC.

The investigation of the thermodynamic integration correction introduced in Sec.~\ref{sec:invernizzi} has revealed that this approach makes the estimation of the energy of the original system even more reliable. Here, the main challenge is given by the construction of a representation of $\phi(\eta)$  with no data points being available below some minimum value. As a solution, we have introduced a suitable empirical fit function [cf.~Eq.~(\ref{eq:eta_fit})] that, remarkably, allows for a highly reliable extrapolation using only a few data points at large $\eta$ as input. At low temperature, this representation allows to accurately estimate the energy of the original system of interest. For higher temperatures, there emerges an additional entropic contribution, which remains \textit{a priori} unknown. Still, we stress that the latter only constitutes a fraction of the full difference between the modified and the original system, and, empirically, exhibits a linear dependence on $\eta$ (cf.~Fig.~\ref{fig:Invernizzi_analysis_N4_high}).

While the present results are certainly encouraging, much can be done to improve the method further. From a technical point of view, implementing the sampling procedure proposed in Ref.~\citenum{Invernizzi2020} would make the method more efficient, requiring only a single simulation for estimating $E(\eta)$ and $\phi(\eta)$ in the whole range of $\eta$ values. Furthermore, we mention that, despite a speed-up of $10^6$, the fermion sign problem has not been completely removed and, while our approach significantly extends the range of accessible parameters, it still suffers from exponential increase in computing time at low enough temperatures. We have also restricted ourselves to the investigation of energy values, and the adaption of the method to other observables such as pair distributions, structure factors, or the different contributions to $E$ like the kinetic or external potential energies is of high interest. Similarly, we have solely used a dipolar repulsive potential $\Psi(r_1,r_2)$ and the optimization of $\Psi$ might substantially improve the approach.

A particularly interesting topic for future research is given by the application to other systems, with warm-dense matter~\cite{new_POP} being a promising candidate. Finally, we stress that our approach is quite general and can readily be adapted to other simulation methods for fermions both in the ground state and at finite temperate.

\section*{Acknowledgments}

This work was partly funded by the Center of Advanced Systems Understanding (CASUS) which is financed by Germany's Federal Ministry of Education and Research (BMBF) and by the Saxon Ministry for Science, Art, and Tourism (SMWK) with tax funds on the basis of the budget approved by the Saxon State Parliament.
M.I. acknowledges support from the Swiss National Science Foundation through the NCCR MARVEL.
The PIMC calculations were carried out at the Norddeutscher Verbund f\"ur Hoch- und H\"ochstleistungsrechnen (HLRN) under grant shp00015 and on a Bull Cluster at the Center for Information Services and High Performace Computing (ZIH) at Technische Universit\"at Dresden.

%%%%%%%%%%%%%%%%%%%%%%%%%%%%%%%%%%%%%%%%%%%%%%%%%%%%%%%%%%%%%%%%%%%%%%%%%%%%%%%%
% literature
%%%%%%%%%%%%%%%%%%%%%%%%%%%%%%%%%%%%%%%%%%%%%%%%%%%%%%%%%%%%%%%%%%%%%%%%%%%%%%%%
\bibliography{bibliography}{}

%merlin.mbs apsrev4-1.bst 2010-07-25 4.21a (PWD, AO, DPC) hacked
%Control: key (0)
%Control: author (0) dotless jnrlst
%Control: editor formatted (1) identically to author
%Control: production of article title (0) allowed
%Control: page (1) range
%Control: year (0) verbatim
%Control: production of eprint (0) enabled
\begin{thebibliography}{89}%
\makeatletter
\providecommand \@ifxundefined [1]{%
 \@ifx{#1\undefined}
}%
\providecommand \@ifnum [1]{%
 \ifnum #1\expandafter \@firstoftwo
 \else \expandafter \@secondoftwo
 \fi
}%
\providecommand \@ifx [1]{%
 \ifx #1\expandafter \@firstoftwo
 \else \expandafter \@secondoftwo
 \fi
}%
\providecommand \natexlab [1]{#1}%
\providecommand \enquote  [1]{``#1''}%
\providecommand \bibnamefont  [1]{#1}%
\providecommand \bibfnamefont [1]{#1}%
\providecommand \citenamefont [1]{#1}%
\providecommand \href@noop [0]{\@secondoftwo}%
\providecommand \href [0]{\begingroup \@sanitize@url \@href}%
\providecommand \@href[1]{\@@startlink{#1}\@@href}%
\providecommand \@@href[1]{\endgroup#1\@@endlink}%
\providecommand \@sanitize@url [0]{\catcode `\\12\catcode `\$12\catcode
  `\&12\catcode `\#12\catcode `\^12\catcode `\_12\catcode `\%12\relax}%
\providecommand \@@startlink[1]{}%
\providecommand \@@endlink[0]{}%
\providecommand \url  [0]{\begingroup\@sanitize@url \@url }%
\providecommand \@url [1]{\endgroup\@href {#1}{\urlprefix }}%
\providecommand \urlprefix  [0]{URL }%
\providecommand \Eprint [0]{\href }%
\providecommand \doibase [0]{http://dx.doi.org/}%
\providecommand \selectlanguage [0]{\@gobble}%
\providecommand \bibinfo  [0]{\@secondoftwo}%
\providecommand \bibfield  [0]{\@secondoftwo}%
\providecommand \translation [1]{[#1]}%
\providecommand \BibitemOpen [0]{}%
\providecommand \bibitemStop [0]{}%
\providecommand \bibitemNoStop [0]{.\EOS\space}%
\providecommand \EOS [0]{\spacefactor3000\relax}%
\providecommand \BibitemShut  [1]{\csname bibitem#1\endcsname}%
\let\auto@bib@innerbib\@empty
%</preamble>
\bibitem [{\citenamefont {Giuliani}\ and\ \citenamefont
  {Vignale}(2008)}]{quantum_theory}%
  \BibitemOpen
  \bibfield  {author} {\bibinfo {author} {\bibfnamefont {G.}~\bibnamefont
  {Giuliani}}\ and\ \bibinfo {author} {\bibfnamefont {G.}~\bibnamefont
  {Vignale}},\ }\href@noop {} {\emph {\bibinfo {title} {Quantum Theory of the
  Electron Liquid}}}\ (\bibinfo  {publisher} {Cambridge University Press},\
  \bibinfo {address} {Cambridge},\ \bibinfo {year} {2008})\BibitemShut
  {NoStop}%
\bibitem [{\citenamefont {Loh}\ \emph {et~al.}(1990)\citenamefont {Loh},
  \citenamefont {Gubernatis}, \citenamefont {Scalettar}, \citenamefont {White},
  \citenamefont {Scalapino},\ and\ \citenamefont
  {Sugar}}]{Loh_sign_problem_PRB_1990}%
  \BibitemOpen
  \bibfield  {author} {\bibinfo {author} {\bibfnamefont {E.~Y.}\ \bibnamefont
  {Loh}}, \bibinfo {author} {\bibfnamefont {J.~E.}\ \bibnamefont {Gubernatis}},
  \bibinfo {author} {\bibfnamefont {R.~T.}\ \bibnamefont {Scalettar}}, \bibinfo
  {author} {\bibfnamefont {S.~R.}\ \bibnamefont {White}}, \bibinfo {author}
  {\bibfnamefont {D.~J.}\ \bibnamefont {Scalapino}}, \ and\ \bibinfo {author}
  {\bibfnamefont {R.~L.}\ \bibnamefont {Sugar}},\ }\bibfield  {title} {\enquote
  {\bibinfo {title} {Sign problem in the numerical simulation of many-electron
  systems},}\ }\href {\doibase 10.1103/PhysRevB.41.9301} {\bibfield  {journal}
  {\bibinfo  {journal} {Phys. Rev. B}\ }\textbf {\bibinfo {volume} {41}},\
  \bibinfo {pages} {9301--9307} (\bibinfo {year} {1990})}\BibitemShut {NoStop}%
\bibitem [{\citenamefont {Troyer}\ and\ \citenamefont {Wiese}(2005)}]{troyer}%
  \BibitemOpen
  \bibfield  {author} {\bibinfo {author} {\bibfnamefont {M.}~\bibnamefont
  {Troyer}}\ and\ \bibinfo {author} {\bibfnamefont {U.~J.}\ \bibnamefont
  {Wiese}},\ }\bibfield  {title} {\enquote {\bibinfo {title} {Computational
  complexity and fundamental limitations to fermionic quantum {M}onte {C}arlo
  simulations},}\ }\href
  {http://link.aps.org/doi/10.1103/PhysRevLett.94.170201} {\bibfield  {journal}
  {\bibinfo  {journal} {Phys. Rev. Lett}\ }\textbf {\bibinfo {volume} {94}},\
  \bibinfo {pages} {170201} (\bibinfo {year} {2005})}\BibitemShut {NoStop}%
\bibitem [{\citenamefont {Dornheim}(2019)}]{dornheim_sign_problem}%
  \BibitemOpen
  \bibfield  {author} {\bibinfo {author} {\bibfnamefont {T.}~\bibnamefont
  {Dornheim}},\ }\bibfield  {title} {\enquote {\bibinfo {title} {Fermion sign
  problem in path integral {M}onte {C}arlo simulations: Quantum dots, ultracold
  atoms, and warm dense matter},}\ }\href
  {https://journals.aps.org/pre/abstract/10.1103/PhysRevE.100.023307}
  {\bibfield  {journal} {\bibinfo  {journal} {Phys. Rev. E}\ }\textbf {\bibinfo
  {volume} {100}},\ \bibinfo {pages} {023307} (\bibinfo {year}
  {2019})}\BibitemShut {NoStop}%
\bibitem [{\citenamefont {Ceperley}\ and\ \citenamefont
  {Alder}(1980)}]{ceperley_alder_UEG}%
  \BibitemOpen
  \bibfield  {author} {\bibinfo {author} {\bibfnamefont {D.~M.}\ \bibnamefont
  {Ceperley}}\ and\ \bibinfo {author} {\bibfnamefont {B.~J.}\ \bibnamefont
  {Alder}},\ }\bibfield  {title} {\enquote {\bibinfo {title} {Ground state of
  the electron gas by a stochastic method},}\ }\href {\doibase
  10.1103/PhysRevLett.45.566} {\bibfield  {journal} {\bibinfo  {journal} {Phys.
  Rev. Lett.}\ }\textbf {\bibinfo {volume} {45}},\ \bibinfo {pages} {566--569}
  (\bibinfo {year} {1980})}\BibitemShut {NoStop}%
\bibitem [{\citenamefont {Perdew}\ and\ \citenamefont
  {Zunger}(1981)}]{perdew_zunger_PRB}%
  \BibitemOpen
  \bibfield  {author} {\bibinfo {author} {\bibfnamefont {J.~P.}\ \bibnamefont
  {Perdew}}\ and\ \bibinfo {author} {\bibfnamefont {Alex}\ \bibnamefont
  {Zunger}},\ }\bibfield  {title} {\enquote {\bibinfo {title} {Self-interaction
  correction to density-functional approximations for many-electron systems},}\
  }\href {\doibase 10.1103/PhysRevB.23.5048} {\bibfield  {journal} {\bibinfo
  {journal} {Phys. Rev. B}\ }\textbf {\bibinfo {volume} {23}},\ \bibinfo
  {pages} {5048--5079} (\bibinfo {year} {1981})}\BibitemShut {NoStop}%
\bibitem [{\citenamefont {Perdew}\ \emph {et~al.}(1996)\citenamefont {Perdew},
  \citenamefont {Burke},\ and\ \citenamefont
  {Ernzerhof}}]{perdew1996generalized}%
  \BibitemOpen
  \bibfield  {author} {\bibinfo {author} {\bibfnamefont {John~P}\ \bibnamefont
  {Perdew}}, \bibinfo {author} {\bibfnamefont {Kieron}\ \bibnamefont {Burke}},
  \ and\ \bibinfo {author} {\bibfnamefont {Matthias}\ \bibnamefont
  {Ernzerhof}},\ }\bibfield  {title} {\enquote {\bibinfo {title} {Generalized
  gradient approximation made simple},}\ }\href
  {https://journals.aps.org/prl/abstract/10.1103/PhysRevLett.77.3865}
  {\bibfield  {journal} {\bibinfo  {journal} {Physical Review Letters}\
  }\textbf {\bibinfo {volume} {77}},\ \bibinfo {pages} {3865} (\bibinfo {year}
  {1996})}\BibitemShut {NoStop}%
\bibitem [{\citenamefont {Burke}(2012)}]{Burke_perspective_JCP_2012}%
  \BibitemOpen
  \bibfield  {author} {\bibinfo {author} {\bibfnamefont {Kieron}\ \bibnamefont
  {Burke}},\ }\bibfield  {title} {\enquote {\bibinfo {title} {Perspective on
  density functional theory},}\ }\href {\doibase 10.1063/1.4704546} {\bibfield
  {journal} {\bibinfo  {journal} {The Journal of Chemical Physics}\ }\textbf
  {\bibinfo {volume} {136}},\ \bibinfo {pages} {150901} (\bibinfo {year}
  {2012})},\ \Eprint {http://arxiv.org/abs/https://doi.org/10.1063/1.4704546}
  {https://doi.org/10.1063/1.4704546} \BibitemShut {NoStop}%
\bibitem [{\citenamefont {Ceperley}(1991)}]{Ceperley1991}%
  \BibitemOpen
  \bibfield  {author} {\bibinfo {author} {\bibfnamefont {D.~M.}\ \bibnamefont
  {Ceperley}},\ }\bibfield  {title} {\enquote {\bibinfo {title} {Fermion
  nodes},}\ }\href {\doibase 10.1007/BF01030009} {\bibfield  {journal}
  {\bibinfo  {journal} {Journal of Statistical Physics}\ }\textbf {\bibinfo
  {volume} {63}},\ \bibinfo {pages} {1237--1267} (\bibinfo {year}
  {1991})}\BibitemShut {NoStop}%
\bibitem [{\citenamefont {Anderson}(1995)}]{Anderson_fixed_node}%
  \BibitemOpen
  \bibfield  {author} {\bibinfo {author} {\bibfnamefont {James~B.}\
  \bibnamefont {Anderson}},\ }\bibfield  {title} {\enquote {\bibinfo {title}
  {Fixed-node quantum monte carlo},}\ }\href {\doibase
  10.1080/01442359509353305} {\bibfield  {journal} {\bibinfo  {journal}
  {International Reviews in Physical Chemistry}\ }\textbf {\bibinfo {volume}
  {14}},\ \bibinfo {pages} {85--112} (\bibinfo {year} {1995})},\ \Eprint
  {http://arxiv.org/abs/https://doi.org/10.1080/01442359509353305}
  {https://doi.org/10.1080/01442359509353305} \BibitemShut {NoStop}%
\bibitem [{\citenamefont {Foulkes}\ \emph {et~al.}(2001)\citenamefont
  {Foulkes}, \citenamefont {Mitas}, \citenamefont {Needs},\ and\ \citenamefont
  {Rajagopal}}]{Foulkes_QMC_RMP}%
  \BibitemOpen
  \bibfield  {author} {\bibinfo {author} {\bibfnamefont {W.~M.~C.}\
  \bibnamefont {Foulkes}}, \bibinfo {author} {\bibfnamefont {L.}~\bibnamefont
  {Mitas}}, \bibinfo {author} {\bibfnamefont {R.~J.}\ \bibnamefont {Needs}}, \
  and\ \bibinfo {author} {\bibfnamefont {G.}~\bibnamefont {Rajagopal}},\
  }\bibfield  {title} {\enquote {\bibinfo {title} {Quantum monte carlo
  simulations of solids},}\ }\href {\doibase 10.1103/RevModPhys.73.33}
  {\bibfield  {journal} {\bibinfo  {journal} {Rev. Mod. Phys.}\ }\textbf
  {\bibinfo {volume} {73}},\ \bibinfo {pages} {33--83} (\bibinfo {year}
  {2001})}\BibitemShut {NoStop}%
\bibitem [{\citenamefont {L\'opez~R\'{\i}os}\ \emph {et~al.}(2006)\citenamefont
  {L\'opez~R\'{\i}os}, \citenamefont {Ma}, \citenamefont {Drummond},
  \citenamefont {Towler},\ and\ \citenamefont {Needs}}]{backflow}%
  \BibitemOpen
  \bibfield  {author} {\bibinfo {author} {\bibfnamefont {P.}~\bibnamefont
  {L\'opez~R\'{\i}os}}, \bibinfo {author} {\bibfnamefont {A.}~\bibnamefont
  {Ma}}, \bibinfo {author} {\bibfnamefont {N.~D.}\ \bibnamefont {Drummond}},
  \bibinfo {author} {\bibfnamefont {M.~D.}\ \bibnamefont {Towler}}, \ and\
  \bibinfo {author} {\bibfnamefont {R.~J.}\ \bibnamefont {Needs}},\ }\bibfield
  {title} {\enquote {\bibinfo {title} {Inhomogeneous backflow transformations
  in quantum monte carlo calculations},}\ }\href {\doibase
  10.1103/PhysRevE.74.066701} {\bibfield  {journal} {\bibinfo  {journal} {Phys.
  Rev. E}\ }\textbf {\bibinfo {volume} {74}},\ \bibinfo {pages} {066701}
  (\bibinfo {year} {2006})}\BibitemShut {NoStop}%
\bibitem [{\citenamefont {Needs}\ \emph {et~al.}(2009)\citenamefont {Needs},
  \citenamefont {Towler}, \citenamefont {Drummond},\ and\ \citenamefont
  {R{\'{\i}}os}}]{Needs_2009}%
  \BibitemOpen
  \bibfield  {author} {\bibinfo {author} {\bibfnamefont {R~J}\ \bibnamefont
  {Needs}}, \bibinfo {author} {\bibfnamefont {M~D}\ \bibnamefont {Towler}},
  \bibinfo {author} {\bibfnamefont {N~D}\ \bibnamefont {Drummond}}, \ and\
  \bibinfo {author} {\bibfnamefont {P~L{\'{o}}pez}\ \bibnamefont
  {R{\'{\i}}os}},\ }\bibfield  {title} {\enquote {\bibinfo {title} {Continuum
  variational and diffusion quantum monte carlo calculations},}\ }\href
  {\doibase 10.1088/0953-8984/22/2/023201} {\bibfield  {journal} {\bibinfo
  {journal} {Journal of Physics: Condensed Matter}\ }\textbf {\bibinfo {volume}
  {22}},\ \bibinfo {pages} {023201} (\bibinfo {year} {2009})}\BibitemShut
  {NoStop}%
\bibitem [{\citenamefont {Booth}\ \emph {et~al.}(2009)\citenamefont {Booth},
  \citenamefont {Thom},\ and\ \citenamefont {Alavi}}]{Booth_FCIQMC_2009}%
  \BibitemOpen
  \bibfield  {author} {\bibinfo {author} {\bibfnamefont {George~H.}\
  \bibnamefont {Booth}}, \bibinfo {author} {\bibfnamefont {Alex J.~W.}\
  \bibnamefont {Thom}}, \ and\ \bibinfo {author} {\bibfnamefont {Ali}\
  \bibnamefont {Alavi}},\ }\bibfield  {title} {\enquote {\bibinfo {title}
  {Fermion monte carlo without fixed nodes: A game of life, death, and
  annihilation in slater determinant space},}\ }\href {\doibase
  10.1063/1.3193710} {\bibfield  {journal} {\bibinfo  {journal} {The Journal of
  Chemical Physics}\ }\textbf {\bibinfo {volume} {131}},\ \bibinfo {pages}
  {054106} (\bibinfo {year} {2009})},\ \Eprint
  {http://arxiv.org/abs/https://aip.scitation.org/doi/pdf/10.1063/1.3193710}
  {https://aip.scitation.org/doi/pdf/10.1063/1.3193710} \BibitemShut {NoStop}%
\bibitem [{\citenamefont {Honma}\ \emph {et~al.}(1995)\citenamefont {Honma},
  \citenamefont {Mizusaki},\ and\ \citenamefont {Otsuka}}]{Auxiliary}%
  \BibitemOpen
  \bibfield  {author} {\bibinfo {author} {\bibfnamefont {Michio}\ \bibnamefont
  {Honma}}, \bibinfo {author} {\bibfnamefont {Takahiro}\ \bibnamefont
  {Mizusaki}}, \ and\ \bibinfo {author} {\bibfnamefont {Takaharu}\ \bibnamefont
  {Otsuka}},\ }\bibfield  {title} {\enquote {\bibinfo {title} {Diagonalization
  of hamiltonians for many-body systems by auxiliary field quantum monte carlo
  technique},}\ }\href {\doibase 10.1103/PhysRevLett.75.1284} {\bibfield
  {journal} {\bibinfo  {journal} {Phys. Rev. Lett.}\ }\textbf {\bibinfo
  {volume} {75}},\ \bibinfo {pages} {1284--1287} (\bibinfo {year}
  {1995})}\BibitemShut {NoStop}%
\bibitem [{\citenamefont {Motta}\ and\ \citenamefont
  {Zhang}(2018)}]{Auxiliary2}%
  \BibitemOpen
  \bibfield  {author} {\bibinfo {author} {\bibfnamefont {Mario}\ \bibnamefont
  {Motta}}\ and\ \bibinfo {author} {\bibfnamefont {Shiwei}\ \bibnamefont
  {Zhang}},\ }\bibfield  {title} {\enquote {\bibinfo {title} {Ab initio
  computations of molecular systems by the auxiliary-field quantum monte carlo
  method},}\ }\href {\doibase 10.1002/wcms.1364} {\bibfield  {journal}
  {\bibinfo  {journal} {WIREs Computational Molecular Science}\ }\textbf
  {\bibinfo {volume} {8}},\ \bibinfo {pages} {e1364} (\bibinfo {year}
  {2018})},\ \Eprint
  {http://arxiv.org/abs/https://onlinelibrary.wiley.com/doi/pdf/10.1002/wcms.1364}
  {https://onlinelibrary.wiley.com/doi/pdf/10.1002/wcms.1364} \BibitemShut
  {NoStop}%
\bibitem [{\citenamefont {LeBlanc}\ \emph {et~al.}(2015)\citenamefont
  {LeBlanc}, \citenamefont {Antipov}, \citenamefont {Becca}, \citenamefont
  {Bulik}, \citenamefont {Chan}, \citenamefont {Chung}, \citenamefont {Deng},
  \citenamefont {Ferrero}, \citenamefont {Henderson}, \citenamefont
  {Jim\'enez-Hoyos}, \citenamefont {Kozik}, \citenamefont {Liu}, \citenamefont
  {Millis}, \citenamefont {Prokof'ev}, \citenamefont {Qin}, \citenamefont
  {Scuseria}, \citenamefont {Shi}, \citenamefont {Svistunov}, \citenamefont
  {Tocchio}, \citenamefont {Tupitsyn}, \citenamefont {White}, \citenamefont
  {Zhang}, \citenamefont {Zheng}, \citenamefont {Zhu},\ and\ \citenamefont
  {Gull}}]{PhysRevX.5.041041}%
  \BibitemOpen
  \bibfield  {author} {\bibinfo {author} {\bibfnamefont {J.~P.~F.}\
  \bibnamefont {LeBlanc}}, \bibinfo {author} {\bibfnamefont {Andrey~E.}\
  \bibnamefont {Antipov}}, \bibinfo {author} {\bibfnamefont {Federico}\
  \bibnamefont {Becca}}, \bibinfo {author} {\bibfnamefont {Ireneusz~W.}\
  \bibnamefont {Bulik}}, \bibinfo {author} {\bibfnamefont {Garnet Kin-Lic}\
  \bibnamefont {Chan}}, \bibinfo {author} {\bibfnamefont {Chia-Min}\
  \bibnamefont {Chung}}, \bibinfo {author} {\bibfnamefont {Youjin}\
  \bibnamefont {Deng}}, \bibinfo {author} {\bibfnamefont {Michel}\ \bibnamefont
  {Ferrero}}, \bibinfo {author} {\bibfnamefont {Thomas~M.}\ \bibnamefont
  {Henderson}}, \bibinfo {author} {\bibfnamefont {Carlos~A.}\ \bibnamefont
  {Jim\'enez-Hoyos}}, \bibinfo {author} {\bibfnamefont {E.}~\bibnamefont
  {Kozik}}, \bibinfo {author} {\bibfnamefont {Xuan-Wen}\ \bibnamefont {Liu}},
  \bibinfo {author} {\bibfnamefont {Andrew~J.}\ \bibnamefont {Millis}},
  \bibinfo {author} {\bibfnamefont {N.~V.}\ \bibnamefont {Prokof'ev}}, \bibinfo
  {author} {\bibfnamefont {Mingpu}\ \bibnamefont {Qin}}, \bibinfo {author}
  {\bibfnamefont {Gustavo~E.}\ \bibnamefont {Scuseria}}, \bibinfo {author}
  {\bibfnamefont {Hao}\ \bibnamefont {Shi}}, \bibinfo {author} {\bibfnamefont
  {B.~V.}\ \bibnamefont {Svistunov}}, \bibinfo {author} {\bibfnamefont
  {Luca~F.}\ \bibnamefont {Tocchio}}, \bibinfo {author} {\bibfnamefont {I.~S.}\
  \bibnamefont {Tupitsyn}}, \bibinfo {author} {\bibfnamefont {Steven~R.}\
  \bibnamefont {White}}, \bibinfo {author} {\bibfnamefont {Shiwei}\
  \bibnamefont {Zhang}}, \bibinfo {author} {\bibfnamefont {Bo-Xiao}\
  \bibnamefont {Zheng}}, \bibinfo {author} {\bibfnamefont {Zhenyue}\
  \bibnamefont {Zhu}}, \ and\ \bibinfo {author} {\bibfnamefont {Emanuel}\
  \bibnamefont {Gull}} (\bibinfo {collaboration} {Simons Collaboration on the
  Many-Electron Problem}),\ }\bibfield  {title} {\enquote {\bibinfo {title}
  {Solutions of the two-dimensional hubbard model: Benchmarks and results from
  a wide range of numerical algorithms},}\ }\href {\doibase
  10.1103/PhysRevX.5.041041} {\bibfield  {journal} {\bibinfo  {journal} {Phys.
  Rev. X}\ }\textbf {\bibinfo {volume} {5}},\ \bibinfo {pages} {041041}
  (\bibinfo {year} {2015})}\BibitemShut {NoStop}%
\bibitem [{\citenamefont {Booth}\ \emph {et~al.}(2013)\citenamefont {Booth},
  \citenamefont {Gr{\"u}neis}, \citenamefont {Kresse},\ and\ \citenamefont
  {Alavi}}]{Booth2013}%
  \BibitemOpen
  \bibfield  {author} {\bibinfo {author} {\bibfnamefont {George~H.}\
  \bibnamefont {Booth}}, \bibinfo {author} {\bibfnamefont {Andreas}\
  \bibnamefont {Gr{\"u}neis}}, \bibinfo {author} {\bibfnamefont {Georg}\
  \bibnamefont {Kresse}}, \ and\ \bibinfo {author} {\bibfnamefont {Ali}\
  \bibnamefont {Alavi}},\ }\bibfield  {title} {\enquote {\bibinfo {title}
  {Towards an exact description of electronic wavefunctions in real solids},}\
  }\href {\doibase 10.1038/nature11770} {\bibfield  {journal} {\bibinfo
  {journal} {Nature}\ }\textbf {\bibinfo {volume} {493}},\ \bibinfo {pages}
  {365--370} (\bibinfo {year} {2013})}\BibitemShut {NoStop}%
\bibitem [{\citenamefont {Brown}\ \emph {et~al.}(2013)\citenamefont {Brown},
  \citenamefont {Clark}, \citenamefont {DuBois},\ and\ \citenamefont
  {Ceperley}}]{Brown_ethan}%
  \BibitemOpen
  \bibfield  {author} {\bibinfo {author} {\bibfnamefont {Ethan~W.}\
  \bibnamefont {Brown}}, \bibinfo {author} {\bibfnamefont {Bryan~K.}\
  \bibnamefont {Clark}}, \bibinfo {author} {\bibfnamefont {Jonathan~L.}\
  \bibnamefont {DuBois}}, \ and\ \bibinfo {author} {\bibfnamefont {David~M.}\
  \bibnamefont {Ceperley}},\ }\bibfield  {title} {\enquote {\bibinfo {title}
  {Path-integral monte carlo simulation of the warm dense homogeneous electron
  gas},}\ }\href {\doibase 10.1103/PhysRevLett.110.146405} {\bibfield
  {journal} {\bibinfo  {journal} {Phys. Rev. Lett.}\ }\textbf {\bibinfo
  {volume} {110}},\ \bibinfo {pages} {146405} (\bibinfo {year}
  {2013})}\BibitemShut {NoStop}%
\bibitem [{\citenamefont {Schoof}\ \emph {et~al.}(2011)\citenamefont {Schoof},
  \citenamefont {Bonitz}, \citenamefont {Filinov}, \citenamefont {Hochstuhl},\
  and\ \citenamefont {Dufty}}]{Schoof_CPP_2011}%
  \BibitemOpen
  \bibfield  {author} {\bibinfo {author} {\bibfnamefont {T.}~\bibnamefont
  {Schoof}}, \bibinfo {author} {\bibfnamefont {M.}~\bibnamefont {Bonitz}},
  \bibinfo {author} {\bibfnamefont {A.}~\bibnamefont {Filinov}}, \bibinfo
  {author} {\bibfnamefont {D.}~\bibnamefont {Hochstuhl}}, \ and\ \bibinfo
  {author} {\bibfnamefont {J.W.}\ \bibnamefont {Dufty}},\ }\bibfield  {title}
  {\enquote {\bibinfo {title} {Configuration path integral monte carlo},}\
  }\href {\doibase 10.1002/ctpp.201100012} {\bibfield  {journal} {\bibinfo
  {journal} {Contributions to Plasma Physics}\ }\textbf {\bibinfo {volume}
  {51}},\ \bibinfo {pages} {687--697} (\bibinfo {year} {2011})}\BibitemShut
  {NoStop}%
\bibitem [{\citenamefont {Dornheim}\ \emph
  {et~al.}(2015{\natexlab{a}})\citenamefont {Dornheim}, \citenamefont {Groth},
  \citenamefont {Filinov},\ and\ \citenamefont {Bonitz}}]{Dornheim_NJP_2015}%
  \BibitemOpen
  \bibfield  {author} {\bibinfo {author} {\bibfnamefont {Tobias}\ \bibnamefont
  {Dornheim}}, \bibinfo {author} {\bibfnamefont {Simon}\ \bibnamefont {Groth}},
  \bibinfo {author} {\bibfnamefont {Alexey}\ \bibnamefont {Filinov}}, \ and\
  \bibinfo {author} {\bibfnamefont {Michael}\ \bibnamefont {Bonitz}},\
  }\bibfield  {title} {\enquote {\bibinfo {title} {Permutation blocking path
  integral monte carlo: a highly efficient approach to the simulation of
  strongly degenerate non-ideal fermions},}\ }\href {\doibase
  10.1088/1367-2630/17/7/073017} {\bibfield  {journal} {\bibinfo  {journal}
  {New Journal of Physics}\ }\textbf {\bibinfo {volume} {17}},\ \bibinfo
  {pages} {073017} (\bibinfo {year} {2015}{\natexlab{a}})}\BibitemShut
  {NoStop}%
\bibitem [{\citenamefont {Dornheim}\ \emph {et~al.}(2017)\citenamefont
  {Dornheim}, \citenamefont {Groth}, \citenamefont {Malone}, \citenamefont
  {Schoof}, \citenamefont {Sjostrom}, \citenamefont {Foulkes},\ and\
  \citenamefont {Bonitz}}]{dornheim_POP}%
  \BibitemOpen
  \bibfield  {author} {\bibinfo {author} {\bibfnamefont {Tobias}\ \bibnamefont
  {Dornheim}}, \bibinfo {author} {\bibfnamefont {Simon}\ \bibnamefont {Groth}},
  \bibinfo {author} {\bibfnamefont {Fionn~D.}\ \bibnamefont {Malone}}, \bibinfo
  {author} {\bibfnamefont {Tim}\ \bibnamefont {Schoof}}, \bibinfo {author}
  {\bibfnamefont {Travis}\ \bibnamefont {Sjostrom}}, \bibinfo {author}
  {\bibfnamefont {W.~M.~C.}\ \bibnamefont {Foulkes}}, \ and\ \bibinfo {author}
  {\bibfnamefont {Michael}\ \bibnamefont {Bonitz}},\ }\bibfield  {title}
  {\enquote {\bibinfo {title} {Ab initio quantum monte carlo simulation of the
  warm dense electron gas},}\ }\href {\doibase 10.1063/1.4977920} {\bibfield
  {journal} {\bibinfo  {journal} {Physics of Plasmas}\ }\textbf {\bibinfo
  {volume} {24}},\ \bibinfo {pages} {056303} (\bibinfo {year} {2017})},\
  \Eprint {http://arxiv.org/abs/https://doi.org/10.1063/1.4977920}
  {https://doi.org/10.1063/1.4977920} \BibitemShut {NoStop}%
\bibitem [{\citenamefont {Blunt}\ \emph {et~al.}(2014)\citenamefont {Blunt},
  \citenamefont {Rogers}, \citenamefont {Spencer},\ and\ \citenamefont
  {Foulkes}}]{Blunt_PRB_2014}%
  \BibitemOpen
  \bibfield  {author} {\bibinfo {author} {\bibfnamefont {N.~S.}\ \bibnamefont
  {Blunt}}, \bibinfo {author} {\bibfnamefont {T.~W.}\ \bibnamefont {Rogers}},
  \bibinfo {author} {\bibfnamefont {J.~S.}\ \bibnamefont {Spencer}}, \ and\
  \bibinfo {author} {\bibfnamefont {W.~M.~C.}\ \bibnamefont {Foulkes}},\
  }\bibfield  {title} {\enquote {\bibinfo {title} {Density-matrix quantum monte
  carlo method},}\ }\href {\doibase 10.1103/PhysRevB.89.245124} {\bibfield
  {journal} {\bibinfo  {journal} {Phys. Rev. B}\ }\textbf {\bibinfo {volume}
  {89}},\ \bibinfo {pages} {245124} (\bibinfo {year} {2014})}\BibitemShut
  {NoStop}%
\bibitem [{\citenamefont {Liu}\ \emph {et~al.}(2018)\citenamefont {Liu},
  \citenamefont {Cho},\ and\ \citenamefont
  {Rubenstein}}]{Rubenstein_auxiliary_finite_T}%
  \BibitemOpen
  \bibfield  {author} {\bibinfo {author} {\bibfnamefont {Yuan}\ \bibnamefont
  {Liu}}, \bibinfo {author} {\bibfnamefont {Minsik}\ \bibnamefont {Cho}}, \
  and\ \bibinfo {author} {\bibfnamefont {Brenda}\ \bibnamefont {Rubenstein}},\
  }\bibfield  {title} {\enquote {\bibinfo {title} {Ab initio finite temperature
  auxiliary field quantum monte carlo},}\ }\href {\doibase
  10.1021/acs.jctc.8b00569} {\bibfield  {journal} {\bibinfo  {journal} {Journal
  of Chemical Theory and Computation}\ }\textbf {\bibinfo {volume} {14}},\
  \bibinfo {pages} {4722--4732} (\bibinfo {year} {2018})}\BibitemShut {NoStop}%
\bibitem [{\citenamefont {Malone}\ \emph {et~al.}(2015)\citenamefont {Malone},
  \citenamefont {Blunt}, \citenamefont {Shepherd}, \citenamefont {Lee},
  \citenamefont {Spencer},\ and\ \citenamefont {Foulkes}}]{Malone_JCP_2015}%
  \BibitemOpen
  \bibfield  {author} {\bibinfo {author} {\bibfnamefont {Fionn~D.}\
  \bibnamefont {Malone}}, \bibinfo {author} {\bibfnamefont {N.~S.}\
  \bibnamefont {Blunt}}, \bibinfo {author} {\bibfnamefont {James~J.}\
  \bibnamefont {Shepherd}}, \bibinfo {author} {\bibfnamefont {D.~K.~K.}\
  \bibnamefont {Lee}}, \bibinfo {author} {\bibfnamefont {J.~S.}\ \bibnamefont
  {Spencer}}, \ and\ \bibinfo {author} {\bibfnamefont {W.~M.~C.}\ \bibnamefont
  {Foulkes}},\ }\bibfield  {title} {\enquote {\bibinfo {title} {Interaction
  picture density matrix quantum monte carlo},}\ }\href {\doibase
  10.1063/1.4927434} {\bibfield  {journal} {\bibinfo  {journal} {The Journal of
  Chemical Physics}\ }\textbf {\bibinfo {volume} {143}},\ \bibinfo {pages}
  {044116} (\bibinfo {year} {2015})},\ \Eprint
  {http://arxiv.org/abs/https://doi.org/10.1063/1.4927434}
  {https://doi.org/10.1063/1.4927434} \BibitemShut {NoStop}%
\bibitem [{\citenamefont {Malone}\ \emph {et~al.}(2016)\citenamefont {Malone},
  \citenamefont {Blunt}, \citenamefont {Brown}, \citenamefont {Lee},
  \citenamefont {Spencer}, \citenamefont {Foulkes},\ and\ \citenamefont
  {Shepherd}}]{Malone_PRL_2016}%
  \BibitemOpen
  \bibfield  {author} {\bibinfo {author} {\bibfnamefont {Fionn~D.}\
  \bibnamefont {Malone}}, \bibinfo {author} {\bibfnamefont {N.~S.}\
  \bibnamefont {Blunt}}, \bibinfo {author} {\bibfnamefont {Ethan~W.}\
  \bibnamefont {Brown}}, \bibinfo {author} {\bibfnamefont {D.~K.~K.}\
  \bibnamefont {Lee}}, \bibinfo {author} {\bibfnamefont {J.~S.}\ \bibnamefont
  {Spencer}}, \bibinfo {author} {\bibfnamefont {W.~M.~C.}\ \bibnamefont
  {Foulkes}}, \ and\ \bibinfo {author} {\bibfnamefont {James~J.}\ \bibnamefont
  {Shepherd}},\ }\bibfield  {title} {\enquote {\bibinfo {title} {Accurate
  exchange-correlation energies for the warm dense electron gas},}\ }\href
  {\doibase 10.1103/PhysRevLett.117.115701} {\bibfield  {journal} {\bibinfo
  {journal} {Phys. Rev. Lett.}\ }\textbf {\bibinfo {volume} {117}},\ \bibinfo
  {pages} {115701} (\bibinfo {year} {2016})}\BibitemShut {NoStop}%
\bibitem [{\citenamefont {Claes}\ and\ \citenamefont
  {Clark}(2017)}]{Clark_PRB_2017}%
  \BibitemOpen
  \bibfield  {author} {\bibinfo {author} {\bibfnamefont {Jahan}\ \bibnamefont
  {Claes}}\ and\ \bibinfo {author} {\bibfnamefont {Bryan~K.}\ \bibnamefont
  {Clark}},\ }\bibfield  {title} {\enquote {\bibinfo {title}
  {Finite-temperature properties of strongly correlated systems via variational
  monte carlo},}\ }\href {\doibase 10.1103/PhysRevB.95.205109} {\bibfield
  {journal} {\bibinfo  {journal} {Phys. Rev. B}\ }\textbf {\bibinfo {volume}
  {95}},\ \bibinfo {pages} {205109} (\bibinfo {year} {2017})}\BibitemShut
  {NoStop}%
\bibitem [{\citenamefont {Dornheim}\ \emph
  {et~al.}(2019{\natexlab{a}})\citenamefont {Dornheim}, \citenamefont {Groth},\
  and\ \citenamefont {Bonitz}}]{Dornheim_CPP_2019}%
  \BibitemOpen
  \bibfield  {author} {\bibinfo {author} {\bibfnamefont {Tobias}\ \bibnamefont
  {Dornheim}}, \bibinfo {author} {\bibfnamefont {Simon}\ \bibnamefont {Groth}},
  \ and\ \bibinfo {author} {\bibfnamefont {Michael}\ \bibnamefont {Bonitz}},\
  }\bibfield  {title} {\enquote {\bibinfo {title} {Permutation blocking path
  integral monte carlo simulations of degenerate electrons at finite
  temperature},}\ }\href {\doibase 10.1002/ctpp.201800157} {\bibfield
  {journal} {\bibinfo  {journal} {Contributions to Plasma Physics}\ }\textbf
  {\bibinfo {volume} {59}},\ \bibinfo {pages} {e201800157} (\bibinfo {year}
  {2019}{\natexlab{a}})},\ \Eprint
  {http://arxiv.org/abs/https://onlinelibrary.wiley.com/doi/pdf/10.1002/ctpp.201800157}
  {https://onlinelibrary.wiley.com/doi/pdf/10.1002/ctpp.201800157} \BibitemShut
  {NoStop}%
\bibitem [{\citenamefont {Yilmaz}\ \emph {et~al.}(2020)\citenamefont {Yilmaz},
  \citenamefont {Hunger}, \citenamefont {Dornheim}, \citenamefont {Groth},\
  and\ \citenamefont {Bonitz}}]{yilmaz2020restricted}%
  \BibitemOpen
  \bibfield  {author} {\bibinfo {author} {\bibfnamefont {A.}~\bibnamefont
  {Yilmaz}}, \bibinfo {author} {\bibfnamefont {K.}~\bibnamefont {Hunger}},
  \bibinfo {author} {\bibfnamefont {T.}~\bibnamefont {Dornheim}}, \bibinfo
  {author} {\bibfnamefont {S.}~\bibnamefont {Groth}}, \ and\ \bibinfo {author}
  {\bibfnamefont {M.}~\bibnamefont {Bonitz}},\ }\href@noop {} {\enquote
  {\bibinfo {title} {Restricted configuration path integral monte carlo},}\ }
  (\bibinfo {year} {2020}),\ \Eprint {http://arxiv.org/abs/2007.12498}
  {arXiv:2007.12498 [physics.comp-ph]} \BibitemShut {NoStop}%
\bibitem [{\citenamefont {Dornheim}\ \emph
  {et~al.}(2020{\natexlab{a}})\citenamefont {Dornheim}, \citenamefont
  {Vorberger},\ and\ \citenamefont {Bonitz}}]{2020arXiv200403229D}%
  \BibitemOpen
  \bibfield  {author} {\bibinfo {author} {\bibfnamefont {Tobias}\ \bibnamefont
  {Dornheim}}, \bibinfo {author} {\bibfnamefont {Jan}\ \bibnamefont
  {Vorberger}}, \ and\ \bibinfo {author} {\bibfnamefont {Michael}\ \bibnamefont
  {Bonitz}},\ }\bibfield  {title} {\enquote {\bibinfo {title} {Nonlinear
  electronic density response in warm dense matter},}\ }\href {\doibase
  10.1103/PhysRevLett.125.085001} {\bibfield  {journal} {\bibinfo  {journal}
  {Phys. Rev. Lett.}\ }\textbf {\bibinfo {volume} {125}},\ \bibinfo {pages}
  {085001} (\bibinfo {year} {2020}{\natexlab{a}})}\BibitemShut {NoStop}%
\bibitem [{\citenamefont {Driver}\ \emph {et~al.}(2018)\citenamefont {Driver},
  \citenamefont {Soubiran},\ and\ \citenamefont {Militzer}}]{Driver_PRE_2018}%
  \BibitemOpen
  \bibfield  {author} {\bibinfo {author} {\bibfnamefont {K.~P.}\ \bibnamefont
  {Driver}}, \bibinfo {author} {\bibfnamefont {F.}~\bibnamefont {Soubiran}}, \
  and\ \bibinfo {author} {\bibfnamefont {B.}~\bibnamefont {Militzer}},\
  }\bibfield  {title} {\enquote {\bibinfo {title} {Path integral monte carlo
  simulations of warm dense aluminum},}\ }\href {\doibase
  10.1103/PhysRevE.97.063207} {\bibfield  {journal} {\bibinfo  {journal} {Phys.
  Rev. E}\ }\textbf {\bibinfo {volume} {97}},\ \bibinfo {pages} {063207}
  (\bibinfo {year} {2018})}\BibitemShut {NoStop}%
\bibitem [{\citenamefont {Dornheim}\ \emph
  {et~al.}(2018{\natexlab{a}})\citenamefont {Dornheim}, \citenamefont {Groth},
  \citenamefont {Vorberger},\ and\ \citenamefont {Bonitz}}]{dornheim_dynamic}%
  \BibitemOpen
  \bibfield  {author} {\bibinfo {author} {\bibfnamefont {T.}~\bibnamefont
  {Dornheim}}, \bibinfo {author} {\bibfnamefont {S.}~\bibnamefont {Groth}},
  \bibinfo {author} {\bibfnamefont {J.}~\bibnamefont {Vorberger}}, \ and\
  \bibinfo {author} {\bibfnamefont {M.}~\bibnamefont {Bonitz}},\ }\bibfield
  {title} {\enquote {\bibinfo {title} {Ab initio path integral {M}onte {C}arlo
  results for the dynamic structure factor of correlated electrons: From the
  electron liquid to warm dense matter},}\ }\href
  {https://journals.aps.org/prl/abstract/10.1103/PhysRevLett.121.255001}
  {\bibfield  {journal} {\bibinfo  {journal} {Phys. Rev. Lett.}\ }\textbf
  {\bibinfo {volume} {121}},\ \bibinfo {pages} {255001} (\bibinfo {year}
  {2018}{\natexlab{a}})}\BibitemShut {NoStop}%
\bibitem [{\citenamefont {Graziani}\ \emph {et~al.}(2014)\citenamefont
  {Graziani}, \citenamefont {Desjarlais}, \citenamefont {Redmer},\ and\
  \citenamefont {Trickey}}]{wdm_book}%
  \BibitemOpen
  \bibinfo {editor} {\bibfnamefont {F.}~\bibnamefont {Graziani}}, \bibinfo
  {editor} {\bibfnamefont {M.~P.}\ \bibnamefont {Desjarlais}}, \bibinfo
  {editor} {\bibfnamefont {R.}~\bibnamefont {Redmer}}, \ and\ \bibinfo {editor}
  {\bibfnamefont {S.~B.}\ \bibnamefont {Trickey}},\ eds.,\ \href@noop {} {\emph
  {\bibinfo {title} {Frontiers and Challenges in Warm Dense Matter}}}\
  (\bibinfo  {publisher} {Springer},\ \bibinfo {address} {International
  Publishing},\ \bibinfo {year} {2014})\BibitemShut {NoStop}%
\bibitem [{\citenamefont {Bonitz}\ \emph {et~al.}(2020)\citenamefont {Bonitz},
  \citenamefont {Dornheim}, \citenamefont {Moldabekov}, \citenamefont {Zhang},
  \citenamefont {Hamann}, \citenamefont {K{\"a}hlert}, \citenamefont {Filinov},
  \citenamefont {Ramakrishna},\ and\ \citenamefont {Vorberger}}]{new_POP}%
  \BibitemOpen
  \bibfield  {author} {\bibinfo {author} {\bibfnamefont {M.}~\bibnamefont
  {Bonitz}}, \bibinfo {author} {\bibfnamefont {T.}~\bibnamefont {Dornheim}},
  \bibinfo {author} {\bibfnamefont {Zh.~A.}\ \bibnamefont {Moldabekov}},
  \bibinfo {author} {\bibfnamefont {S.}~\bibnamefont {Zhang}}, \bibinfo
  {author} {\bibfnamefont {P.}~\bibnamefont {Hamann}}, \bibinfo {author}
  {\bibfnamefont {H.}~\bibnamefont {K{\"a}hlert}}, \bibinfo {author}
  {\bibfnamefont {A.}~\bibnamefont {Filinov}}, \bibinfo {author} {\bibfnamefont
  {K.}~\bibnamefont {Ramakrishna}}, \ and\ \bibinfo {author} {\bibfnamefont
  {J.}~\bibnamefont {Vorberger}},\ }\bibfield  {title} {\enquote {\bibinfo
  {title} {Ab initio simulation of warm dense matter},}\ }\href {\doibase
  10.1063/1.5143225} {\bibfield  {journal} {\bibinfo  {journal} {Physics of
  Plasmas}\ }\textbf {\bibinfo {volume} {27}},\ \bibinfo {pages} {042710}
  (\bibinfo {year} {2020})}\BibitemShut {NoStop}%
\bibitem [{\citenamefont {Dornheim}\ \emph
  {et~al.}(2018{\natexlab{b}})\citenamefont {Dornheim}, \citenamefont {Groth},\
  and\ \citenamefont {Bonitz}}]{review}%
  \BibitemOpen
  \bibfield  {author} {\bibinfo {author} {\bibfnamefont {T.}~\bibnamefont
  {Dornheim}}, \bibinfo {author} {\bibfnamefont {S.}~\bibnamefont {Groth}}, \
  and\ \bibinfo {author} {\bibfnamefont {M.}~\bibnamefont {Bonitz}},\
  }\bibfield  {title} {\enquote {\bibinfo {title} {The uniform electron gas at
  warm dense matter conditions},}\ }\href
  {https://www.sciencedirect.com/science/article/abs/pii/S0370157318300516}
  {\bibfield  {journal} {\bibinfo  {journal} {Phys. Reports}\ }\textbf
  {\bibinfo {volume} {744}},\ \bibinfo {pages} {1--86} (\bibinfo {year}
  {2018}{\natexlab{b}})}\BibitemShut {NoStop}%
\bibitem [{\citenamefont {Fortov}(2009)}]{fortov_review}%
  \BibitemOpen
  \bibfield  {author} {\bibinfo {author} {\bibfnamefont {V.~E.}\ \bibnamefont
  {Fortov}},\ }\bibfield  {title} {\enquote {\bibinfo {title} {Extreme states
  of matter on earth and in space},}\ }\href
  {https://www.turpion.org/php/paper.phtml?journal_id=pu&paper_id=6821}
  {\bibfield  {journal} {\bibinfo  {journal} {Phys.-Usp}\ }\textbf {\bibinfo
  {volume} {52}},\ \bibinfo {pages} {615--647} (\bibinfo {year}
  {2009})}\BibitemShut {NoStop}%
\bibitem [{\citenamefont {Pribram-Jones}\ \emph {et~al.}(2014)\citenamefont
  {Pribram-Jones}, \citenamefont {Pittalis}, \citenamefont {Gross},\ and\
  \citenamefont {Burke}}]{Pribram-Jones2014}%
  \BibitemOpen
  \bibfield  {author} {\bibinfo {author} {\bibfnamefont {Aurora}\ \bibnamefont
  {Pribram-Jones}}, \bibinfo {author} {\bibfnamefont {Stefano}\ \bibnamefont
  {Pittalis}}, \bibinfo {author} {\bibfnamefont {E.~K.~U.}\ \bibnamefont
  {Gross}}, \ and\ \bibinfo {author} {\bibfnamefont {Kieron}\ \bibnamefont
  {Burke}},\ }\bibfield  {title} {\enquote {\bibinfo {title} {{Thermal Density
  Functional Theory in Context}},}\ \ }(\bibinfo {year} {2014})\ pp.\ \bibinfo
  {pages} {25--60}\BibitemShut {NoStop}%
\bibitem [{\citenamefont {Smith}\ \emph {et~al.}(2018)\citenamefont {Smith},
  \citenamefont {Sagredo},\ and\ \citenamefont {Burke}}]{Smith2018}%
  \BibitemOpen
  \bibfield  {author} {\bibinfo {author} {\bibfnamefont {Justin~C.}\
  \bibnamefont {Smith}}, \bibinfo {author} {\bibfnamefont {Francisca}\
  \bibnamefont {Sagredo}}, \ and\ \bibinfo {author} {\bibfnamefont {Kieron}\
  \bibnamefont {Burke}},\ }\bibfield  {title} {\enquote {\bibinfo {title}
  {{Warming Up Density Functional Theory}},}\ }in\ \href {\doibase
  10.1007/978-981-10-5651-2_11} {\emph {\bibinfo {booktitle} {Frontiers of
  Quantum Chemistry}}}\ (\bibinfo  {publisher} {Springer Singapore},\ \bibinfo
  {address} {Singapore},\ \bibinfo {year} {2018})\ pp.\ \bibinfo {pages}
  {249--271}\BibitemShut {NoStop}%
\bibitem [{\citenamefont {Cytter}\ \emph {et~al.}(2018)\citenamefont {Cytter},
  \citenamefont {Rabani}, \citenamefont {Neuhauser},\ and\ \citenamefont
  {Baer}}]{Cytter2018}%
  \BibitemOpen
  \bibfield  {author} {\bibinfo {author} {\bibfnamefont {Yael}\ \bibnamefont
  {Cytter}}, \bibinfo {author} {\bibfnamefont {Eran}\ \bibnamefont {Rabani}},
  \bibinfo {author} {\bibfnamefont {Daniel}\ \bibnamefont {Neuhauser}}, \ and\
  \bibinfo {author} {\bibfnamefont {Roi}\ \bibnamefont {Baer}},\ }\bibfield
  {title} {\enquote {\bibinfo {title} {Stochastic density functional theory at
  finite temperatures},}\ }\href {\doibase 10.1103/PhysRevB.97.115207}
  {\bibfield  {journal} {\bibinfo  {journal} {Phys. Rev. B}\ }\textbf {\bibinfo
  {volume} {97}},\ \bibinfo {pages} {115207} (\bibinfo {year}
  {2018})}\BibitemShut {NoStop}%
\bibitem [{\citenamefont {Karasiev}\ \emph {et~al.}(2016)\citenamefont
  {Karasiev}, \citenamefont {Calderin},\ and\ \citenamefont
  {Trickey}}]{karasiev_importance}%
  \BibitemOpen
  \bibfield  {author} {\bibinfo {author} {\bibfnamefont {V.~V.}\ \bibnamefont
  {Karasiev}}, \bibinfo {author} {\bibfnamefont {L.}~\bibnamefont {Calderin}},
  \ and\ \bibinfo {author} {\bibfnamefont {S.~B.}\ \bibnamefont {Trickey}},\
  }\bibfield  {title} {\enquote {\bibinfo {title} {Importance of
  finite-temperature exchange correlation for warm dense matter
  calculations},}\ }\href
  {https://journals.aps.org/pre/abstract/10.1103/PhysRevE.93.063207} {\bibfield
   {journal} {\bibinfo  {journal} {Phys. Rev. E}\ }\textbf {\bibinfo {volume}
  {93}},\ \bibinfo {pages} {063207} (\bibinfo {year} {2016})}\BibitemShut
  {NoStop}%
\bibitem [{\citenamefont {Ramakrishna}\ \emph {et~al.}(2020)\citenamefont
  {Ramakrishna}, \citenamefont {Dornheim},\ and\ \citenamefont
  {Vorberger}}]{kushal}%
  \BibitemOpen
  \bibfield  {author} {\bibinfo {author} {\bibfnamefont {Kushal}\ \bibnamefont
  {Ramakrishna}}, \bibinfo {author} {\bibfnamefont {Tobias}\ \bibnamefont
  {Dornheim}}, \ and\ \bibinfo {author} {\bibfnamefont {Jan}\ \bibnamefont
  {Vorberger}},\ }\bibfield  {title} {\enquote {\bibinfo {title} {Influence of
  finite temperature exchange-correlation effects in hydrogen},}\ }\href
  {\doibase 10.1103/PhysRevB.101.195129} {\bibfield  {journal} {\bibinfo
  {journal} {Phys. Rev. B}\ }\textbf {\bibinfo {volume} {101}},\ \bibinfo
  {pages} {195129} (\bibinfo {year} {2020})}\BibitemShut {NoStop}%
\bibitem [{\citenamefont {Dornheim}\ \emph
  {et~al.}(2016{\natexlab{a}})\citenamefont {Dornheim}, \citenamefont {Groth},
  \citenamefont {Sjostrom}, \citenamefont {Malone}, \citenamefont {Foulkes},\
  and\ \citenamefont {Bonitz}}]{dornheim_prl}%
  \BibitemOpen
  \bibfield  {author} {\bibinfo {author} {\bibfnamefont {T.}~\bibnamefont
  {Dornheim}}, \bibinfo {author} {\bibfnamefont {S.}~\bibnamefont {Groth}},
  \bibinfo {author} {\bibfnamefont {T.}~\bibnamefont {Sjostrom}}, \bibinfo
  {author} {\bibfnamefont {F.~D.}\ \bibnamefont {Malone}}, \bibinfo {author}
  {\bibfnamefont {W.~M.~C.}\ \bibnamefont {Foulkes}}, \ and\ \bibinfo {author}
  {\bibfnamefont {M.}~\bibnamefont {Bonitz}},\ }\bibfield  {title} {\enquote
  {\bibinfo {title} {Ab initio quantum {M}onte {C}arlo simulation of the warm
  dense electron gas in the thermodynamic limit},}\ }\href
  {http://link.aps.org/doi/10.1103/PhysRevLett.117.156403} {\bibfield
  {journal} {\bibinfo  {journal} {Phys. Rev. Lett.}\ }\textbf {\bibinfo
  {volume} {117}},\ \bibinfo {pages} {156403} (\bibinfo {year}
  {2016}{\natexlab{a}})}\BibitemShut {NoStop}%
\bibitem [{\citenamefont {Groth}\ \emph {et~al.}(2017)\citenamefont {Groth},
  \citenamefont {Dornheim}, \citenamefont {Sjostrom}, \citenamefont {Malone},
  \citenamefont {Foulkes},\ and\ \citenamefont {Bonitz}}]{groth_prl}%
  \BibitemOpen
  \bibfield  {author} {\bibinfo {author} {\bibfnamefont {S.}~\bibnamefont
  {Groth}}, \bibinfo {author} {\bibfnamefont {T.}~\bibnamefont {Dornheim}},
  \bibinfo {author} {\bibfnamefont {T.}~\bibnamefont {Sjostrom}}, \bibinfo
  {author} {\bibfnamefont {F.~D.}\ \bibnamefont {Malone}}, \bibinfo {author}
  {\bibfnamefont {W.~M.~C.}\ \bibnamefont {Foulkes}}, \ and\ \bibinfo {author}
  {\bibfnamefont {M.}~\bibnamefont {Bonitz}},\ }\bibfield  {title} {\enquote
  {\bibinfo {title} {Ab initio exchange--correlation free energy of the uniform
  electron gas at warm dense matter conditions},}\ }\href
  {https://journals.aps.org/prl/abstract/10.1103/PhysRevLett.119.135001}
  {\bibfield  {journal} {\bibinfo  {journal} {Phys. Rev. Lett.}\ }\textbf
  {\bibinfo {volume} {119}},\ \bibinfo {pages} {135001} (\bibinfo {year}
  {2017})}\BibitemShut {NoStop}%
\bibitem [{\citenamefont {Dornheim}(2020)}]{Dornheim_PRA_2020}%
  \BibitemOpen
  \bibfield  {author} {\bibinfo {author} {\bibfnamefont {Tobias}\ \bibnamefont
  {Dornheim}},\ }\bibfield  {title} {\enquote {\bibinfo {title} {Path-integral
  monte carlo simulations of quantum dipole systems in traps: Superfluidity,
  quantum statistics, and structural properties},}\ }\href {\doibase
  10.1103/PhysRevA.102.023307} {\bibfield  {journal} {\bibinfo  {journal}
  {Phys. Rev. A}\ }\textbf {\bibinfo {volume} {102}},\ \bibinfo {pages}
  {023307} (\bibinfo {year} {2020})}\BibitemShut {NoStop}%
\bibitem [{\citenamefont {Filinov}(2016)}]{dynamic_Alex_2}%
  \BibitemOpen
  \bibfield  {author} {\bibinfo {author} {\bibfnamefont {A.}~\bibnamefont
  {Filinov}},\ }\bibfield  {title} {\enquote {\bibinfo {title} {Correlation
  effects and collective excitations in bosonic bilayers: Role of quantum
  statistics, superfluidity, and the dimerization transition},}\ }\href
  {\doibase 10.1103/PhysRevA.94.013603} {\bibfield  {journal} {\bibinfo
  {journal} {Phys. Rev. A}\ }\textbf {\bibinfo {volume} {94}},\ \bibinfo
  {pages} {013603} (\bibinfo {year} {2016})}\BibitemShut {NoStop}%
\bibitem [{\citenamefont {Schleede}\ \emph {et~al.}(2012)\citenamefont
  {Schleede}, \citenamefont {Filinov}, \citenamefont {Bonitz},\ and\
  \citenamefont {Fehske}}]{Schleede_bilayer_2012}%
  \BibitemOpen
  \bibfield  {author} {\bibinfo {author} {\bibfnamefont {J.}~\bibnamefont
  {Schleede}}, \bibinfo {author} {\bibfnamefont {A.}~\bibnamefont {Filinov}},
  \bibinfo {author} {\bibfnamefont {M.}~\bibnamefont {Bonitz}}, \ and\ \bibinfo
  {author} {\bibfnamefont {H.}~\bibnamefont {Fehske}},\ }\bibfield  {title}
  {\enquote {\bibinfo {title} {Phase diagram of bilayer electron-hole
  plasmas},}\ }\href {\doibase 10.1002/ctpp.201200045} {\bibfield  {journal}
  {\bibinfo  {journal} {Contributions to Plasma Physics}\ }\textbf {\bibinfo
  {volume} {52}},\ \bibinfo {pages} {819--826} (\bibinfo {year} {2012})},\
  \Eprint
  {http://arxiv.org/abs/https://onlinelibrary.wiley.com/doi/pdf/10.1002/ctpp.201200045}
  {https://onlinelibrary.wiley.com/doi/pdf/10.1002/ctpp.201200045} \BibitemShut
  {NoStop}%
\bibitem [{\citenamefont {Dornheim}\ \emph
  {et~al.}(2016{\natexlab{b}})\citenamefont {Dornheim}, \citenamefont
  {Thomsen}, \citenamefont {Ludwig}, \citenamefont {Filinov},\ and\
  \citenamefont {Bonitz}}]{Dornheim_CPP_2016}%
  \BibitemOpen
  \bibfield  {author} {\bibinfo {author} {\bibfnamefont {T.}~\bibnamefont
  {Dornheim}}, \bibinfo {author} {\bibfnamefont {H.}~\bibnamefont {Thomsen}},
  \bibinfo {author} {\bibfnamefont {P.}~\bibnamefont {Ludwig}}, \bibinfo
  {author} {\bibfnamefont {A.}~\bibnamefont {Filinov}}, \ and\ \bibinfo
  {author} {\bibfnamefont {M.}~\bibnamefont {Bonitz}},\ }\bibfield  {title}
  {\enquote {\bibinfo {title} {Analyzing quantum correlations made simple},}\
  }\href {\doibase 10.1002/ctpp.201500120} {\bibfield  {journal} {\bibinfo
  {journal} {Contributions to Plasma Physics}\ }\textbf {\bibinfo {volume}
  {56}},\ \bibinfo {pages} {371--379} (\bibinfo {year} {2016}{\natexlab{b}})},\
  \Eprint
  {http://arxiv.org/abs/https://onlinelibrary.wiley.com/doi/pdf/10.1002/ctpp.201500120}
  {https://onlinelibrary.wiley.com/doi/pdf/10.1002/ctpp.201500120} \BibitemShut
  {NoStop}%
\bibitem [{\citenamefont {Kyl\"anp\"a\"a}\ and\ \citenamefont
  {R\"as\"anen}(2017)}]{Ilkka_PRB_2017}%
  \BibitemOpen
  \bibfield  {author} {\bibinfo {author} {\bibfnamefont {Ilkka}\ \bibnamefont
  {Kyl\"anp\"a\"a}}\ and\ \bibinfo {author} {\bibfnamefont {Esa}\ \bibnamefont
  {R\"as\"anen}},\ }\bibfield  {title} {\enquote {\bibinfo {title} {Path
  integral monte carlo benchmarks for two-dimensional quantum dots},}\ }\href
  {\doibase 10.1103/PhysRevB.96.205445} {\bibfield  {journal} {\bibinfo
  {journal} {Phys. Rev. B}\ }\textbf {\bibinfo {volume} {96}},\ \bibinfo
  {pages} {205445} (\bibinfo {year} {2017})}\BibitemShut {NoStop}%
\bibitem [{\citenamefont {Egger}\ \emph {et~al.}(1999)\citenamefont {Egger},
  \citenamefont {H\"ausler}, \citenamefont {Mak},\ and\ \citenamefont
  {Grabert}}]{Egger_PRL_1999}%
  \BibitemOpen
  \bibfield  {author} {\bibinfo {author} {\bibfnamefont {R.}~\bibnamefont
  {Egger}}, \bibinfo {author} {\bibfnamefont {W.}~\bibnamefont {H\"ausler}},
  \bibinfo {author} {\bibfnamefont {C.~H.}\ \bibnamefont {Mak}}, \ and\
  \bibinfo {author} {\bibfnamefont {H.}~\bibnamefont {Grabert}},\ }\bibfield
  {title} {\enquote {\bibinfo {title} {Crossover from fermi liquid to wigner
  molecule behavior in quantum dots},}\ }\href {\doibase
  10.1103/PhysRevLett.82.3320} {\bibfield  {journal} {\bibinfo  {journal}
  {Phys. Rev. Lett.}\ }\textbf {\bibinfo {volume} {82}},\ \bibinfo {pages}
  {3320--3323} (\bibinfo {year} {1999})}\BibitemShut {NoStop}%
\bibitem [{\citenamefont {EGGER}\ and\ \citenamefont {MAK}(2001)}]{Egger_2001}%
  \BibitemOpen
  \bibfield  {author} {\bibinfo {author} {\bibfnamefont {R.}~\bibnamefont
  {EGGER}}\ and\ \bibinfo {author} {\bibfnamefont {C.~H.}\ \bibnamefont
  {MAK}},\ }\bibfield  {title} {\enquote {\bibinfo {title} {Multilevel blocking
  monte carlo simulations for quantum dots},}\ }\href {\doibase
  10.1142/S021797920100591X} {\bibfield  {journal} {\bibinfo  {journal}
  {International Journal of Modern Physics B}\ }\textbf {\bibinfo {volume}
  {15}},\ \bibinfo {pages} {1416--1425} (\bibinfo {year} {2001})},\ \Eprint
  {http://arxiv.org/abs/https://doi.org/10.1142/S021797920100591X}
  {https://doi.org/10.1142/S021797920100591X} \BibitemShut {NoStop}%
\bibitem [{\citenamefont {Filinov}\ \emph {et~al.}(2001)\citenamefont
  {Filinov}, \citenamefont {Bonitz},\ and\ \citenamefont
  {Lozovik}}]{Filinov_PRL_2001}%
  \BibitemOpen
  \bibfield  {author} {\bibinfo {author} {\bibfnamefont {A.~V.}\ \bibnamefont
  {Filinov}}, \bibinfo {author} {\bibfnamefont {M.}~\bibnamefont {Bonitz}}, \
  and\ \bibinfo {author} {\bibfnamefont {Yu.~E.}\ \bibnamefont {Lozovik}},\
  }\bibfield  {title} {\enquote {\bibinfo {title} {Wigner crystallization in
  mesoscopic 2d electron systems},}\ }\href {\doibase
  10.1103/PhysRevLett.86.3851} {\bibfield  {journal} {\bibinfo  {journal}
  {Phys. Rev. Lett.}\ }\textbf {\bibinfo {volume} {86}},\ \bibinfo {pages}
  {3851--3854} (\bibinfo {year} {2001})}\BibitemShut {NoStop}%
\bibitem [{\citenamefont {Filinov}\ \emph {et~al.}(2013)\citenamefont
  {Filinov}, \citenamefont {Ivanov}, \citenamefont {Fortov}, \citenamefont
  {Bonitz},\ and\ \citenamefont {Levashov}}]{Filinov_PRC_2013}%
  \BibitemOpen
  \bibfield  {author} {\bibinfo {author} {\bibfnamefont {V.~S.}\ \bibnamefont
  {Filinov}}, \bibinfo {author} {\bibfnamefont {Yu.~B.}\ \bibnamefont
  {Ivanov}}, \bibinfo {author} {\bibfnamefont {V.~E.}\ \bibnamefont {Fortov}},
  \bibinfo {author} {\bibfnamefont {M.}~\bibnamefont {Bonitz}}, \ and\ \bibinfo
  {author} {\bibfnamefont {P.~R.}\ \bibnamefont {Levashov}},\ }\bibfield
  {title} {\enquote {\bibinfo {title} {Color path-integral monte-carlo
  simulations of quark-gluon plasma: Thermodynamic and transport properties},}\
  }\href {\doibase 10.1103/PhysRevC.87.035207} {\bibfield  {journal} {\bibinfo
  {journal} {Phys. Rev. C}\ }\textbf {\bibinfo {volume} {87}},\ \bibinfo
  {pages} {035207} (\bibinfo {year} {2013})}\BibitemShut {NoStop}%
\bibitem [{\citenamefont {Filinov}\ \emph {et~al.}(2015)\citenamefont
  {Filinov}, \citenamefont {Bonitz}, \citenamefont {Ivanov}, \citenamefont
  {Ilgenfritz},\ and\ \citenamefont {Fortov}}]{Filinov_quark_2015}%
  \BibitemOpen
  \bibfield  {author} {\bibinfo {author} {\bibfnamefont {V.S.}\ \bibnamefont
  {Filinov}}, \bibinfo {author} {\bibfnamefont {M.}~\bibnamefont {Bonitz}},
  \bibinfo {author} {\bibfnamefont {Y.B.}\ \bibnamefont {Ivanov}}, \bibinfo
  {author} {\bibfnamefont {E.-M.}\ \bibnamefont {Ilgenfritz}}, \ and\ \bibinfo
  {author} {\bibfnamefont {V.E.}\ \bibnamefont {Fortov}},\ }\bibfield  {title}
  {\enquote {\bibinfo {title} {Thermodynamics of the quark-gluon plasma at
  finite chemical potential: Color path integral monte carlo results},}\ }\href
  {\doibase 10.1002/ctpp.201400056} {\bibfield  {journal} {\bibinfo  {journal}
  {Contributions to Plasma Physics}\ }\textbf {\bibinfo {volume} {55}},\
  \bibinfo {pages} {203--208} (\bibinfo {year} {2015})},\ \Eprint
  {http://arxiv.org/abs/https://onlinelibrary.wiley.com/doi/pdf/10.1002/ctpp.201400056}
  {https://onlinelibrary.wiley.com/doi/pdf/10.1002/ctpp.201400056} \BibitemShut
  {NoStop}%
\bibitem [{\citenamefont {Yan}\ and\ \citenamefont
  {Blume}(2014)}]{Blume_PRL_2014}%
  \BibitemOpen
  \bibfield  {author} {\bibinfo {author} {\bibfnamefont {Yangqian}\
  \bibnamefont {Yan}}\ and\ \bibinfo {author} {\bibfnamefont {D.}~\bibnamefont
  {Blume}},\ }\bibfield  {title} {\enquote {\bibinfo {title} {Abnormal
  superfluid fraction of harmonically trapped few-fermion systems},}\ }\href
  {\doibase 10.1103/PhysRevLett.112.235301} {\bibfield  {journal} {\bibinfo
  {journal} {Phys. Rev. Lett.}\ }\textbf {\bibinfo {volume} {112}},\ \bibinfo
  {pages} {235301} (\bibinfo {year} {2014})}\BibitemShut {NoStop}%
\bibitem [{\citenamefont {Filinov}\ \emph {et~al.}(2000)\citenamefont
  {Filinov}, \citenamefont {Lozovik},\ and\ \citenamefont
  {Bonitz}}]{Filinov_crystal2}%
  \BibitemOpen
  \bibfield  {author} {\bibinfo {author} {\bibfnamefont {A.V.}\ \bibnamefont
  {Filinov}}, \bibinfo {author} {\bibfnamefont {Yu.E.}\ \bibnamefont
  {Lozovik}}, \ and\ \bibinfo {author} {\bibfnamefont {M.}~\bibnamefont
  {Bonitz}},\ }\bibfield  {title} {\enquote {\bibinfo {title} {Path integral
  simulations of crystallization of quantum confined electrons},}\ }\href
  {\doibase 10.1002/1521-3951(200009)221:1<231::AID-PSSB231>3.0.CO;2-D}
  {\bibfield  {journal} {\bibinfo  {journal} {physica status solidi (b)}\
  }\textbf {\bibinfo {volume} {221}},\ \bibinfo {pages} {231--234} (\bibinfo
  {year} {2000})}\BibitemShut {NoStop}%
\bibitem [{\citenamefont {Zenker}\ \emph {et~al.}(2012)\citenamefont {Zenker},
  \citenamefont {Ihle}, \citenamefont {Bronold},\ and\ \citenamefont
  {Fehske}}]{Fehske_BCS_2012}%
  \BibitemOpen
  \bibfield  {author} {\bibinfo {author} {\bibfnamefont {B.}~\bibnamefont
  {Zenker}}, \bibinfo {author} {\bibfnamefont {D.}~\bibnamefont {Ihle}},
  \bibinfo {author} {\bibfnamefont {F.~X.}\ \bibnamefont {Bronold}}, \ and\
  \bibinfo {author} {\bibfnamefont {H.}~\bibnamefont {Fehske}},\ }\bibfield
  {title} {\enquote {\bibinfo {title} {Electron-hole pair condensation at the
  semimetal-semiconductor transition: A bcs-bec crossover scenario},}\ }\href
  {\doibase 10.1103/PhysRevB.85.121102} {\bibfield  {journal} {\bibinfo
  {journal} {Phys. Rev. B}\ }\textbf {\bibinfo {volume} {85}},\ \bibinfo
  {pages} {121102} (\bibinfo {year} {2012})}\BibitemShut {NoStop}%
\bibitem [{\citenamefont {Ohashi}\ and\ \citenamefont
  {Griffin}(2002)}]{BCS_2002}%
  \BibitemOpen
  \bibfield  {author} {\bibinfo {author} {\bibfnamefont {Y.}~\bibnamefont
  {Ohashi}}\ and\ \bibinfo {author} {\bibfnamefont {A.}~\bibnamefont
  {Griffin}},\ }\bibfield  {title} {\enquote {\bibinfo {title} {Bcs-bec
  crossover in a gas of fermi atoms with a feshbach resonance},}\ }\href
  {\doibase 10.1103/PhysRevLett.89.130402} {\bibfield  {journal} {\bibinfo
  {journal} {Phys. Rev. Lett.}\ }\textbf {\bibinfo {volume} {89}},\ \bibinfo
  {pages} {130402} (\bibinfo {year} {2002})}\BibitemShut {NoStop}%
\bibitem [{\citenamefont {Groth}\ \emph {et~al.}(2019)\citenamefont {Groth},
  \citenamefont {Dornheim},\ and\ \citenamefont
  {Vorberger}}]{dynamic_folgepaper}%
  \BibitemOpen
  \bibfield  {author} {\bibinfo {author} {\bibfnamefont {S.}~\bibnamefont
  {Groth}}, \bibinfo {author} {\bibfnamefont {T.}~\bibnamefont {Dornheim}}, \
  and\ \bibinfo {author} {\bibfnamefont {J.}~\bibnamefont {Vorberger}},\
  }\bibfield  {title} {\enquote {\bibinfo {title} {Ab initio path integral
  {M}onte {C}arlo approach to the static and dynamic density response of the
  uniform electron gas},}\ }\href
  {https://link.aps.org/doi/10.1103/PhysRevB.99.235122} {\bibfield  {journal}
  {\bibinfo  {journal} {Phys. Rev. B}\ }\textbf {\bibinfo {volume} {99}},\
  \bibinfo {pages} {235122} (\bibinfo {year} {2019})}\BibitemShut {NoStop}%
\bibitem [{\citenamefont {Dornheim}\ \emph
  {et~al.}(2020{\natexlab{b}})\citenamefont {Dornheim}, \citenamefont
  {Moldabekov}, \citenamefont {Vorberger},\ and\ \citenamefont
  {Groth}}]{dornheim_HEDP}%
  \BibitemOpen
  \bibfield  {author} {\bibinfo {author} {\bibfnamefont {Tobias}\ \bibnamefont
  {Dornheim}}, \bibinfo {author} {\bibfnamefont {Zhandos~A}\ \bibnamefont
  {Moldabekov}}, \bibinfo {author} {\bibfnamefont {Jan}\ \bibnamefont
  {Vorberger}}, \ and\ \bibinfo {author} {\bibfnamefont {Simon}\ \bibnamefont
  {Groth}},\ }\bibfield  {title} {\enquote {\bibinfo {title} {Ab initio path
  integral monte carlo simulation of the uniform electron gas in the high
  energy density regime},}\ }\href {\doibase 10.1088/1361-6587/ab8bb4}
  {\bibfield  {journal} {\bibinfo  {journal} {Plasma Physics and Controlled
  Fusion}\ }\textbf {\bibinfo {volume} {62}},\ \bibinfo {pages} {075003}
  (\bibinfo {year} {2020}{\natexlab{b}})}\BibitemShut {NoStop}%
\bibitem [{\citenamefont {Hamann}\ \emph {et~al.}(2020)\citenamefont {Hamann},
  \citenamefont {Dornheim}, \citenamefont {Vorberger}, \citenamefont
  {Moldabekov},\ and\ \citenamefont {Bonitz}}]{hamann2020dynamic}%
  \BibitemOpen
  \bibfield  {author} {\bibinfo {author} {\bibfnamefont {Paul}\ \bibnamefont
  {Hamann}}, \bibinfo {author} {\bibfnamefont {Tobias}\ \bibnamefont
  {Dornheim}}, \bibinfo {author} {\bibfnamefont {Jan}\ \bibnamefont
  {Vorberger}}, \bibinfo {author} {\bibfnamefont {Zhandos~A.}\ \bibnamefont
  {Moldabekov}}, \ and\ \bibinfo {author} {\bibfnamefont {Michael}\
  \bibnamefont {Bonitz}},\ }\href@noop {} {\enquote {\bibinfo {title} {Dynamic
  properties of the warm dense electron gas: an ab initio path integral monte
  carlo approach},}\ } (\bibinfo {year} {2020}),\ \Eprint
  {http://arxiv.org/abs/2007.15471} {arXiv:2007.15471 [physics.comp-ph]}
  \BibitemShut {NoStop}%
\bibitem [{\citenamefont {Karasiev}\ \emph {et~al.}(2019)\citenamefont
  {Karasiev}, \citenamefont {Trickey},\ and\ \citenamefont {Dufty}}]{status}%
  \BibitemOpen
  \bibfield  {author} {\bibinfo {author} {\bibfnamefont {V.~V.}\ \bibnamefont
  {Karasiev}}, \bibinfo {author} {\bibfnamefont {S.~B.}\ \bibnamefont
  {Trickey}}, \ and\ \bibinfo {author} {\bibfnamefont {J.~W.}\ \bibnamefont
  {Dufty}},\ }\bibfield  {title} {\enquote {\bibinfo {title} {Status of
  free-energy representations for the homogeneous electron gas},}\ }\href
  {https://journals.aps.org/prb/abstract/10.1103/PhysRevB.99.195134} {\bibfield
   {journal} {\bibinfo  {journal} {Phys. Rev. B}\ }\textbf {\bibinfo {volume}
  {99}},\ \bibinfo {pages} {195134} (\bibinfo {year} {2019})}\BibitemShut
  {NoStop}%
\bibitem [{\citenamefont {Ceperley}(1995)}]{cep}%
  \BibitemOpen
  \bibfield  {author} {\bibinfo {author} {\bibfnamefont {D.~M.}\ \bibnamefont
  {Ceperley}},\ }\bibfield  {title} {\enquote {\bibinfo {title} {Path integrals
  in the theory of condensed helium},}\ }\href
  {https://journals.aps.org/rmp/abstract/10.1103/RevModPhys.67.279} {\bibfield
  {journal} {\bibinfo  {journal} {Rev. Mod. Phys}\ }\textbf {\bibinfo {volume}
  {67}},\ \bibinfo {pages} {279} (\bibinfo {year} {1995})}\BibitemShut
  {NoStop}%
\bibitem [{\citenamefont {Verbeure}(2010)}]{verbeure2010many}%
  \BibitemOpen
  \bibfield  {author} {\bibinfo {author} {\bibfnamefont {A.F.}\ \bibnamefont
  {Verbeure}},\ }\href {https://books.google.de/books?id=IZ2USeCXtHkC} {\emph
  {\bibinfo {title} {Many-Body Boson Systems: Half a Century Later}}},\
  Theoretical and Mathematical Physics\ (\bibinfo  {publisher} {Springer
  London},\ \bibinfo {year} {2010})\BibitemShut {NoStop}%
\bibitem [{\citenamefont {Hirshberg}\ \emph {et~al.}(2020)\citenamefont
  {Hirshberg}, \citenamefont {Invernizzi},\ and\ \citenamefont
  {Parrinello}}]{Hirshberg_JCP_2020}%
  \BibitemOpen
  \bibfield  {author} {\bibinfo {author} {\bibfnamefont {Barak}\ \bibnamefont
  {Hirshberg}}, \bibinfo {author} {\bibfnamefont {Michele}\ \bibnamefont
  {Invernizzi}}, \ and\ \bibinfo {author} {\bibfnamefont {Michele}\
  \bibnamefont {Parrinello}},\ }\bibfield  {title} {\enquote {\bibinfo {title}
  {Path integral molecular dynamics for fermions: Alleviating the sign problem
  with the bogoliubov inequality},}\ }\href {\doibase 10.1063/5.0008720}
  {\bibfield  {journal} {\bibinfo  {journal} {The Journal of Chemical Physics}\
  }\textbf {\bibinfo {volume} {152}},\ \bibinfo {pages} {171102} (\bibinfo
  {year} {2020})},\ \Eprint
  {http://arxiv.org/abs/https://doi.org/10.1063/5.0008720}
  {https://doi.org/10.1063/5.0008720} \BibitemShut {NoStop}%
\bibitem [{\citenamefont {Hirshberg}\ \emph {et~al.}(2019)\citenamefont
  {Hirshberg}, \citenamefont {Rizzi},\ and\ \citenamefont
  {Parrinello}}]{Hirshberg_PIMD}%
  \BibitemOpen
  \bibfield  {author} {\bibinfo {author} {\bibfnamefont {Barak}\ \bibnamefont
  {Hirshberg}}, \bibinfo {author} {\bibfnamefont {Valerio}\ \bibnamefont
  {Rizzi}}, \ and\ \bibinfo {author} {\bibfnamefont {Michele}\ \bibnamefont
  {Parrinello}},\ }\bibfield  {title} {\enquote {\bibinfo {title} {Path
  integral molecular dynamics for bosons},}\ }\href {\doibase
  10.1073/pnas.1913365116} {\bibfield  {journal} {\bibinfo  {journal}
  {Proceedings of the National Academy of Sciences}\ }\textbf {\bibinfo
  {volume} {116}},\ \bibinfo {pages} {21445--21449} (\bibinfo {year} {2019})},\
  \Eprint
  {http://arxiv.org/abs/https://www.pnas.org/content/116/43/21445.full.pdf}
  {https://www.pnas.org/content/116/43/21445.full.pdf} \BibitemShut {NoStop}%
\bibitem [{\citenamefont {Frenkel}\ and\ \citenamefont
  {Smit}(2001)}]{FrenkelBook}%
  \BibitemOpen
  \bibfield  {author} {\bibinfo {author} {\bibfnamefont {Daan}\ \bibnamefont
  {Frenkel}}\ and\ \bibinfo {author} {\bibfnamefont {Berend}\ \bibnamefont
  {Smit}},\ }\href
  {https://www.elsevier.com/books/understanding-molecular-simulation/frenkel/978-0-12-267351-1}
  {\emph {\bibinfo {title} {{Understanding Molecular Simulation: from
  algorithms to applications}}}}\ (\bibinfo  {publisher} {Academic Press},\
  \bibinfo {year} {2001})\ p.\ \bibinfo {pages} {664}\BibitemShut {NoStop}%
\bibitem [{\citenamefont {De~Raedt}\ and\ \citenamefont
  {De~Raedt}(1983)}]{Trotter}%
  \BibitemOpen
  \bibfield  {author} {\bibinfo {author} {\bibfnamefont {Hans}\ \bibnamefont
  {De~Raedt}}\ and\ \bibinfo {author} {\bibfnamefont {Bart}\ \bibnamefont
  {De~Raedt}},\ }\bibfield  {title} {\enquote {\bibinfo {title} {Applications
  of the generalized trotter formula},}\ }\href {\doibase
  10.1103/PhysRevA.28.3575} {\bibfield  {journal} {\bibinfo  {journal} {Phys.
  Rev. A}\ }\textbf {\bibinfo {volume} {28}},\ \bibinfo {pages} {3575--3580}
  (\bibinfo {year} {1983})}\BibitemShut {NoStop}%
\bibitem [{\citenamefont {Metropolis}\ \emph {et~al.}(1953)\citenamefont
  {Metropolis}, \citenamefont {Rosenbluth}, \citenamefont {Rosenbluth},
  \citenamefont {Teller},\ and\ \citenamefont {Teller}}]{metropolis}%
  \BibitemOpen
  \bibfield  {author} {\bibinfo {author} {\bibfnamefont {Nicholas}\
  \bibnamefont {Metropolis}}, \bibinfo {author} {\bibfnamefont {Arianna~W.}\
  \bibnamefont {Rosenbluth}}, \bibinfo {author} {\bibfnamefont {Marshall~N.}\
  \bibnamefont {Rosenbluth}}, \bibinfo {author} {\bibfnamefont {Augusta~H.}\
  \bibnamefont {Teller}}, \ and\ \bibinfo {author} {\bibfnamefont {Edward}\
  \bibnamefont {Teller}},\ }\bibfield  {title} {\enquote {\bibinfo {title}
  {Equation of state calculations by fast computing machines},}\ }\href
  {\doibase 10.1063/1.1699114} {\bibfield  {journal} {\bibinfo  {journal} {The
  Journal of Chemical Physics}\ }\textbf {\bibinfo {volume} {21}},\ \bibinfo
  {pages} {1087--1092} (\bibinfo {year} {1953})},\ \Eprint
  {http://arxiv.org/abs/https://doi.org/10.1063/1.1699114}
  {https://doi.org/10.1063/1.1699114} \BibitemShut {NoStop}%
\bibitem [{\citenamefont {Dornheim}\ \emph
  {et~al.}(2019{\natexlab{b}})\citenamefont {Dornheim}, \citenamefont {Groth},
  \citenamefont {Filinov},\ and\ \citenamefont
  {Bonitz}}]{dornheim_permutation_cycles}%
  \BibitemOpen
  \bibfield  {author} {\bibinfo {author} {\bibfnamefont {T.}~\bibnamefont
  {Dornheim}}, \bibinfo {author} {\bibfnamefont {S.}~\bibnamefont {Groth}},
  \bibinfo {author} {\bibfnamefont {A.~V.}\ \bibnamefont {Filinov}}, \ and\
  \bibinfo {author} {\bibfnamefont {M.}~\bibnamefont {Bonitz}},\ }\bibfield
  {title} {\enquote {\bibinfo {title} {Path integral monte carlo simulation of
  degenerate electrons: Permutation-cycle properties},}\ }\href {\doibase
  10.1063/1.5093171} {\bibfield  {journal} {\bibinfo  {journal} {The Journal of
  Chemical Physics}\ }\textbf {\bibinfo {volume} {151}},\ \bibinfo {pages}
  {014108} (\bibinfo {year} {2019}{\natexlab{b}})}\BibitemShut {NoStop}%
\bibitem [{\citenamefont {Boninsegni}\ \emph
  {et~al.}(2006{\natexlab{a}})\citenamefont {Boninsegni}, \citenamefont
  {Prokofev},\ and\ \citenamefont {Svistunov}}]{boninsegni1}%
  \BibitemOpen
  \bibfield  {author} {\bibinfo {author} {\bibfnamefont {M.}~\bibnamefont
  {Boninsegni}}, \bibinfo {author} {\bibfnamefont {N.~V.}\ \bibnamefont
  {Prokofev}}, \ and\ \bibinfo {author} {\bibfnamefont {B.~V.}\ \bibnamefont
  {Svistunov}},\ }\bibfield  {title} {\enquote {\bibinfo {title} {Worm
  algorithm and diagrammatic {M}onte {C}arlo: A new approach to
  continuous-space path integral {M}onte {C}arlo simulations},}\ }\href
  {https://journals.aps.org/pre/abstract/10.1103/PhysRevE.74.036701} {\bibfield
   {journal} {\bibinfo  {journal} {Phys. Rev. E}\ }\textbf {\bibinfo {volume}
  {74}},\ \bibinfo {pages} {036701} (\bibinfo {year}
  {2006}{\natexlab{a}})}\BibitemShut {NoStop}%
\bibitem [{\citenamefont {Boninsegni}\ \emph
  {et~al.}(2006{\natexlab{b}})\citenamefont {Boninsegni}, \citenamefont
  {Prokofev},\ and\ \citenamefont {Svistunov}}]{boninsegni2}%
  \BibitemOpen
  \bibfield  {author} {\bibinfo {author} {\bibfnamefont {M.}~\bibnamefont
  {Boninsegni}}, \bibinfo {author} {\bibfnamefont {N.~V.}\ \bibnamefont
  {Prokofev}}, \ and\ \bibinfo {author} {\bibfnamefont {B.~V.}\ \bibnamefont
  {Svistunov}},\ }\bibfield  {title} {\enquote {\bibinfo {title} {Worm
  algorithm for continuous-space path integral {M}onte {C}arlo simulations},}\
  }\href {https://journals.aps.org/prl/abstract/10.1103/PhysRevLett.96.070601}
  {\bibfield  {journal} {\bibinfo  {journal} {Phys. Rev. Lett}\ }\textbf
  {\bibinfo {volume} {96}},\ \bibinfo {pages} {070601} (\bibinfo {year}
  {2006}{\natexlab{b}})}\BibitemShut {NoStop}%
\bibitem [{\citenamefont {Filinov}\ \emph {et~al.}(2010)\citenamefont
  {Filinov}, \citenamefont {Prokof'ev},\ and\ \citenamefont
  {Bonitz}}]{Filinov_PRL_2010}%
  \BibitemOpen
  \bibfield  {author} {\bibinfo {author} {\bibfnamefont {A.}~\bibnamefont
  {Filinov}}, \bibinfo {author} {\bibfnamefont {N.~V.}\ \bibnamefont
  {Prokof'ev}}, \ and\ \bibinfo {author} {\bibfnamefont {M.}~\bibnamefont
  {Bonitz}},\ }\bibfield  {title} {\enquote {\bibinfo {title}
  {Berezinskii-kosterlitz-thouless transition in two-dimensional dipole
  systems},}\ }\href {\doibase 10.1103/PhysRevLett.105.070401} {\bibfield
  {journal} {\bibinfo  {journal} {Phys. Rev. Lett.}\ }\textbf {\bibinfo
  {volume} {105}},\ \bibinfo {pages} {070401} (\bibinfo {year}
  {2010})}\BibitemShut {NoStop}%
\bibitem [{\citenamefont {Boninsegni}\ and\ \citenamefont
  {Prokof'ev}(2012)}]{Boninsegni_supersolid}%
  \BibitemOpen
  \bibfield  {author} {\bibinfo {author} {\bibfnamefont {Massimo}\ \bibnamefont
  {Boninsegni}}\ and\ \bibinfo {author} {\bibfnamefont {Nikolay~V.}\
  \bibnamefont {Prokof'ev}},\ }\bibfield  {title} {\enquote {\bibinfo {title}
  {Colloquium: Supersolids: What and where are they?}}\ }\href {\doibase
  10.1103/RevModPhys.84.759} {\bibfield  {journal} {\bibinfo  {journal} {Rev.
  Mod. Phys.}\ }\textbf {\bibinfo {volume} {84}},\ \bibinfo {pages} {759--776}
  (\bibinfo {year} {2012})}\BibitemShut {NoStop}%
\bibitem [{\citenamefont {Dornheim}\ \emph
  {et~al.}(2015{\natexlab{b}})\citenamefont {Dornheim}, \citenamefont
  {Filinov},\ and\ \citenamefont {Bonitz}}]{dornheim_superfluid}%
  \BibitemOpen
  \bibfield  {author} {\bibinfo {author} {\bibfnamefont {T.}~\bibnamefont
  {Dornheim}}, \bibinfo {author} {\bibfnamefont {A.}~\bibnamefont {Filinov}}, \
  and\ \bibinfo {author} {\bibfnamefont {M.}~\bibnamefont {Bonitz}},\
  }\bibfield  {title} {\enquote {\bibinfo {title} {Superfluidity of strongly
  correlated bosons in two- and three-dimensional traps},}\ }\href {\doibase
  10.1103/PhysRevB.91.054503} {\bibfield  {journal} {\bibinfo  {journal} {Phys.
  Rev. B}\ }\textbf {\bibinfo {volume} {91}},\ \bibinfo {pages} {054503}
  (\bibinfo {year} {2015}{\natexlab{b}})}\BibitemShut {NoStop}%
\bibitem [{\citenamefont {Pollet}\ \emph {et~al.}(2007)\citenamefont {Pollet},
  \citenamefont {Boninsegni}, \citenamefont {Kuklov}, \citenamefont
  {Prokof'ev}, \citenamefont {Svistunov},\ and\ \citenamefont
  {Troyer}}]{Pollet_PRL_superfluid}%
  \BibitemOpen
  \bibfield  {author} {\bibinfo {author} {\bibfnamefont {L.}~\bibnamefont
  {Pollet}}, \bibinfo {author} {\bibfnamefont {M.}~\bibnamefont {Boninsegni}},
  \bibinfo {author} {\bibfnamefont {A.~B.}\ \bibnamefont {Kuklov}}, \bibinfo
  {author} {\bibfnamefont {N.~V.}\ \bibnamefont {Prokof'ev}}, \bibinfo {author}
  {\bibfnamefont {B.~V.}\ \bibnamefont {Svistunov}}, \ and\ \bibinfo {author}
  {\bibfnamefont {M.}~\bibnamefont {Troyer}},\ }\bibfield  {title} {\enquote
  {\bibinfo {title} {Superfluidity of grain boundaries in solid
  $^{4}\mathrm{He}$},}\ }\href {\doibase 10.1103/PhysRevLett.98.135301}
  {\bibfield  {journal} {\bibinfo  {journal} {Phys. Rev. Lett.}\ }\textbf
  {\bibinfo {volume} {98}},\ \bibinfo {pages} {135301} (\bibinfo {year}
  {2007})}\BibitemShut {NoStop}%
\bibitem [{\citenamefont {Boninsegni}\ and\ \citenamefont
  {Ceperley}(1996)}]{Boninsegni1996}%
  \BibitemOpen
  \bibfield  {author} {\bibinfo {author} {\bibfnamefont {Massimo}\ \bibnamefont
  {Boninsegni}}\ and\ \bibinfo {author} {\bibfnamefont {David~M.}\ \bibnamefont
  {Ceperley}},\ }\bibfield  {title} {\enquote {\bibinfo {title} {Density
  fluctuations in liquid4he. path integrals and maximum entropy},}\ }\href
  {\doibase 10.1007/BF00751861} {\bibfield  {journal} {\bibinfo  {journal}
  {Journal of Low Temperature Physics}\ }\textbf {\bibinfo {volume} {104}},\
  \bibinfo {pages} {339--357} (\bibinfo {year} {1996})}\BibitemShut {NoStop}%
\bibitem [{\citenamefont {Filinov}\ and\ \citenamefont
  {Bonitz}(2012)}]{Filinov_PRA_2012}%
  \BibitemOpen
  \bibfield  {author} {\bibinfo {author} {\bibfnamefont {A.}~\bibnamefont
  {Filinov}}\ and\ \bibinfo {author} {\bibfnamefont {M.}~\bibnamefont
  {Bonitz}},\ }\bibfield  {title} {\enquote {\bibinfo {title} {Collective and
  single-particle excitations in two-dimensional dipolar bose gases},}\ }\href
  {\doibase 10.1103/PhysRevA.86.043628} {\bibfield  {journal} {\bibinfo
  {journal} {Phys. Rev. A}\ }\textbf {\bibinfo {volume} {86}},\ \bibinfo
  {pages} {043628} (\bibinfo {year} {2012})}\BibitemShut {NoStop}%
\bibitem [{\citenamefont {{Dornheim}}\ and\ \citenamefont
  {{Vorberger}}(2020)}]{Dornheim_Vorberger_finite_size_2020}%
  \BibitemOpen
  \bibfield  {author} {\bibinfo {author} {\bibfnamefont {Tobias}\ \bibnamefont
  {{Dornheim}}}\ and\ \bibinfo {author} {\bibfnamefont {Jan}\ \bibnamefont
  {{Vorberger}}},\ }\bibfield  {title} {\enquote {\bibinfo {title}
  {{Finite-size effects in the reconstruction of dynamic properties from ab
  initio path integral Monte-Carlo simulations}},}\ }\href@noop {} {\bibfield
  {journal} {\bibinfo  {journal} {arXiv e-prints}\ ,\ \bibinfo {eid}
  {arXiv:2004.13429}} (\bibinfo {year} {2020})},\ \Eprint
  {http://arxiv.org/abs/2004.13429} {arXiv:2004.13429 [cond-mat.str-el]}
  \BibitemShut {NoStop}%
\bibitem [{\citenamefont {Kora}\ and\ \citenamefont
  {Boninsegni}(2018)}]{Boninsegni_PRB_2018}%
  \BibitemOpen
  \bibfield  {author} {\bibinfo {author} {\bibfnamefont {Youssef}\ \bibnamefont
  {Kora}}\ and\ \bibinfo {author} {\bibfnamefont {Massimo}\ \bibnamefont
  {Boninsegni}},\ }\bibfield  {title} {\enquote {\bibinfo {title} {Dynamic
  structure factor of superfluid $^{4}\mathrm{He}$ from quantum monte carlo:
  Maximum entropy revisited},}\ }\href {\doibase 10.1103/PhysRevB.98.134509}
  {\bibfield  {journal} {\bibinfo  {journal} {Phys. Rev. B}\ }\textbf {\bibinfo
  {volume} {98}},\ \bibinfo {pages} {134509} (\bibinfo {year}
  {2018})}\BibitemShut {NoStop}%
\bibitem [{\citenamefont {Groth}\ \emph {et~al.}(2016)\citenamefont {Groth},
  \citenamefont {Schoof}, \citenamefont {Dornheim},\ and\ \citenamefont
  {Bonitz}}]{Groth_PRB_2016}%
  \BibitemOpen
  \bibfield  {author} {\bibinfo {author} {\bibfnamefont {S.}~\bibnamefont
  {Groth}}, \bibinfo {author} {\bibfnamefont {T.}~\bibnamefont {Schoof}},
  \bibinfo {author} {\bibfnamefont {T.}~\bibnamefont {Dornheim}}, \ and\
  \bibinfo {author} {\bibfnamefont {M.}~\bibnamefont {Bonitz}},\ }\bibfield
  {title} {\enquote {\bibinfo {title} {Ab initio quantum monte carlo
  simulations of the uniform electron gas without fixed nodes},}\ }\href
  {\doibase 10.1103/PhysRevB.93.085102} {\bibfield  {journal} {\bibinfo
  {journal} {Phys. Rev. B}\ }\textbf {\bibinfo {volume} {93}},\ \bibinfo
  {pages} {085102} (\bibinfo {year} {2016})}\BibitemShut {NoStop}%
\bibitem [{\citenamefont {Prokof'ev}\ and\ \citenamefont
  {Svistunov}(2008)}]{sign_blessing}%
  \BibitemOpen
  \bibfield  {author} {\bibinfo {author} {\bibfnamefont {Nikolay}\ \bibnamefont
  {Prokof'ev}}\ and\ \bibinfo {author} {\bibfnamefont {Boris}\ \bibnamefont
  {Svistunov}},\ }\bibfield  {title} {\enquote {\bibinfo {title} {Fermi-polaron
  problem: Diagrammatic monte carlo method for divergent sign-alternating
  series},}\ }\href {\doibase 10.1103/PhysRevB.77.020408} {\bibfield  {journal}
  {\bibinfo  {journal} {Phys. Rev. B}\ }\textbf {\bibinfo {volume} {77}},\
  \bibinfo {pages} {020408} (\bibinfo {year} {2008})}\BibitemShut {NoStop}%
\bibitem [{\citenamefont {Ellenberger}\ \emph {et~al.}(2006)\citenamefont
  {Ellenberger}, \citenamefont {Ihn}, \citenamefont {Yannouleas}, \citenamefont
  {Landman}, \citenamefont {Ensslin}, \citenamefont {Driscoll},\ and\
  \citenamefont {Gossard}}]{Ellenberger96}%
  \BibitemOpen
  \bibfield  {author} {\bibinfo {author} {\bibfnamefont {C.}~\bibnamefont
  {Ellenberger}}, \bibinfo {author} {\bibfnamefont {T.}~\bibnamefont {Ihn}},
  \bibinfo {author} {\bibfnamefont {C.}~\bibnamefont {Yannouleas}}, \bibinfo
  {author} {\bibfnamefont {U.}~\bibnamefont {Landman}}, \bibinfo {author}
  {\bibfnamefont {K.}~\bibnamefont {Ensslin}}, \bibinfo {author} {\bibfnamefont
  {D.}~\bibnamefont {Driscoll}}, \ and\ \bibinfo {author} {\bibfnamefont
  {A.~C.}\ \bibnamefont {Gossard}},\ }\bibfield  {title} {\enquote {\bibinfo
  {title} {Excitation spectrum of two correlated electrons in a lateral quantum
  dot with negligible zeeman splitting},}\ }\href {\doibase
  10.1103/PhysRevLett.96.126806} {\bibfield  {journal} {\bibinfo  {journal}
  {Phys. Rev. Lett.}\ }\textbf {\bibinfo {volume} {96}},\ \bibinfo {pages}
  {126806} (\bibinfo {year} {2006})}\BibitemShut {NoStop}%
\bibitem [{\citenamefont {Mezzacapo}\ and\ \citenamefont
  {Boninsegni}(2007)}]{mezza}%
  \BibitemOpen
  \bibfield  {author} {\bibinfo {author} {\bibfnamefont {F.}~\bibnamefont
  {Mezzacapo}}\ and\ \bibinfo {author} {\bibfnamefont {M.}~\bibnamefont
  {Boninsegni}},\ }\bibfield  {title} {\enquote {\bibinfo {title} {Structure,
  superfluidity, and quantum melting of hydrogen clusters},}\ }\href
  {https://journals.aps.org/pra/abstract/10.1103/PhysRevA.75.033201} {\bibfield
   {journal} {\bibinfo  {journal} {Phys. Rev. A}\ }\textbf {\bibinfo {volume}
  {75}},\ \bibinfo {pages} {033201} (\bibinfo {year} {2007})}\BibitemShut
  {NoStop}%
\bibitem [{\citenamefont {Sakkos}\ \emph {et~al.}(2009)\citenamefont {Sakkos},
  \citenamefont {Casulleras},\ and\ \citenamefont {Boronat}}]{Sakkos_JCP_2009}%
  \BibitemOpen
  \bibfield  {author} {\bibinfo {author} {\bibfnamefont {K.}~\bibnamefont
  {Sakkos}}, \bibinfo {author} {\bibfnamefont {J.}~\bibnamefont {Casulleras}},
  \ and\ \bibinfo {author} {\bibfnamefont {J.}~\bibnamefont {Boronat}},\
  }\bibfield  {title} {\enquote {\bibinfo {title} {High order chin actions in
  path integral monte carlo},}\ }\href {\doibase 10.1063/1.3143522} {\bibfield
  {journal} {\bibinfo  {journal} {The Journal of Chemical Physics}\ }\textbf
  {\bibinfo {volume} {130}},\ \bibinfo {pages} {204109} (\bibinfo {year}
  {2009})},\ \Eprint {http://arxiv.org/abs/https://doi.org/10.1063/1.3143522}
  {https://doi.org/10.1063/1.3143522} \BibitemShut {NoStop}%
\bibitem [{\citenamefont {Brualla}\ \emph {et~al.}(2004)\citenamefont
  {Brualla}, \citenamefont {Sakkos}, \citenamefont {Boronat},\ and\
  \citenamefont {Casulleras}}]{Brualla_JCP_2004}%
  \BibitemOpen
  \bibfield  {author} {\bibinfo {author} {\bibfnamefont {L.}~\bibnamefont
  {Brualla}}, \bibinfo {author} {\bibfnamefont {K.}~\bibnamefont {Sakkos}},
  \bibinfo {author} {\bibfnamefont {J.}~\bibnamefont {Boronat}}, \ and\
  \bibinfo {author} {\bibfnamefont {J.}~\bibnamefont {Casulleras}},\ }\bibfield
   {title} {\enquote {\bibinfo {title} {Higher order and infinite
  trotter-number extrapolations in path integral monte carlo},}\ }\href
  {\doibase 10.1063/1.1760512} {\bibfield  {journal} {\bibinfo  {journal} {The
  Journal of Chemical Physics}\ }\textbf {\bibinfo {volume} {121}},\ \bibinfo
  {pages} {636--643} (\bibinfo {year} {2004})},\ \Eprint
  {http://arxiv.org/abs/https://doi.org/10.1063/1.1760512}
  {https://doi.org/10.1063/1.1760512} \BibitemShut {NoStop}%
\bibitem [{\citenamefont {Mak}\ \emph {et~al.}(1998)\citenamefont {Mak},
  \citenamefont {Egger},\ and\ \citenamefont {Weber-Gottschick}}]{MLB_1}%
  \BibitemOpen
  \bibfield  {author} {\bibinfo {author} {\bibfnamefont {C.~H.}\ \bibnamefont
  {Mak}}, \bibinfo {author} {\bibfnamefont {R.}~\bibnamefont {Egger}}, \ and\
  \bibinfo {author} {\bibfnamefont {H.}~\bibnamefont {Weber-Gottschick}},\
  }\bibfield  {title} {\enquote {\bibinfo {title} {Multilevel blocking approach
  to the fermion sign problem in path-integral monte carlo simulations},}\
  }\href {\doibase 10.1103/PhysRevLett.81.4533} {\bibfield  {journal} {\bibinfo
   {journal} {Phys. Rev. Lett.}\ }\textbf {\bibinfo {volume} {81}},\ \bibinfo
  {pages} {4533--4536} (\bibinfo {year} {1998})}\BibitemShut {NoStop}%
\bibitem [{\citenamefont {Dikovsky}\ and\ \citenamefont {Mak}(2001)}]{MLB_2}%
  \BibitemOpen
  \bibfield  {author} {\bibinfo {author} {\bibfnamefont {Mikhail~V.}\
  \bibnamefont {Dikovsky}}\ and\ \bibinfo {author} {\bibfnamefont {C.~H.}\
  \bibnamefont {Mak}},\ }\bibfield  {title} {\enquote {\bibinfo {title}
  {Analysis of the multilevel blocking approach to the fermion sign problem:
  Accuracy, errors, and practice},}\ }\href {\doibase
  10.1103/PhysRevB.63.235105} {\bibfield  {journal} {\bibinfo  {journal} {Phys.
  Rev. B}\ }\textbf {\bibinfo {volume} {63}},\ \bibinfo {pages} {235105}
  (\bibinfo {year} {2001})}\BibitemShut {NoStop}%
\bibitem [{\citenamefont {Schoof}(2016)}]{Schoof_PHD}%
  \BibitemOpen
  \bibfield  {author} {\bibinfo {author} {\bibfnamefont {Tim}\ \bibnamefont
  {Schoof}},\ }\emph {\bibinfo {title} {Configuration Path Integral Monte
  Carlo: Ab initio simulations of fermions in the warm dense matter regime}},\
  \href@noop {} {Ph.D. thesis},\ \bibinfo  {school}
  {Christian-Albrechts-Universit\"at zu Kiel}, \bibinfo {address} {Kiel,
  Germany} (\bibinfo {year} {2016})\BibitemShut {NoStop}%
\bibitem [{\citenamefont {Invernizzi}\ \emph {et~al.}(2020)\citenamefont
  {Invernizzi}, \citenamefont {Piaggi},\ and\ \citenamefont
  {Parrinello}}]{Invernizzi2020}%
  \BibitemOpen
  \bibfield  {author} {\bibinfo {author} {\bibfnamefont {Michele}\ \bibnamefont
  {Invernizzi}}, \bibinfo {author} {\bibfnamefont {Pablo~M.}\ \bibnamefont
  {Piaggi}}, \ and\ \bibinfo {author} {\bibfnamefont {Michele}\ \bibnamefont
  {Parrinello}},\ }\bibfield  {title} {\enquote {\bibinfo {title} {{A Unified
  Approach to Enhanced Sampling}},}\ }\href {http://arxiv.org/abs/2007.03055}
  {\bibfield  {journal} {\bibinfo  {journal} {arXiv e-prints}\ } (\bibinfo
  {year} {2020})},\ \Eprint {http://arxiv.org/abs/2007.03055}
  {arXiv:2007.03055} \BibitemShut {NoStop}%
\end{thebibliography}%

\end{document}